\documentclass[
 twocolumn,
 nofootinbib,
 amsmath,
 amssymb,
 aps,
]{revtex4}

\usepackage{graphicx} 
\usepackage{dcolumn} 
\usepackage{bm} 

\usepackage{hyperref} 

\usepackage{xcolor}
\allowdisplaybreaks

\usepackage{mathtools}
\DeclareMathOperator*{\argmin}{arg\,min}

\usepackage{amsthm}
\theoremstyle{plain}
\newtheorem{theorem}{Theorem}

\newtheorem{lemma}[theorem]{Lemma}

\theoremstyle{definition}

\theoremstyle{remark}

\usepackage{tikz}
\usetikzlibrary{automata, arrows.meta, positioning}
\usetikzlibrary{cd}
\usetikzlibrary{quantikz}

\usepackage{coffeestains}

\usepackage{xr}
\makeatletter
\@addtoreset{equation}{section}
\makeatother

\usepackage{chngcntr}
\makeatletter
\g@addto@macro\appendix{
  \counterwithin{equation}{section}
  
  \counterwithin{figure}{section}
  
}
\makeatother

\begin{document}

\preprint{APS/123-QED}

\title{Information geometry of nonmonotonic quantum natural gradient}

\author{Hideyuki Miyahara}

\email{miyahara@ist.hokudai.ac.jp, hmiyahara512@gmail.com}


\affiliation{
	Graduate School of Information Science and Technology,
	Hokkaido University, Sapporo, Hokkaido 060-0814, Japan
}




\date{\today}

\begin{abstract}
	Natural gradient is an advanced optimization method based on information geometry, where the Fisher metric plays a crucial role.
	Its quantum counterpart, known as quantum natural gradient (QNG), employs the symmetric logarithmic derivative (SLD) metric, one of the quantum Fisher metrics.
	While quantization in physics is typically well-defined via the canonical commutation relations, the quantization of information-theoretic quantities introduces inherent arbitrariness.
	To resolve this ambiguity, monotonicity has been used as a guiding principle for constructing geometries in physics, as it aligns with physical intuition.
	Recently, a variant of QNG, which we refer to as nonmonotonic QNG in this paper, was proposed by relaxing the monotonicity condition.
	It was shown to achieve faster convergence compared to conventional QNG.
	In this paper, we investigate the properties of nonmonotonic QNG.
	To ensure the paper is self-contained, we first demonstrate that the SLD metric is locally optimal under the monotonicity condition and that non-monotone quantum Fisher metrics can lead to faster convergence in QNG.
	Previous studies primarily relied on a specific type of quantum divergence and assumed that density operators are full-rank.
	Here, we explicitly consider an alternative quantum divergence and extend the analysis to non-full-rank cases.
	Additionally, we explore how geometries can be designed using Petz functions, given that quantum Fisher metrics are characterized through them.
	Finally, we present numerical simulations comparing different quantum Fisher metrics in the context of parameter estimation problems in quantum circuit learning.
\end{abstract}


\maketitle



\section{Introduction}

Information geometry (IG) is a branch of differential geometry that studies probability distributions and positive measures, providing new perspectives on probability and information theory, such as dually flat geometry~\cite{Amari_002, Amari_004}.
Moreover, IG and physics are becoming increasingly intertwined, with applications in stochastic thermodynamics and chemical reaction networks~\cite{Schmiedl_001, Ito_001, Yoshimura_001, Mizohata_001}.
The quantum counterpart of IG, known as quantum information geometry (QIG), extends these ideas to quantum states, analyzing their geometry in a manner analogous to the information geometric treatment of probability distributions~\cite{Amari_002, Hayashi_003, Lambert_001}.
In quantum mechanics, non-commutativity is a fundamental distinction from classical mechanics.
While canonical quantization is naturally determined in quantum mechanics, QIG lacks a uniquely defined quantization procedure, leading to multiple possible geometries.
Interestingly, when QIG is based on canonical quantization, its geometric structure closely resembles that of classical IG.
The quantum nature of QIG arises from the noncommutativity of parameterized density operators and their derivatives, expressed as $[\hat{\rho}_\theta, X \hat{\rho}_\theta] \ne 0$.
The arbitrariness in QIG stems from the nonuniqueness of defining the logarithmic derivative of density operators, which influences the resulting geometric framework.

The scientific impact of information geometry (IG) extends beyond theoretical mathematics and it has practical applications, notably in the development of the natural gradient (NG), a powerful continuous optimization method widely used in applied sciences~\cite{Amari_005, Martens_001}.
In NG, the Fisher metric plays a crucial role.
A quantum extension of NG, known as quantum natural gradient (QNG)~\cite{Stokes_001, Koczor_001}, has been introduced and is gaining popularity for parameter estimation in quantum states across various applications, including the variational Monte Carlo (VMC) method~\cite{Scherer_001}, quantum circuit learning~\cite{Mitarai_001, Schuld_001}, and tensor networks~\cite{Orus_001}.
Similar to how the Fisher metric underpins NG, quantum Fisher metrics play a fundamental role in QNG.
Specifically, the symmetric logarithmic derivative (SLD) metric, one of the quantum Fisher metrics, is commonly employed in QNG.
As mentioned earlier, the quantization of the Fisher metric is not unique, leading to the identification of alternative quantum Fisher metrics such as the real right logarithmic derivative (rRLD) and Bogoliubov-Kubo-Mori (BKM) metrics.
Notably, the variety of quantum Fisher metrics can be experimentally determined using linear response theory~\cite{Shitara_001}.
QNG with the SLD metric is also known as the stochastic reconfiguration (SR) algorithm, which stabilizes the VMC method in condensed matter physics~\cite{Sorella_001, Sorella_002, Sorella_003, Becca_001, Mazzola_001, Park_001, Xie_001}.
In addition, the time-dependent VMC method, used for simulating the dynamics of quantum systems, has been recently generalized using QNG~\cite{Ido_001, Stokes_002}.
Beyond quantum physics, quantum mechanically designed metrics have also found applications in classical machine learning~\cite{Lopatnikova_001, Miyahara_001, Miyahara_002, Miyahara_003}.
Recently, QNG has been further generalized within the framework of quantum information geometry (QIG)~\cite{Sasaki_001}.
In Ref.~\cite{Sasaki_001}, it was shown that SLD-based QNG, which is equivalent to the method proposed in Refs.~\cite{Stokes_001, Koczor_001}, is optimal under the monotonicity condition.
Furthermore, QNG without the monotonicity constraint demonstrated better performance than its SLD-based counterpart.

In this paper, we generalize SLD-based QNG within the framework of quantum information geometry (QIG) and propose a variety of QNG approaches.
Monotonicity ensures that information is not gained under completely positive and trace-preserving (CPTP) maps, making it a widely accepted guiding principle for defining geometries in physics.
However, its necessity from an optimization perspective remains unclear.
We first establish that monotonicity serves as a condition for the SLD metric to be optimal in terms of the convergence speed of QNG.
Next, we extend QNG by incorporating quantum divergences, leading to non-monotone quantum Fisher metrics.
We theoretically demonstrate that QNG based on non-monotone quantum Fisher metrics converges faster than its SLD-based counterpart.
In Ref.~\cite{Sasaki_001}, the full-rank condition for density operators was assumed.
However, in practical applications, pure states or non-full-rank density operators are often of interest.
We explicitly derive the quantum Fisher metric for general density operators and show that an additional condition on the Petz function is required for non-full-rank cases.
Moreover, in numerical simulations, the diagonal or block-diagonal approximation of the Fisher metric is commonly used in neural networks (NNs) due to the high computational cost of inverting the full Fisher metric~\cite{Gupta_001, Hennig_001}.
We demonstrate that under this approximation, our findings remain valid.
Additionally, we propose a flexible method for designing geometries that optimize QNG efficiency by directly constructing Petz functions.
Finally, numerical simulations confirm that non-monotone quantum Fisher metrics outperform the SLD metric in terms of convergence speed.

The remainder of this paper is organized as follows.
Sections~\ref{main_sec_natural_gradient_001_001} and \ref{main_sec_quantum_natural_gradient_001_001} review NG and QNG, respectively.
Section~\ref{main_sec_bra-ket_notation_001_001} reformulates quantum Fisher metrics using bra-ket notation.
Section~\ref{main_sec_monotonicity_quantum_Fisher_metric_001_001} explains the monotonicity of metrics, divergences, and operator functions.
Section~\ref{main_sec_speed_QNG_001_001} elucidates the relationship between quantum Fisher metrics and the speed of QNG.
Section~\ref{main_sec_design_of_geometry_via_Petz_functions_001_001} demonstrates a design method for geometries suitable for QNG.
Section~\ref{main_sec_numerical_simulations_001_001} presents numerical simulations comparing various quantum Fisher metrics.
Section~\ref{main_sec_discussions_001_001} discusses the advantages of QNG over Newton's method.
Finally, Sec.~\ref{main_sec_conclusions_001_001} concludes this paper.

\section{Natural gradient} \label{main_sec_natural_gradient_001_001}

This section aims to review NG~\cite{Amari_005}.
We first define the Fisher metric and introduce classical divergences between probability distributions: the Kullback-Leibler (KL) divergence and the rescaled classical R\'enyi divergence.
Then, we derive NG by considering an optimization problem under a constraint with respect to the KL divergence.

\subsection{Discrete probability distribution}

In IG, probability distributions that belong to the exponential family are intensively studied because of their elegant structure called dually flatness~\cite{Amari_002, Amari_004}.
For example, the canonical distribution, which is of special importance in physics since it enables us to compute expected values at equilibrium, also belongs to the exponential family.
Among members of the exponential family, the following form of a parameterized probability distribution is the simplest one:
\begin{align}
	p_\theta (x) & \coloneqq \sum_{i = 1}^N p_i \delta_{i, x}, \label{main_eq_def_parameterized_probability_distribution_001_001}
\end{align}
where $\delta_{i, x}$ is the Kronecker delta function defined as
\begin{align}
	\delta_{i, x} & \coloneqq
	\begin{cases}
		1 & (i = x),   \\
		0 & (i \ne x).
	\end{cases}
\end{align}
Here, $\theta \in \mathbb{R}^{N_\theta}$ corresponds to $\{ p_i \}_{i = 1}^N$ with
\begin{align}
	\sum_{i = 1}^N p_i = 1. \label{main_eq_condition_parameter_discrete_distribution_001_001}
\end{align}
Equation~\eqref{main_eq_def_parameterized_probability_distribution_001_001} is often called a discrete probability distribution and will play an important role in introducing a quantum counterpart.
In most cases, $N_\theta$, which is the number of dimensions of $\theta$, is much smaller than that of $N$, which is the number of states that a stochastic variable of interest may take, while $N_\theta = N-1$ in Eq.~\eqref{main_eq_def_parameterized_probability_distribution_001_001}.
Note that $N_\theta$ is the degrees of freedom in parameter space, and not the number of parameters since in general parameters can be introduced redundantly.
For example, we can impose $p_N = 1 - \sum_{i=1}^{N-1} p_i$, which satisfies Eq.~\eqref{main_eq_condition_parameter_discrete_distribution_001_001} to simplify Eq.~\eqref{main_eq_def_parameterized_probability_distribution_001_001}; then, Eq.~\eqref{main_eq_def_parameterized_probability_distribution_001_001} becomes
\begin{align}
	p_\theta (x) & = \sum_{i = 1}^{N-1} p_i \delta_{i, x} + \bigg( 1 - \sum_{i=1}^{N-1} p_i \bigg) \delta_{N, x}. \label{main_eq_parameterized_probability_distribution_001_001}
\end{align}

\subsection{Fisher metric}

Here, we introduce the Fisher metric, which plays a crucial role in NG.
Assume that we have $\theta \in \mathbb{R}^{N_\theta}$.
We define tangent vectors as
\begin{align}
	X & \coloneqq X^i \partial_i,
\end{align}
where, for $i = 1, 2, \dots, N_\theta$,
\begin{align}
	\partial_i & \coloneqq \frac{\partial}{\partial \theta^i}. \label{main_eq_def_partial_differential_001_001}
\end{align}
Then, the Fisher metric between two tangent vectors $X$ and $Y$ is defined as
\begin{align}
	g_{p_\theta (\cdot)} (X, Y) & \coloneqq \sum_{x=1}^N X_{p_\theta (\cdot)}^\mathrm{m} (x) Y_{p_\theta (\cdot)}^\mathrm{e} (x), \label{main_eq_def_Fisher_metric_001_001}
\end{align}
where the e- and m-representations of $X$ are given, respectively, by
\begin{subequations} \label{main_eq_classical_X_e_X_m_001_001}
	\begin{align}
		X_{p_\theta (\cdot)}^\mathrm{m} (\cdot) & \coloneqq X p_\theta (\cdot),                                           \\
		X_{p_\theta (\cdot)}^\mathrm{e} (\cdot) & \coloneqq X \ln p_\theta (\cdot). \label{main_eq_classical_X_e_001_001}
	\end{align}
\end{subequations}
Furthermore, the Fisher metric, Eq.~\eqref{main_eq_def_Fisher_metric_001_001}, is also expressed as
\begin{align}
	g_{p_\theta (\cdot)} (X, Y) & = \sum_{x=1}^N p_\theta (x) (X \ln p_\theta (x)) (Y \ln p_\theta (x))                                              \\
	                            & = \sum_{x=1}^N \frac{1}{p_\theta (x)} (X p_\theta (x)) (Y p_\theta (x)). \label{main_eq_def_Fisher_metric_001_002}
\end{align}

\subsection{Classical divergences}

Divergences between probability distributions are pseudodistances that quantify the difference between two probability distributions and ubiquitously appear in optimization and machine learning.
Note that, in general, divergences do not satisfy the symmetric property while distances do, and we call divergences between probability distributions \textit{classical divergences} since we also deal with differences between density operators.
Here, we review the KL divergence and the rescaled classical R\'enyi divergence for the preparation of NG.

\subsubsection{KL divergence}

The definition of the KL divergence reads~\cite{Kullback_001, Csiszar_001}
\begin{align}
	D_\mathrm{KL} (p_{\bar{\theta}} (\cdot) \| p_\theta (\cdot)) & \coloneqq \sum_{x=1}^N p_{\bar{\theta}} (x) \ln \frac{p_{\bar{\theta}} (x)}{p_\theta (x)}. \label{main_eq_def_KL_divergence_001_001}
\end{align}
Note that the axiomatic characterization of the KL divergence is discussed in Ref.~\cite{Csiszar_002}.
To clarify the relation between the KL divergence, Eq.~\eqref{main_eq_def_KL_divergence_001_001}, and the Fisher metric, Eq.~\eqref{main_eq_def_Fisher_metric_001_001}, we consider the infinitesimal shift of $p_\theta (x)$ by $\Delta \theta$:
\begin{align}
	p_{\theta + \Delta \theta} (x) & \approx p_\theta (x) + \nabla_\theta p_\theta (x) \cdot \Delta \theta,
\end{align}
where
\begin{align}
	\nabla_\theta & \coloneqq \bigg[ \frac{\partial}{\partial \theta^1}, \frac{\partial}{\partial \theta^2}, \dots, \frac{\partial}{\partial \theta^{N_\theta}} \bigg]^\intercal, \\
	\Delta \theta & \coloneqq [\Delta \theta^1, \Delta \theta^2, \dots, \Delta \theta^{N_\theta}]^\intercal.
\end{align}
Hereafter, $\frac{\partial}{\partial \theta^i}$ is often written as $\partial_i$.
The KL divergence between $p_{\theta + \Delta \theta} (\cdot)$ and $p_\theta (\cdot)$ is given by~\footnote{See Appendix~\ref{main_appendix_Fisher_metric_KL_001_001} for the derivation of Eq.~\eqref{main_eq_Delta_theta_shift_KL_001_001}.}
\begin{align}
	D_\mathrm{KL} (p_{\theta + \Delta \theta} (\cdot) \| p_\theta (\cdot)) & = \sum_{x=1}^N p_{\theta + \Delta \theta} (x) \ln \frac{p_{\theta + \Delta \theta} (x)}{p_\theta (x)}                            \\
	                                                                       & = \frac{1}{2} \sum_{i, j = 1}^{N_\theta} g_{p_\theta (\cdot)} (\partial_i, \partial_j) \Delta \theta^i \Delta \theta^j \nonumber \\
	                                                                       & \quad + \mathcal{O} (\| \Delta \theta \|^3). \label{main_eq_Delta_theta_shift_KL_001_001}
\end{align}
From Eq.~\eqref{main_eq_Delta_theta_shift_KL_001_001}, we have the following relation:
\begin{align}
	g_{p_\theta (\cdot)} (X, Y) & = \bar{X} \bar{Y} D_\mathrm{KL} (p_{\bar{\theta}} (\cdot) \| p_\theta (\cdot)) |_{\bar{\theta} = \theta},
\end{align}
where, similarly to Eq.~\eqref{main_eq_def_partial_differential_001_001}, for $i = 1, 2, \dots, N_\theta$,
\begin{align}
	\bar{\partial}_i \coloneqq \frac{\partial}{\partial \bar{\theta}^i}.
\end{align}
In the case of Eq.~\eqref{main_eq_parameterized_probability_distribution_001_001}, we have
\begin{align}
	D_\mathrm{KL} (p_{\bar{\theta}} (\cdot) \| p_\theta (\cdot))                                                             & = \sum_{i=1}^N \bar{p}_i \ln \frac{\bar{p}_i}{p_i},                                                   \\
	\bar{\partial}_i \bar{\partial}_j D_\mathrm{KL} (p_{\bar{\theta}} (\cdot) \| p_\theta (\cdot)) |_{\bar{\theta} = \theta} & = \delta_{i, j} p_i^{-1} + p_N^{-1}. \label{main_eq_second_derivative_rescaled_KL_divergence_001_001}
\end{align}
Similarly to Eq.~\eqref{main_eq_parameterized_probability_distribution_001_001}, we have defined
\begin{align}
	p_{\bar{\theta}} (x) & \coloneqq \sum_{i = 1}^N \bar{p}_i \delta_{i, x}, \label{main_eq_def_parameterized_probability_distribution_001_002}
\end{align}
where $\bar{\theta} \in \mathbb{R}^{N_\theta}$ corresponds to $\{ \bar{p}_i \}_{i = 1}^{N-1}$ and $\bar{p}_N \coloneqq 1 - \sum_{i=1}^{N-1} \bar{p}_i$.

\subsubsection{Rescaled classical R\'enyi divergence}

We turn our attention to the classical rescaled R\'enyi divergence~\cite{Erven_001}, which will be quantized later.
The rescaled classical R\'enyi divergence is given by
\begin{align}
	R_\alpha (p_{\bar{\theta}} (\cdot) \| p_\theta (\cdot)) & \coloneqq \frac{1}{\alpha (\alpha - 1)} \ln \sum_{x=1}^N p_{\bar{\theta}}^\alpha (x) p_\theta^{1 - \alpha} (x). \label{main_eq_def_rescaled_classical_Renyi_divergence_001_001}
\end{align}
Note that ``rescaled" refers to the additional factor $1 / \alpha$ in Eq.~\eqref{main_eq_def_rescaled_classical_Renyi_divergence_001_001} and the conventional R\'enyi divergence does not include this factor~\cite{Erven_001}.
The rescaled classical R\'enyi divergence, Eq.~\eqref{main_eq_def_rescaled_classical_Renyi_divergence_001_001}, in the limit $\alpha \to 1$ corresponds to the KL divergence, Eq.~\eqref{main_eq_def_KL_divergence_001_001}~\cite{Erven_001}:
\begin{align}
	\lim_{\alpha \to 1} R_\alpha (p_{\bar{\theta}} (\cdot) \| p_\theta (\cdot)) & = D_\mathrm{KL} (p_{\bar{\theta}} (\cdot) \| p_\theta (\cdot)).
\end{align}
Next, let us focus on the metric induced by the rescaled classical R\'enyi divergence, Eq.~\eqref{main_eq_def_rescaled_classical_Renyi_divergence_001_001}.
The rescaled classical R\'enyi divergence between $p_{\theta + \Delta \theta} (\cdot)$ and $p_\theta (\cdot)$ is computed as follows~\footnote{See Appendix~\ref{main_appendix_Fisher_metric_rescaled_classical_Renyi_001_001} for the derivation of Eq.~\eqref{main_eq_Delta_theta_shift_rescaled_classical_Renyi_001_001}.}:
\begin{align}
	R_\alpha (p_{\theta + \Delta \theta} (\cdot) \| p_\theta (\cdot)) & = \frac{1}{\alpha (\alpha - 1)} \ln \sum_{x=1}^N p_{\theta + \Delta \theta}^\alpha (x) p_\theta^{1 - \alpha} (x)                  \\
	                                                                  & = \frac{1}{2} \sum_{i, j = 1}^{N_\theta} g_{p_\theta (\cdot)} (\partial_i, \partial_j) \Delta \theta^i \Delta \theta^j \nonumber  \\
	                                                                  & \quad + \mathcal{O} (\| \Delta \theta \|^3). \label{main_eq_Delta_theta_shift_rescaled_classical_Renyi_001_001}
\end{align}
From Eq.~\eqref{main_eq_Delta_theta_shift_rescaled_classical_Renyi_001_001}, the second derivative of Eq.~\eqref{main_eq_def_rescaled_classical_Renyi_divergence_001_001} is computed as
\begin{align}
	\bar{X} \bar{Y} R_\alpha (p_{\bar{\theta}} (\cdot) \| p_\theta (\cdot)) |_{\bar{\theta} = \theta} & = \bar{X} \bar{Y} D_\mathrm{KL} (p_{\bar{\theta}} (\cdot) \| p_\theta (\cdot)) |_{\bar{\theta} = \theta} \\
	                                                                                                  & = g_{p_\theta (\cdot)} (X, Y). \label{main_eq_second_derivative_rescaled_classical_Renyi_001_001}
\end{align}
Equation~\eqref{main_eq_second_derivative_rescaled_classical_Renyi_001_001} implies that the second derivative of Eq.~\eqref{main_eq_def_rescaled_classical_Renyi_divergence_001_001} does not depend on $\alpha$.
In the case of Eqs.~\eqref{main_eq_parameterized_probability_distribution_001_001} and \eqref{main_eq_def_parameterized_probability_distribution_001_002}, Eq.~\eqref{main_eq_def_rescaled_classical_Renyi_divergence_001_001} reads
\begin{align}
	R_\alpha (p_{\bar{\theta}} (\cdot) \| p_\theta (\cdot)) & = \frac{1}{\alpha (\alpha - 1)} \ln \sum_{i = 1}^N \bar{p}_i^\alpha p_i^{1 - \alpha}. \label{main_eq_def_rescaled_classical_Renyi_divergence_discrete_distribution_001_001}
\end{align}
In addition, Eq.~\eqref{main_eq_second_derivative_rescaled_classical_Renyi_001_001} becomes~\footnote{See Appendix~\ref{main_sec_Fisher_metric_rescaled_Renyi_divergence_001_001} for details.}
\begin{align}
	\bar{\partial}_i \bar{\partial}_j R_\alpha (p_{\bar{\theta}} (\cdot) \| p_\theta (\cdot)) |_{\bar{\theta} = \theta} & = \delta_{i, j} p_i^{-1} + p_N^{-1}. \label{main_eq_derivative_rescaled_classical_Renyi_divergence_discrete_distribution_001_001}
\end{align}

In the following section, we will see that the quantum counterpart of Eq.~\eqref{main_eq_def_rescaled_classical_Renyi_divergence_001_001} yields a different metric in contrast to the classical case in which the induced Fisher metric does not depend on $\alpha$ as shown in Eq.~\eqref{main_eq_second_derivative_rescaled_classical_Renyi_001_001}, and this fact will play an important role in QNG.

\subsection{Natural gradient}

We here review NG, which is one of the optimization methods for continuous variables motivated by IG~\cite{Amari_005}.
Let us consider the minimization problem of $L (\theta)$:
\begin{align}
	\min_{\theta \in \mathbb{R}^{N_\theta}} L (\theta). \label{main_eq_def_optimization_problem_001_001}
\end{align}
Then, the dynamics of NG is given by
\begin{subequations} \label{main_optimization_problem_NG_001_001}
	\begin{align}
		\theta_{\tau+1}               & = \theta_\tau + \Delta \theta_\tau (\epsilon), \label{main_optimization_problem_NG_001_011} \\
		\Delta \theta_\tau (\epsilon) & = \argmin_{\substack{\Delta \theta \in \mathbb{R}^{N_\theta}:                               \\ R_\alpha (p_{\theta_\tau + \Delta \theta} (\cdot) \| p_{\theta_\tau} (\cdot)) \le \epsilon}} [L (\theta_\tau + \Delta \theta) - L (\theta_\tau)], \label{main_optimization_problem_NG_001_012}
	\end{align}
\end{subequations}
where $\epsilon$ is a positive number.
Note that Eq.~\eqref{main_optimization_problem_NG_001_001} is not the only approach to formulating NG; rather, it is just one of the formulations that lead to NG.
Furthermore, we focus on Eq.~\eqref{main_eq_def_rescaled_classical_Renyi_divergence_001_001}, but we do not need to limit ourselves to Eq.~\eqref{main_eq_def_rescaled_classical_Renyi_divergence_001_001}.

We derive the update equation of NG by solving Eq.~\eqref{main_optimization_problem_NG_001_001}.
First, we transform Eq.~\eqref{main_optimization_problem_NG_001_012} by using the lowest-order approximation:
\begin{align}
	\Delta \theta_\tau (\epsilon) & \approx \argmin_{\substack{\Delta \theta \in \mathbb{R}^{N_\theta}: \\ \frac{1}{2} \Delta \theta^\intercal G (\theta_\tau) \Delta \theta \le \epsilon}} \nabla_\theta L (\theta_\tau)^\intercal \Delta \theta, \label{main_optimization_problem_NG_002_001}
\end{align}
where
\begin{align}
	\nabla_\theta L (\theta) & \coloneqq [\partial_1 L (\theta), \partial_2 L (\theta), \dots, \partial_{N_\theta} L (\theta)]^\intercal,
\end{align}
$\nabla_\theta L (\theta_\tau) \coloneqq \nabla_\theta L (\theta) |_{\theta = \theta_\tau}$, and $G (\theta_\tau)$ is the matrix representation of the Fisher metric $g_{p_{\theta_\tau} (\cdot)} (\partial_i, \partial_j)$ induced by $R_\alpha (p_{\theta_\tau + \Delta \theta} (\cdot) \| p_{\theta_\tau} (\cdot))$, Eq.~\eqref{main_eq_second_derivative_rescaled_classical_Renyi_001_001}.
To solve Eq.~\eqref{main_optimization_problem_NG_002_001}, we use the method of Lagrange multipliers~\cite{Mizohata_001}.
The Lagrange function for Eq.~\eqref{main_optimization_problem_NG_002_001} reads
\begin{align}
	\tilde{L} (\Delta \theta, \lambda) & \coloneqq \nabla_\theta L (\theta_\tau)^\intercal \Delta \theta - \lambda \bigg( \frac{1}{2} \Delta \theta^\intercal G (\theta_\tau) \Delta \theta - \epsilon \bigg). \label{main_eq_Lagrange_function_NG_001_001}
\end{align}
Then we compute the derivative of Eq.~\eqref{main_eq_Lagrange_function_NG_001_001} with respect to $\Delta \theta$:
\begin{align}
	\frac{\partial}{\partial \Delta \theta} \tilde{L} (\Delta \theta, \lambda) & = \nabla_\theta L (\theta_\tau) - \lambda G (\theta_\tau) \Delta \theta. \label{main_eq_Lagrange_function_NG_001_002}
\end{align}
Solving Eq.~\eqref{main_eq_Lagrange_function_NG_001_002}, we get:
\begin{align}
	\Delta \theta & = \frac{1}{\lambda} G^{-1} (\theta_\tau) \nabla_\theta L (\theta_\tau). \label{main_eq_Lagrange_function_NG_001_003}
\end{align}
We also compute the derivative of Eq.~\eqref{main_eq_Lagrange_function_NG_001_001} with respect to $\lambda$:
\begin{align}
	\frac{\partial}{\partial \lambda} \tilde{L} (\Delta \theta, \lambda) & = \frac{1}{2} \Delta \theta^\intercal G (\theta_\tau) \Delta \theta - \epsilon. \label{main_eq_Lagrange_function_NG_001_004}
\end{align}
Similarly, solving Eq.~\eqref{main_eq_Lagrange_function_NG_001_004}, we also get the following relation:
\begin{align}
	\lambda^2 & = \frac{\nabla_\theta L (\theta_\tau)^\intercal G^{-1} (\theta_\tau) \nabla_\theta L (\theta_\tau)}{2 \epsilon}. \label{main_eq_Lagrange_function_NG_001_005}
\end{align}
From Eqs.~\eqref{main_eq_Lagrange_function_NG_001_003} and \eqref{main_eq_Lagrange_function_NG_001_005}, Eq.~\eqref{main_optimization_problem_NG_002_001} is rewritten as
\begin{align}
	\Delta \theta_\tau (\epsilon) & = - \sqrt{\frac{2 \epsilon}{\nabla_\theta L (\theta_\tau)^\intercal G^{-1} (\theta_\tau) \nabla_\theta L (\theta_\tau)}} G^{-1} (\theta_\tau) \nabla_\theta L (\theta_\tau). \label{main_eq_update_theta_001_001}
\end{align}
Note that we have added a negative sign to Eq.~\eqref{main_eq_update_theta_001_001} to decrease the value of $L (\theta)$ by Eq.~\eqref{main_optimization_problem_NG_001_011}.
This procedure is required because the method of Lagrange multipliers provides us with only necessary conditions.
In practical applications of NG, we may use the following update rule instead of Eq.~\eqref{main_eq_update_theta_001_001}:
\begin{align}
	\Delta \theta_\tau (\epsilon) & = - \eta G^{-1} (\theta_\tau) \nabla_\theta L (\theta_\tau), \label{main_eq_update_theta_001_002}
\end{align}
where $\eta$ is a positive number.
In the following section, we extend NG to the quantum regime.

\section{Quantum natural gradient} \label{main_sec_quantum_natural_gradient_001_001}

QNG is the quantum extension of NG~\cite{Stokes_001, Koczor_001}, and the SLD metric, which corresponds to the classical Fisher metric in the classical limit, is used in its standard formulation.
In this section, we introduce the quantum Fisher metric and QNG, and generalize SLD-based QNG by considering a variety of quantum Fisher metrics.

\subsection{Parameterized quantum states}

Similarly to Eq.~\eqref{main_eq_parameterized_probability_distribution_001_001}, we consider a quantum state parameterized by $\theta \in \mathbb{R}^{N_\theta}$:
\begin{align}
	\hat{\rho}_\theta & = \sum_{i = 1}^N p_i | \psi_i \rangle \langle \psi_i |, \label{main_eq_def_parameterized_quantum_state_001_001}
\end{align}
where $\theta \in \mathbb{R}^{N_\theta}$ corresponds to $\{ p_i \}_{i = 1, 2, \dots, N-1}$ and $\{ \psi_i \}_{i=1, 2, \dots, N-1}$.
We have assumed $\mathrm{Tr} [\hat{\rho}_\theta] = 1$ and the orthonormality condition of $\langle \psi_i | \psi_j \rangle = \delta_{i, j}$, and these conditions lead, respectively, to
\begin{align}
	p_N                               & = 1 - \sum_{i=1}^{N-1} p_i,                                     \\
	| \psi_N \rangle \langle \psi_N | & = \hat{1} - \sum_{i=1}^{N-1} | \psi_i \rangle \langle \psi_i |.
\end{align}
The most important point in the quantum case is that $\{ | \psi_i \rangle \}_{i = 1}^N$ in Eq.~\eqref{main_eq_def_parameterized_quantum_state_001_001} varies depending on $\theta \in \mathbb{R}^{N_\theta}$; otherwise, there is no difference between Eqs.~\eqref{main_eq_parameterized_probability_distribution_001_001} and \eqref{main_eq_def_parameterized_quantum_state_001_001} from the viewpoint of divergences.

\subsection{Quantumness of QIG and arbitrariness of quantization}

Quantization in quantum mechanics involves imposing the canonical commutation relation (e.g., $[\hat{x}, \hat{p}] = i \hbar$ and $[\hat{\sigma}_p, \hat{\sigma}_q] = 2 i \hbar \epsilon_{pqr} \hat{\sigma}_r$ for $p, q, r = x, y, z$).
On the other hand, quantization in QIG becomes critical when $[\hat{\rho}_\theta, X \hat{\rho}_\theta] \ne 0$, where $X$ is a tangent vector~\cite{Amari_002, Hayashi_003, Lambert_001}.
That is, how a quantum state is labeled by parameters is essential.
However, $[\hat{\rho}_\theta, X \hat{\rho}_\theta] \ne 0$ leaves an arbitrariness of quantization in QIG.
The arbitrariness of quantization in QIG is fully specified by how one chooses to define the logarithmic derivatives of density operators, and it is known that the choice of these logarithmic derivatives of density operators yields different geometries~\cite{Amari_002, Hayashi_003, Lambert_001}.
In the rest of this section, we describe the difference between geometries arising from different definitions of the logarithmic derivatives of density operators.

\subsection{Quantum Fisher metric}

Here, we give the definition of the quantum Fisher metric and present its examples.

\subsubsection{General framework}

To describe the quantum extension of the Fisher metric, Eq.~\eqref{main_eq_def_Fisher_metric_001_001}, we introduce some operators~\cite{Petz_001, Petz_002, Petz_003, Sagawa_001}.
We first define $\Delta_{\hat{\rho}_\theta}$, which is often called the modular operator, as follows:
\begin{align}
	\Delta_{\hat{\rho}_\theta} \hat{A} & \coloneqq \hat{\rho}_\theta \hat{A} \hat{\rho}_\theta^{-1}. \label{main_eq_def_Delta_operator_001_001}
\end{align}
Next, we also introduce the nonlinear transformation of Eq.~\eqref{main_eq_def_Delta_operator_001_001}:
\begin{align}
	f^{-1} (\Delta_{\hat{\rho}_\theta}) \hat{A} & \coloneqq \sum_{i, j = 1}^N \frac{\langle \psi_i | \hat{A} | \psi_j \rangle}{f (p_i / p_j)} | \psi_i \rangle \langle \psi_j |. \label{main_eq_def_f_inverse_Delta_001_001}
\end{align}
Using Eq.~\eqref{main_eq_def_f_inverse_Delta_001_001}, we define the following quantities as the quantum counterpart of Eq.~\eqref{main_eq_classical_X_e_X_m_001_001}:
\begin{subequations} \label{main_eq_quantum_m_e_representations_001_001}
	\begin{align}
		\hat{X}_{\hat{\rho}_\theta}^\mathrm{m}            & \coloneqq X \hat{\rho}_\theta, \label{main_eq_quantum_m_representation_001_001}                                                                \\
		\hat{X}_{\hat{\rho}_\theta, f (\cdot)}^\mathrm{e} & \coloneqq f^{-1} (\Delta_{\hat{\rho}_\theta}) ([X \hat{\rho}_\theta] \hat{\rho}_\theta^{-1}), \label{main_eq_quantum_e_representation_001_001}
	\end{align}
\end{subequations}
where $f (\cdot): \mathbb{R}_{\ge 0} \to \mathbb{R}_{\ge 0}$ is called the Petz function, which satisfies
\begin{subequations} \label{main_eq_condition_Petz_function_001_001}
	\begin{align}
		f (1) & = 1, \label{main_eq_condition_Petz_function_001_011}            \\
		f (t) & = t f (t^{-1}). \label{main_eq_condition_Petz_function_001_012}
	\end{align}
\end{subequations}
It is known that the Petz function fully characterizes the quantum Fisher metric~\cite{Petz_001, Petz_002, Petz_003, Sagawa_001}.
Note that Eqs.~\eqref{main_eq_quantum_m_representation_001_001} and \eqref{main_eq_quantum_e_representation_001_001} satisfy
\begin{align}
	\langle \psi_i | \hat{X}_{\hat{\rho}_\theta, f (\cdot)}^\mathrm{e} | \psi_j \rangle & = \frac{\langle \psi_i | \hat{X}_{\hat{\rho}_\theta}^\mathrm{m} | \psi_j \rangle}{p_j f (p_i / p_j)}. \label{main_eq_quantum_e_representation_001_002}
\end{align}
Using Eq.~\eqref{main_eq_quantum_m_e_representations_001_001}, the quantum Fisher metric is given by
\begin{align}
	g_{\hat{\rho}_\theta, f (\cdot)} (X, Y) & \coloneqq \mathrm{Tr} [\hat{X}_{\hat{\rho}_\theta}^\mathrm{m} \hat{Y}_{\hat{\rho}_\theta, f (\cdot)}^\mathrm{e}]                                                                                                                \\
	                                        & = \sum_{i, j = 1}^N \langle \psi_i | \hat{X}_{\hat{\rho}_\theta}^\mathrm{m} | \psi_j \rangle \langle \psi_j | \hat{Y}_{\hat{\rho}_\theta, f (\cdot)}^\mathrm{e} | \psi_i \rangle. \label{main_eq_quantum_Fisher_metric_001_001}
\end{align}
The key point of the quantum Fisher metric, Eq.~\eqref{main_eq_quantum_Fisher_metric_001_001}, is its dependence on $f (\cdot)$.
Using Eq.~\eqref{main_eq_quantum_e_representation_001_002}, the quantum Fisher metric, Eq.~\eqref{main_eq_quantum_Fisher_metric_001_001}, can be transformed into
\begin{align}
	g_{\hat{\rho}_\theta, f (\cdot)} (X, Y) & = \sum_{i, j = 1}^N \frac{1}{p_j f (p_i / p_j)} \langle \psi_j | \hat{X}_{\hat{\rho}_\theta}^\mathrm{m} | \psi_i \rangle \langle \psi_i | \hat{Y}_{\hat{\rho}_\theta}^\mathrm{m} | \psi_j \rangle. \label{main_eq_quantum_Fisher_metric_001_002}
\end{align}

Finally, we discuss the significance of Eq.~\eqref{main_eq_condition_Petz_function_001_001}.
In general, the classical limit is important for understanding the relation between a quantum notion and its classical counterpart.
Equation~\eqref{main_eq_condition_Petz_function_001_011} guarantees that the classical Fisher metric, Eq.~\eqref{main_eq_def_Fisher_metric_001_002}, and the quantum Fisher metric, Eq.~\eqref{main_eq_quantum_Fisher_metric_001_002}, are identical in the classical limit.
Let us consider the classical limit, that is, the case where $\hat{\rho}_\theta$ is diagonal.
Since $f (p_i / p_i) = f (1) = 1$, we have
\begin{align}
	\hat{X}_{\hat{\rho}_\theta, f (\cdot)}^\mathrm{e} & = X \ln \hat{\rho}_\theta. \label{main_eq_quantum_X_e_classical_limit_001_001}
\end{align}
Eq.~\eqref{main_eq_quantum_X_e_classical_limit_001_001} implies that $\hat{X}_{\hat{\rho}_\theta, f (\cdot)}^\mathrm{e}$ does not depend on the choice of $f (\cdot)$, and this fact is consistent with the fact that Eq.~\eqref{main_eq_classical_X_e_001_001} does not have the degrees of freedom with respect to $f (\cdot)$.
Another condition of $f (\cdot)$, Eq.~\eqref{main_eq_condition_Petz_function_001_012}, guarantees the realness of the quantum Fisher metric:
\begin{align}
	\overline{g_{\hat{\rho}_\theta, f (\cdot)} (X, Y)} & = \overline{\sum_{i, j = 1}^N \frac{1}{p_j f (p_i / p_j)} \langle \psi_j | \hat{X}_{\hat{\rho}_\theta}^\mathrm{m} | \psi_i \rangle \langle \psi_i | \hat{Y}_{\hat{\rho}_\theta}^\mathrm{m} | \psi_j \rangle}      \\
	                                                   & = \sum_{i, j = 1}^N \frac{1}{p_j f (p_i / p_j)} \overline{\langle \psi_j | \hat{X}_{\hat{\rho}_\theta}^\mathrm{m} | \psi_i \rangle \langle \psi_i | \hat{Y}_{\hat{\rho}_\theta}^\mathrm{m} | \psi_j \rangle}      \\
	                                                   & = \sum_{i, j = 1}^N \frac{1}{p_j f (p_i / p_j)} \langle \psi_i | \hat{X}_{\hat{\rho}_\theta}^\mathrm{m} | \psi_j \rangle \langle \psi_j | \hat{Y}_{\hat{\rho}_\theta}^\mathrm{m} | \psi_i \rangle                 \\
	                                                   & = \sum_{i, j = 1}^N \frac{1}{p_j \frac{p_i}{p_j} f (p_j / p_i)} \langle \psi_i | \hat{X}_{\hat{\rho}_\theta}^\mathrm{m} | \psi_j \rangle \langle \psi_j | \hat{Y}_{\hat{\rho}_\theta}^\mathrm{m} | \psi_i \rangle \\
	                                                   & = \sum_{i, j = 1}^N \frac{1}{p_i f (p_j / p_i)} \langle \psi_i | \hat{X}_{\hat{\rho}_\theta}^\mathrm{m} | \psi_j \rangle \langle \psi_j | \hat{Y}_{\hat{\rho}_\theta}^\mathrm{m} | \psi_i \rangle                 \\
	                                                   & = g_{\hat{\rho}_\theta, f (\cdot)} (X, Y).
\end{align}
The realness of a metric is important to ensure that the parameter space remains closed within real values under addition and multiplication with the metric.
In other words, Eq.~\eqref{main_eq_condition_Petz_function_001_012} may not be necessary when parameters take complex values.

\subsubsection{Bogoliubov-Kubo-Mori metric}

We explain the BKM metric, which often appears in statistical mechanics, such as linear response theory.
The Petz function for the BKM metric is given by
\begin{align}
	f_\mathrm{BKM} (t) & = \frac{t-1}{\ln t}. \label{main_eq_def_f_BKM_001_001}
\end{align}

From Eqs.~\eqref{main_eq_quantum_e_representation_001_001} and \eqref{main_eq_def_f_BKM_001_001}, we have
\begin{align}
	(X \hat{\rho}_\theta) \hat{\rho}_\theta^{-1} & = f_\mathrm{BKM} (\Delta_{\hat{\rho}_\theta}) \hat{X}_{\hat{\rho}_\theta, f_\mathrm{BKM} (\cdot)}^\mathrm{e}                                                                                    \\
	                                             & = \int_0^1 \mathrm{d}\lambda \, \hat{\rho}_\theta^\lambda \hat{X}_{\hat{\rho}_\theta, f_\mathrm{BKM} (\cdot)}^\mathrm{e} \hat{\rho}_\theta^{-\lambda}. \label{main_eq_X_rho_rho^-1_BKM_001_001}
\end{align}
where we have used the following equality:
\begin{align}
	\int_0^1 \mathrm{d}\lambda \, t^\lambda & = \int_0^1 \mathrm{d}\lambda \, \mathrm{e}^{(\ln t) \lambda} \\
	                                        & = \frac{1}{\ln t} [\mathrm{e}^{(\ln t) \lambda}]_0^1         \\
	                                        & = \frac{t-1}{\ln t}.
\end{align}
Then, m- and e-representations of $X$ for Eq.~\eqref{main_eq_def_f_BKM_001_001} are linked via
\begin{align}
	\hat{X}_{\hat{\rho}_\theta}^\mathrm{m} & = X \hat{\rho}_\theta                                                                                                                                   \\
	                                       & = \int_0^1 \mathrm{d}\lambda \, \hat{\rho}_\theta^\lambda \hat{X}_{\hat{\rho}_\theta, f_\mathrm{BKM} (\cdot)}^\mathrm{e} \hat{\rho}_\theta^{1-\lambda}.
\end{align}
Finally, we obtain the well-known form of the BKM metric:
\begin{align}
	 & g_{\hat{\rho}_\theta, f_\mathrm{BKM} (\cdot)} (X, Y) \nonumber                                                                                                                                                                                                                 \\
	 & \quad = \mathrm{Tr} [\hat{X}_{\hat{\rho}_\theta}^\mathrm{m} \hat{Y}_{\hat{\rho}_\theta, f_\mathrm{BKM} (\cdot)}^\mathrm{e}]                                                                                                                                                    \\
	 & \quad = \int_0^1 \mathrm{d}\lambda \, \mathrm{Tr} [ \hat{\rho}_\theta^\lambda \hat{X}_{\hat{\rho}_\theta, f_\mathrm{BKM} (\cdot)}^\mathrm{e} \hat{\rho}_\theta^{1-\lambda} \hat{Y}_{\hat{\rho}_\theta, f_\mathrm{BKM} (\cdot)}^\mathrm{e}]. \label{main_eq_BKM_metric_001_001}
\end{align}

\subsubsection{SLD metric}

Then, we discuss the SLD metric, which appears in quantum estimation theory~\cite{Liu_001, Liu_002, Safranek_001, Zanardi_001, Liu_003, Liu_004}.
The Petz function for the SLD metric is given by
\begin{align}
	f_\mathrm{SLD} (t) & = \frac{1+t}{2}. \label{main_eq_def_f_SLD_001_001}
\end{align}
From Eqs.~\eqref{main_eq_quantum_e_representation_001_001} and \eqref{main_eq_def_f_SLD_001_001}, we have
\begin{align}
	(X \hat{\rho}_\theta) \hat{\rho}_\theta^{-1} & = f_\mathrm{SLD} (\Delta_{\hat{\rho}_\theta}) \hat{X}_{\hat{\rho}_\theta, f_\mathrm{SLD} (\cdot)}^\mathrm{e}                                                                              \\
	                                             & = \frac{1 + \Delta_{\hat{\rho}_\theta}}{2} \hat{X}_{\hat{\rho}_\theta, f_\mathrm{SLD} (\cdot)}^\mathrm{e}                                                                                 \\
	                                             & = \frac{1}{2} (\hat{X}_{\hat{\rho}_\theta, f_\mathrm{SLD} (\cdot)}^\mathrm{e} + \hat{\rho}_\theta \hat{X}_{\hat{\rho}_\theta, f_\mathrm{SLD} (\cdot)}^\mathrm{e} \hat{\rho}_\theta^{-1}).
\end{align}
Then, the m- and e-representations of $X$ in the case of the SLD metric are related via
\begin{align}
	\hat{X}_{\hat{\rho}_\theta}^\mathrm{m} & = X \hat{\rho}_\theta                                                                                                                                                                \\
	                                       & = \frac{1}{2} (\hat{X}_{\hat{\rho}_\theta, f_\mathrm{SLD} (\cdot)}^\mathrm{e} \hat{\rho}_\theta + \hat{\rho}_\theta \hat{X}_{\hat{\rho}_\theta, f_\mathrm{SLD} (\cdot)}^\mathrm{e}).
\end{align}
Then, the SLD metric has the following form:
\begin{align}
	 & g_{\hat{\rho}_\theta, f_\mathrm{SLD} (\cdot)} (X, Y) \nonumber                                                                                                                                                                                                                                                                                          \\
	 & \quad = \mathrm{Tr} [\hat{X}_{\hat{\rho}_\theta}^\mathrm{m} \hat{Y}_{\hat{\rho}_\theta, f_\mathrm{SLD} (\cdot)}^\mathrm{e}]                                                                                                                                                                                                                             \\
	 & \quad = \mathrm{Tr} \bigg[ \bigg( \frac{1}{2} (\hat{X}_{\hat{\rho}_\theta, f_\mathrm{SLD} (\cdot)}^\mathrm{e} + \hat{\rho}_\theta \hat{X}_{\hat{\rho}_\theta, f_\mathrm{SLD} (\cdot)}^\mathrm{e} \hat{\rho}_\theta^{-1}) \hat{\rho}_\theta \bigg) \hat{Y}_{\hat{\rho}_\theta, f_\mathrm{SLD} (\cdot)}^\mathrm{e} \bigg]                                 \\
	 & \quad = \frac{1}{2} \mathrm{Tr} [\hat{\rho}_\theta (\hat{X}_{\hat{\rho}_\theta, f_\mathrm{SLD} (\cdot)}^\mathrm{e} \hat{Y}_{\hat{\rho}_\theta, f_\mathrm{SLD} (\cdot)}^\mathrm{e} + \hat{Y}_{\hat{\rho}_\theta, f_\mathrm{SLD} (\cdot)}^\mathrm{e} \hat{X}_{\hat{\rho}_\theta, f_\mathrm{SLD} (\cdot)}^\mathrm{e})]. \label{main_eq_SLD_metric_001_001}
\end{align}

More directly, the SLD metric is often defined as follows:
\begin{align}
	 & g_{\hat{\rho}_\theta, \mathrm{SLD}} (X, Y) \nonumber                                                                                                                                                        \\
	 & \quad \coloneqq \frac{1}{2} (\hat{\rho}_\theta \hat{L}_\mathrm{SLD} (X) \hat{L}_\mathrm{SLD} (Y) + \hat{\rho}_\theta \hat{L}_\mathrm{SLD} (Y) \hat{L}_\mathrm{SLD} (X)), \label{main_eq_SLD_metric_002_001}
\end{align}
where $\hat{L}_\mathrm{SLD} (X)$ is the operator such that
\begin{align}
	X \hat{\rho}_\theta & = \frac{1}{2} (\hat{L}_\mathrm{SLD} (X) \hat{\rho}_\theta + \hat{\rho}_\theta \hat{L}_\mathrm{SLD} (X)). \label{main_eq_differential_equation_SLD_L_X_001_001}
\end{align}
By imposing $\hat{X}_{\hat{\rho}_\theta, f_\mathrm{SLD} (\cdot)}^\mathrm{e} = \hat{L}_\mathrm{SLD} (X)$, Eqs.~\eqref{main_eq_SLD_metric_001_001} and \eqref{main_eq_SLD_metric_002_001} become identical.

\subsubsection{rRLD metric}

Next, we focus on the rRLD metric.
The Petz function for the rRLD metric is given by
\begin{align}
	f_\mathrm{rRLD} (t) & = \frac{2t}{1+t}. \label{main_eq_def_f_rRLD_001_001}
\end{align}
In the case of Eq.~\eqref{main_eq_def_f_rRLD_001_001}, the e-representation of $X$ as defined in Eq.~\eqref{main_eq_quantum_e_representation_001_001} is computed as
\begin{align}
	\hat{X}_{\hat{\rho}_\theta, f_\mathrm{rRLD} (\cdot)}^\mathrm{e} & = \sum_{i, j = 1}^N \frac{\langle \psi_i | \hat{X}_{\hat{\rho}_\theta}^\mathrm{m} | \psi_j \rangle}{p_j f_\mathrm{rRLD} (p_i / p_j)} | \psi_i \rangle \langle \psi_j |                                                \\
	                                                                & = \sum_{i, j = 1}^N \frac{\langle \psi_i | \hat{X}_{\hat{\rho}_\theta}^\mathrm{m} | \psi_j \rangle}{p_j \frac{2 (p_i / p_j)}{1 + p_i / p_j}} | \psi_i \rangle \langle \psi_j |                                            \\
	                                                                & = \sum_{i, j = 1}^N \frac{\langle \psi_i | \hat{X}_{\hat{\rho}_\theta}^\mathrm{m} | \psi_j \rangle}{\frac{2 p_i p_j}{p_i + p_j}} | \psi_i \rangle \langle \psi_j |                                                    \\
	                                                                & = \sum_{i, j = 1}^N \frac{p_i + p_j}{2 p_i p_j} \langle \psi_i | \hat{X}_{\hat{\rho}_\theta}^\mathrm{m} | \psi_j \rangle | \psi_i \rangle \langle \psi_j |                                                            \\
	                                                                & = \frac{1}{2} \sum_{i, j = 1}^N \langle \psi_i | (\hat{\rho}^{-1} \hat{X}_{\hat{\rho}_\theta}^\mathrm{m} + \hat{X}_{\hat{\rho}_\theta}^\mathrm{m} \hat{\rho}^{-1}) | \psi_j \rangle | \psi_i \rangle \langle \psi_j | \\
	                                                                & = \frac{1}{2} (\hat{\rho}^{-1} \hat{X}_{\hat{\rho}_\theta}^\mathrm{m} + \hat{X}_{\hat{\rho}_\theta}^\mathrm{m} \hat{\rho}^{-1}).
\end{align}
Then the rRLD metric reads
\begin{align}
	g_{\hat{\rho}_\theta, f_\mathrm{rRLD} (\cdot)} (X, Y) & = \mathrm{Tr} [\hat{X}_{\hat{\rho}_\theta}^\mathrm{m} \hat{Y}_{\hat{\rho}_\theta, f_\mathrm{rRLD} (\cdot)}^\mathrm{e}]                                                                                                                                                                   \\
	                                                      & = \mathrm{Tr} \bigg[ \hat{X}_{\hat{\rho}_\theta}^\mathrm{m} \frac{1}{2} (\hat{\rho}^{-1} \hat{Y}_{\hat{\rho}_\theta}^\mathrm{m} + \hat{Y}_{\hat{\rho}_\theta}^\mathrm{m} \hat{\rho}^{-1}) \bigg]                                                                                         \\
	                                                      & = \frac{1}{2} \mathrm{Tr} [\hat{\rho}^{-1} \hat{X}_{\hat{\rho}_\theta}^\mathrm{m} \hat{Y}_{\hat{\rho}_\theta}^\mathrm{m}] + \frac{1}{2} \mathrm{Tr} [\hat{\rho}^{-1} \hat{Y}_{\hat{\rho}_\theta}^\mathrm{m} \hat{X}_{\hat{\rho}_\theta}^\mathrm{m}]. \label{main_eq_rRLD_metric_001_001}
\end{align}

In the literature, the rRLD metric is often defined as follows:
\begin{align}
	g_{\hat{\rho}_\theta, \mathrm{rRLD}} (X, Y) & \coloneqq \Re \mathrm{Tr} [\hat{\rho}_\theta \hat{L}_\mathrm{RLD} (X) \hat{L}_\mathrm{RLD} (Y)], \label{main_eq_rRLD_metric_002_001}
\end{align}
where $\hat{L}_\mathrm{RLD} (X)$ is the operator such that
\begin{align}
	X \hat{\rho}_\theta & = \hat{\rho}_\theta \hat{L}_\mathrm{RLD} (X).
\end{align}
Imposing $\hat{X}_{\hat{\rho}_\theta, f_\mathrm{rRLD} (\cdot)}^\mathrm{e} = \hat{L}_\mathrm{RLD} (X)$, Eqs.~\eqref{main_eq_rRLD_metric_001_001} and \eqref{main_eq_rRLD_metric_002_001} become identical.

\subsubsection{$1/2$-metric}

The last example of the monotone metric is the $1/2$-metric.
The Petz function for the $1/2$-metric is given by
\begin{align}
	f_{(1/2)} (t) & = t^\frac{1}{2}. \label{main_eq_def_f_1/2_001_001}
\end{align}
From Eq.~\eqref{main_eq_def_f_1/2_001_001}, the e-representation of $X$ as defined in Eq.~\eqref{main_eq_quantum_e_representation_001_001} is computed as
\begin{align}
	\hat{X}_{\hat{\rho}_\theta, f_{(1/2)} (\cdot)}^\mathrm{e} & = \sum_{i, j = 1}^N \frac{\langle \psi_i | \hat{X}_{\hat{\rho}_\theta}^\mathrm{m} | \psi_j \rangle}{p_j \sqrt{\frac{p_i}{p_j}}} | \psi_i \rangle \langle \psi_j |                                           \\
	                                                          & = \sum_{i, j = 1}^N \frac{\langle \psi_i | \hat{X}_{\hat{\rho}_\theta}^\mathrm{m} | \psi_j \rangle}{\sqrt{p_i p_j}} | \psi_i \rangle \langle \psi_j |. \label{main_eq_quantum_m_representation_1/2_001_001}
\end{align}
Using Eq.~\eqref{main_eq_quantum_m_representation_1/2_001_001}, the $1/2$-metric reads
\begin{align}
	g_{\hat{\rho}_\theta, f_\mathrm{(1/2)} (\cdot)} (X, Y) & = \mathrm{Tr} [\hat{X}_{\hat{\rho}_\theta}^\mathrm{m} \hat{Y}_{\hat{\rho}_\theta, f_{(1/2)} (\cdot)}^\mathrm{e}]                                                                             \\
	                                                       & = \sum_{i, j = 1}^N \frac{\langle \psi_j | \hat{X}_{\hat{\rho}_\theta}^\mathrm{m} | \psi_i \rangle \langle \psi_i | \hat{Y}_{\hat{\rho}_\theta}^\mathrm{m} | \psi_j \rangle}{\sqrt{p_i p_j}}.
\end{align}

\subsubsection{Table and Figure of several Petz functions}

\begin{figure}[t]
	\centering
	\includegraphics[scale=0.60]{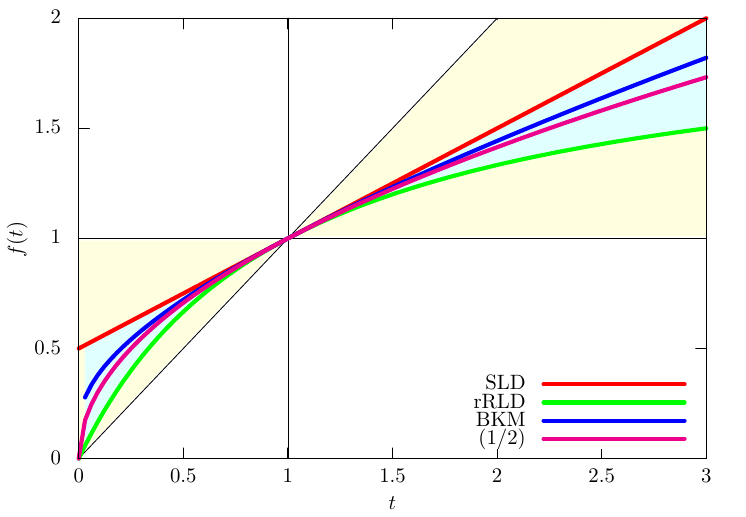}
	\caption{$f_\mathrm{SLD} (t) = \frac{1 + t}{2}$, $f_{(1/2)} (t) = t^\frac{1}{2}$, $f_\mathrm{BKM} (t) = \frac{t - 1}{\ln t}$, $f_\mathrm{rRLD} (t) = \frac{2 t}{1 + t}$. The regimes of the monotone Petz functions and the Petz functions associated with the rescaled sandwiched quantum R\'enyi divergence are highlighted by light cyan and light yellow, respectively.}
	\label{main_fig_gnuplot_operator_monotone_functions_001_001}
\end{figure}

At the end of this subsection, we present the plots of the Petz functions discussed above, Eqs.~\eqref{main_eq_def_f_BKM_001_001}, \eqref{main_eq_def_f_SLD_001_001}, \eqref{main_eq_def_f_rRLD_001_001}, and \eqref{main_eq_def_f_1/2_001_001}, in Fig.~\ref{main_fig_gnuplot_operator_monotone_functions_001_001}.
In addition, we summarize them in Table~\ref{main_table_operator_monotone_functions_003_001}.
\begin{table}[t]
	\caption{Summary of operator-monotone functions} \label{main_table_operator_monotone_functions_003_001}
	\begin{ruledtabular}
		\begin{tabular}{cc}
			Metric       & Petz function                          \\
			\hline
			BKM metric   & Eq.~\eqref{main_eq_def_f_BKM_001_001}  \\
			SLD metric   & Eq.~\eqref{main_eq_def_f_SLD_001_001}  \\
			rRLD metric  & Eq.~\eqref{main_eq_def_f_rRLD_001_001} \\
			$1/2$-metric & Eq.~\eqref{main_eq_def_f_1/2_001_001}
		\end{tabular}
	\end{ruledtabular}
\end{table}

\subsection{Quantum divergences}

For the preparation of QNG, we introduce quantum extensions of the KL divergence, Eq.~\eqref{main_eq_def_KL_divergence_001_001}, and the rescaled classical R\'enyi divergence, Eq.~\eqref{main_eq_def_rescaled_classical_Renyi_divergence_001_001}.
We see that, due to the noncommutative nature of density operators, we cannot define the quantum extension of the rescaled classical R\'enyi divergence, Eq.~\eqref{main_eq_def_rescaled_classical_Renyi_divergence_001_001}, uniquely.
However, from the viewpoint of QNG, the nonuniqueness of the rescaled sandwiched quantum R\'enyi divergence leads to the richness of QNG that does not exist in \textit{classical} NG.

\subsubsection{Quantum KL divergence}

We introduce the quantum KL divergence, which is the quantum extension of the KL divergence, Eq.~\eqref{main_eq_def_KL_divergence_001_001}.
The quantum KL divergence is defined via Refs.~\cite{Umegaki_001, Sagawa_001}
\begin{align}
	D_\mathrm{qKL} (\hat{\rho}_{\bar{\theta}} \| \hat{\rho}_\theta ) & \coloneqq \mathrm{Tr} [\hat{\rho}_{\bar{\theta}} (\ln \hat{\rho}_{\bar{\theta}} - \ln \hat{\rho}_\theta)]. \label{main_eq_def_quantum_KL_divergence_001_001}
\end{align}
Similarly to Eq.~\eqref{main_eq_def_parameterized_quantum_state_001_001}, we define
\begin{align}
	\hat{\rho}_{\bar{\theta}} & \coloneqq \sum_{i = 1}^N \bar{p}_i | \bar{\psi}_i \rangle \langle \bar{\psi}_i |.
\end{align}
Eq.~\eqref{main_eq_def_quantum_KL_divergence_001_001} is widely used to study quantum systems and is also referred to as the quantum relative entropy or the Umegaki entropy~\cite{Umegaki_001}.

The quantum KL divergence, Eq.~\eqref{main_eq_def_quantum_KL_divergence_001_001}, is equivalent to the KL divergence, Eq.~\eqref{main_eq_def_KL_divergence_001_001}, in the classical limit.
That is, the following relationship holds when $| \bar{\psi}_i \rangle = | \psi_i \rangle$ for $i = 1, 2, \dots, N$:
\begin{align}
	D_\mathrm{qKL} (\hat{\rho}_{\bar{\theta}} \| \hat{\rho}_\theta) & = D_\mathrm{KL} (p_{\bar{\theta}} (\cdot) \| p_\theta (\cdot)). \label{main_eq_equality_quantum_KL_divergence_classical_KL_divergence_001_001}
\end{align}

\subsubsection{Rescaled standard quantum R\'enyi divergence}

We describe the rescaled standard quantum R\'enyi divergence, which is a quantum extension of the rescaled classical R\'enyi divergence, Eq.~\eqref{main_eq_def_rescaled_classical_Renyi_divergence_001_001}.
The rescaled standard quantum R\'enyi divergence is given by~\cite{Amari_002}
\begin{align}
	R_\alpha^\mathrm{st} (\hat{\rho}_{\bar{\theta}} \| \hat{\rho}_\theta) & \coloneqq \frac{1}{\alpha (\alpha - 1)} \ln \mathrm{Tr} [\hat{\rho}_{\bar{\theta}}^\alpha \hat{\rho}_\theta^{1 - \alpha}]. \label{main_eq_def_rescaled_standard_quantum_Renyi_divergence_001_001}
\end{align}
The rescaled standard quantum R\'enyi divergence, Eq.~\eqref{main_eq_def_rescaled_standard_quantum_Renyi_divergence_001_001}, also reduces to the quantum KL divergence, Eq.~\eqref{main_eq_def_quantum_KL_divergence_001_001}, in the limit $\alpha \to 1$:
\begin{align}
	\lim_{\alpha \to 1} R_\alpha^\mathrm{st} (\hat{\rho}_{\bar{\theta}} \| \hat{\rho}_\theta) & = D_\mathrm{qKL} (\hat{\rho}_{\bar{\theta}} \| \hat{\rho}_\theta).
\end{align}
The rescaled standard quantum R\'enyi divergence, Eq.~\eqref{main_eq_def_rescaled_standard_quantum_Renyi_divergence_001_001}, is also equivalent to the rescaled classical R\'enyi divergence, Eq.~\eqref{main_eq_def_rescaled_classical_Renyi_divergence_001_001}, in the classical limit.
Analogously to Eq.~\eqref{main_eq_equality_quantum_KL_divergence_classical_KL_divergence_001_001}, the following relationship holds when $| \bar{\psi}_i \rangle = | \psi_i \rangle$ for $i = 1, 2, \dots, N$:
\begin{align}
	R_\alpha^\mathrm{st} (\hat{\rho}_{\bar{\theta}} \| \hat{\rho}_\theta) & = R_\alpha (p_{\bar{\theta}} (\cdot) \| p_\theta (\cdot)).
\end{align}

\begin{figure}[t]
	\centering
	\includegraphics[scale=0.60]{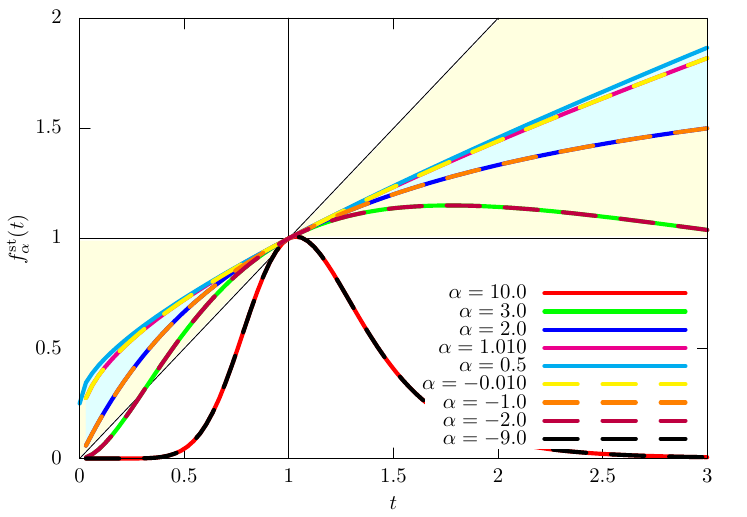}
	\caption{$f_\alpha^\mathrm{st} (t)$, Eq.~\eqref{main_eq_def_tilde_f_alpha_001_001}, with $\alpha = 4.0, 2.0, 1.010, 0.50, -0.010, -1.0, -3.0$. The regimes of the monotone Petz functions and the Petz functions associated with the rescaled sandwiched quantum R\'enyi divergence are highlighted by light cyan and light yellow, respectively.}
	\label{main_fig_gnuplot_operator_monotone_functions_002_002}
\end{figure}

The Petz function for the rescaled standard quantum R\'enyi divergence, Eq.~\eqref{main_eq_def_rescaled_standard_quantum_Renyi_divergence_001_001}, reads~\footnote{See Appendices~\ref{main_sec_quantum_F_divergence_001_001} and \ref{main_sec_derivatin_tilde_f_001_001} for the detailed derivation of Eq.~\eqref{main_eq_def_tilde_f_alpha_001_001}.}
\begin{align}
	f_\alpha^\mathrm{st} (t) & = \frac{\alpha (1 - \alpha) (t - 1)^2}{1 + t - t^{1 - \alpha} - t^\alpha}. \label{main_eq_def_tilde_f_alpha_001_001}
\end{align}
For $\alpha \in [-1, 2]$, $f_\alpha^\mathrm{st} (\cdot)$ is operator-monotone.
Furthermore, we have, for $t \in \mathbb{R}$,
\begin{align}
	f_\frac{1}{2}^\mathrm{st} (t) & = \frac{(t - 1)^2}{4 (t^\frac{1}{2} - 1)^2}, \\
	f_1^\mathrm{st} (\cdot)       & = f_0^\mathrm{st} (\cdot)                    \\
	                              & = f_\mathrm{BKM} (\cdot),                    \\
	f_2^\mathrm{st} (\cdot)       & = f_{-1}^\mathrm{st} (\cdot)                 \\
	                              & = f_\mathrm{rRLD} (\cdot).
\end{align}
We also note that Eq.~\eqref{main_eq_def_tilde_f_alpha_001_001} satisfies the following relationship for $\alpha \in \mathbb{R}$:
\begin{align}
	f_{- \alpha + \frac{1}{2}}^\mathrm{st} (t) & = f_{\alpha + \frac{1}{2}}^\mathrm{st} (t).
\end{align}
In Fig.~\ref{main_fig_gnuplot_operator_monotone_functions_002_002}, we plot Eq.~\eqref{main_eq_def_tilde_f_alpha_001_001} with $\alpha = 3.0, 2.0, 1.010, 0.50, -0.010, -1.0, -2.0$.

\subsubsection{Rescaled sandwiched quantum R\'enyi divergence}

We consider another quantum extension of the rescaled classical R\'enyi divergence, Eq.~\eqref{main_eq_def_rescaled_classical_Renyi_divergence_001_001}, which is called the rescaled sandwiched quantum R\'enyi divergence~\cite{Muller-Lennert_001, Wilde_001, Takahashi_001}:
\begin{align}
	R_\alpha^\mathrm{sw} (\hat{\rho}_{\bar{\theta}} \| \hat{\rho}_\theta) & \coloneqq \frac{1}{\alpha (\alpha - 1)} \ln \mathrm{Tr} \Big[ \Big( \hat{\rho}_\theta^\frac{1 - \alpha}{2 \alpha} \hat{\rho}_{\bar{\theta}} \hat{\rho}_\theta^\frac{1 - \alpha}{2 \alpha} \Big)^\alpha \Big]. \label{main_eq_def_rescaled_sandwiched_quantum_Renyi_divergence_001_001}
\end{align}
Note that ``sandwiched", in this context, means that $\hat{\rho}_{\bar{\theta}}$ is sandwiched by $\hat{\rho}_\theta$, and ``rescaled" means that Eq.~\eqref{main_eq_def_rescaled_sandwiched_quantum_Renyi_divergence_001_001} has the additional factor of $1 / \alpha$.

The rescaled sandwiched quantum R\'enyi divergence, Eq.~\eqref{main_eq_def_rescaled_sandwiched_quantum_Renyi_divergence_001_001}, becomes the quantum KL divergence, Eq.~\eqref{main_eq_def_quantum_KL_divergence_001_001}, in the limit $\alpha \to 1$:
\begin{align}
	\lim_{\alpha \to 1} R_\alpha^\mathrm{sw} (\hat{\rho}_{\bar{\theta}} \| \hat{\rho}_\theta) & = D_\mathrm{qKL} (\hat{\rho}_{\bar{\theta}} \| \hat{\rho}_\theta).
\end{align}
The rescaled sandwiched quantum R\'enyi divergence, Eq.~\eqref{main_eq_def_rescaled_sandwiched_quantum_Renyi_divergence_001_001}, is also equivalent to the rescaled classical R\'enyi divergence, Eq.~\eqref{main_eq_def_rescaled_classical_Renyi_divergence_001_001}, in the classical limit.
Analogously to Eq.~\eqref{main_eq_equality_quantum_KL_divergence_classical_KL_divergence_001_001}, the following relationship holds when $| \bar{\psi}_i \rangle = | \psi_i \rangle$ for $i = 1, 2, \dots, N$:
\begin{align}
	R_\alpha^\mathrm{sw} (\hat{\rho}_{\bar{\theta}} \| \hat{\rho}_\theta) & = R_\alpha (p_{\bar{\theta}} (\cdot) \| p_\theta (\cdot)).
\end{align}

\begin{figure}[t]
	\centering
	\includegraphics[scale=0.60]{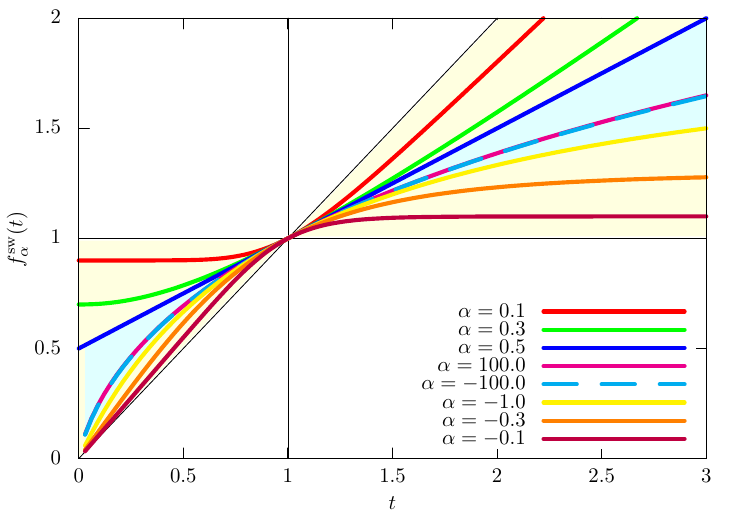}
	\caption{$f_\alpha^\mathrm{sw} (t)$, Eq.~\eqref{main_eq_def_f_alpha_001_001}, with $\alpha = 0.1, 0.3, 0.5, 100.0, -100.0, -1.0, -0.3, -0.1$. Note that $\alpha = 0.5$ and $\alpha = -1.0$ yield the SLD and rRLD metrics, respectively. The regimes of the monotone Petz functions and the Petz functions associated with the rescaled sandwiched quantum R\'enyi divergence are highlighted by light cyan and light yellow, respectively.}
	\label{main_fig_gnuplot_operator_monotone_functions_002_001}
\end{figure}

\begin{figure}[t]
	\centering
	\includegraphics[scale=0.60]{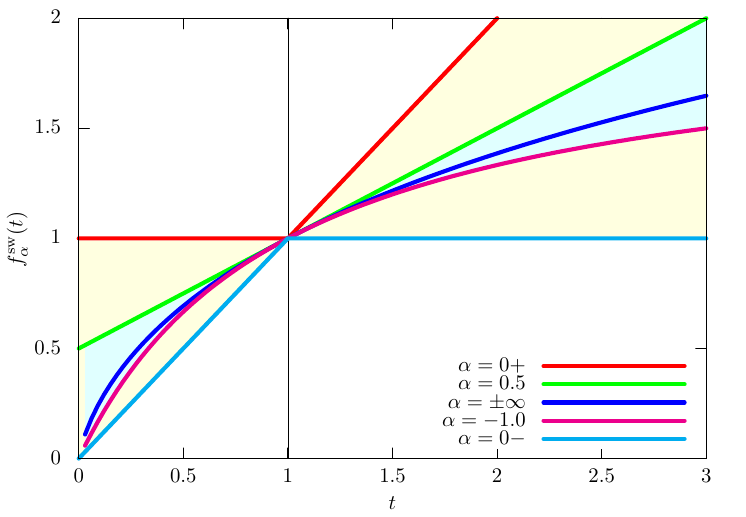}
	\caption{$f_\alpha^\mathrm{sw} (t)$, Eq.~\eqref{main_eq_def_f_alpha_001_001}, with $\alpha = 0+, 0.5, \pm \infty, -1.0, -0.3, 0-$. Note that $\alpha = 0.5$ and $\alpha = -1.0$ yield the SLD and rRLD metrics, respectively. The regimes of the monotone Petz functions and the Petz functions associated with the rescaled sandwiched quantum R\'enyi divergence are highlighted by light cyan and light yellow, respectively.}
	\label{main_fig_gnuplot_operator_monotone_functions_003_002}
\end{figure}

Next, we consider the quantum Fisher metric induced by the rescaled sandwiched quantum R\'enyi divergence, Eq.~\eqref{main_eq_def_rescaled_sandwiched_quantum_Renyi_divergence_001_001}.
Introducing~\footnote{See Appendices~\ref{main_sec_dervation_f_alpha_preliminary_001_001} and \ref{main_sec_dervation_f_alpha_main_001_001} for the derivation of Eq.~\eqref{main_eq_def_f_alpha_001_001}.}
\begin{align}
	f_\alpha^\mathrm{sw} (t) & \coloneqq (1 - \alpha) \frac{1 - t^\frac{1}{\alpha}}{1 - t^\frac{1 - \alpha}{\alpha}}, \label{main_eq_def_f_alpha_001_001}
\end{align}
Eq.~\eqref{main_eq_quantum_Fisher_metric_001_001} and Eq.~\eqref{main_eq_def_rescaled_sandwiched_quantum_Renyi_divergence_001_001} satisfy the following relation:
\begin{align}
	g_{\hat{\rho}_\theta, f_\alpha^\mathrm{sw} (\cdot)} (X, Y) & = \bar{X} \bar{Y} R_\alpha^\mathrm{sw} (\hat{\rho}_{\bar{\theta}} \| \hat{\rho}_\theta) |_{\bar{\theta} = \theta}, \label{main_eq_quantum_Fisher_metric_derivative_alpha_Renyi_divergence_001_001}
\end{align}
where $\bar{X} \coloneqq X^i \bar{\partial}_i$ and $\bar{Y} \coloneqq Y^i \bar{\partial}_i$.
Furthermore, Eqs.~\eqref{main_eq_def_f_BKM_001_001}, \eqref{main_eq_def_f_SLD_001_001}, \eqref{main_eq_def_f_rRLD_001_001}, and \eqref{main_eq_def_f_1/2_001_001} are equivalent to Eq.~\eqref{main_eq_def_f_alpha_001_001} for specific $\alpha$~\footnote{See Appendix~\ref{main_sec_derivation_f1_BKM_001_001} for the derivation of Eq.~\eqref{main_eq_f1_BKM_001_001}.}:
\begin{align}
	f_\frac{1}{2}^\mathrm{sw} (\cdot) & = f_\mathrm{SLD} (\cdot),                                \\
	f_2^\mathrm{sw} (\cdot)           & = f_{(1/2)} (\cdot),                                     \\
	f_{-1}^\mathrm{sw} (\cdot)        & = f_\mathrm{rRLD} (\cdot),                               \\
	f_1^\mathrm{sw} (\cdot)           & = f_\mathrm{BKM} (\cdot). \label{main_eq_f1_BKM_001_001}
\end{align}
In the limits $\alpha \to \pm \infty$, Eq.~\eqref{main_eq_def_f_alpha_001_001} reads~\footnote{See Ref.~\cite{Takahashi_001} and Appendix~\ref{main_sec_derivation_f_pm_infty_001_001} for the derivation of Eq.~\eqref{main_eq_def_f_pm_infty_001_001}.}
\begin{align}
	\lim_{\alpha \to \pm \infty} f_\alpha^\mathrm{sw} (t) & = \frac{t \ln t}{t - 1}. \label{main_eq_def_f_pm_infty_001_001}
\end{align}
In the limits $\alpha \to 0\pm$, Eq.~\eqref{main_eq_def_f_alpha_001_001} becomes~\footnote{See Appendix~\ref{main_sec_derivation_f_0pm_001_001} for the derivation of Eq.~\eqref{main_eq_f0pm_001_001}.}
\begin{subequations} \label{main_eq_f0pm_001_001}
	\begin{align}
		\lim_{\alpha \to 0+} f_\alpha^\mathrm{sw} (t) & =
		\begin{cases}
			t & (t \ge 1), \\
			1 & (t \le 1),
		\end{cases} \label{main_eq_f0+_001_001}           \\
		\lim_{\alpha \to 0-} f_\alpha^\mathrm{sw} (t) & =
		\begin{cases}
			1 & (t \ge 1), \\
			t & (t \le 1).
		\end{cases} \label{main_eq_f0-_001_001}
	\end{align}
\end{subequations}
In Fig.~\ref{main_fig_gnuplot_operator_monotone_functions_002_001}, we show $f_\alpha^\mathrm{sw} (t)$, Eq.~\eqref{main_eq_def_f_alpha_001_001}, for several different $\alpha$.
As shown in Eq.~\eqref{main_eq_def_f_pm_infty_001_001}, the Petz functions for $\alpha = \pm 100.0$ are almost identical in Fig.~\ref{main_fig_gnuplot_operator_monotone_functions_002_001}.
In Fig.~\ref{main_fig_gnuplot_operator_monotone_functions_003_002}, we also plot $f_\alpha^\mathrm{sw} (t)$, Eq.~\eqref{main_eq_def_f_alpha_001_001}, with $\alpha = 0+, 0.5, \pm \infty, -1.0, -0.3, 0-$.

\subsection{Quantum natural gradient}

To formulate the quantum counterpart of Eq.~\eqref{main_optimization_problem_NG_001_001}, let us go back to the minimization problem of $L (\theta)$, Eq.~\eqref{main_eq_def_optimization_problem_001_001}.
Similarly to the case of NG, we consider the following dynamics:
\begin{subequations} \label{main_optimization_problem_QNG_001_001}
	\begin{align}
		\theta_{\tau+1}               & = \theta_\tau + \Delta \theta_\tau (\epsilon),                \\
		\Delta \theta_\tau (\epsilon) & = \argmin_{\substack{\Delta \theta \in \mathbb{R}^{N_\theta}: \\ R_\alpha^\mathrm{sw} (\hat{\rho}_{\theta_\tau + \Delta \theta} \| \hat{\rho}_{\theta_\tau}) \le \epsilon}} [L (\theta_\tau + \Delta \theta) - L (\theta_\tau)]. \label{main_optimization_problem_QNG_001_012}
	\end{align}
\end{subequations}
In Eq.~\eqref{main_optimization_problem_QNG_001_001}, $\hat{\rho}_\theta$ appears in the constraint instead of $p_\theta (\cdot)$; as a result, the rescaled sandwiched quantum R\'enyi divergence, Eq.~\eqref{main_eq_def_rescaled_sandwiched_quantum_Renyi_divergence_001_001}, is used instead of the classical one.
By using the lowest-order approximation, Eq.~\eqref{main_optimization_problem_QNG_001_012} can be transformed into
\begin{align}
	\Delta \theta_\tau (\epsilon) & \approx \argmin_{\substack{\Delta \theta \in \mathbb{R}^{N_\theta}: \\ \frac{1}{2} \Delta \theta^\intercal G_\alpha (\theta_\tau) \Delta \theta \le \epsilon}} \nabla_\theta L (\theta_\tau)^\intercal \Delta \theta,
\end{align}
where $G_\alpha (\theta)$ is the quantum Fisher metric, Eq.~\eqref{main_eq_quantum_Fisher_metric_001_001}, induced by $R_\alpha^\mathrm{sw} (\hat{\rho}_{\theta + \Delta \theta} \| \hat{\rho}_\theta)$.
In other words, it can be rewritten as $[G_\alpha (\theta)]_{i, j} = g_{\hat{\rho}_\theta, f_\alpha^\mathrm{sw} (\cdot)} (\partial_i, \partial_j)$, where $[\cdot]_{i, j}$ is the $i, j$-th element of a matrix because of Eq.~\eqref{main_eq_quantum_Fisher_metric_derivative_alpha_Renyi_divergence_001_001}.

To solve Eq.~\eqref{main_optimization_problem_QNG_001_001}, we employ almost the same technique used in the previous section.
Then, we reach the same update equations as in the classical case, Eqs.~\eqref{main_eq_update_theta_001_001} and \eqref{main_eq_update_theta_001_002}, but $G_\alpha^{-1} (\theta_\tau)$ is computed from Eq.~\eqref{main_eq_quantum_Fisher_metric_derivative_alpha_Renyi_divergence_001_001}.
Note that in this paper, we focus on the rescaled sandwiched quantum R\'enyi divergence, Eq.~\eqref{main_eq_def_rescaled_sandwiched_quantum_Renyi_divergence_001_001}, but we do not need to limit ourselves to it.

\section{Quantum Fisher metrics in the bra-ket notation} \label{main_sec_bra-ket_notation_001_001}

The bra-ket notation of the quantum Fisher metric, Eq.~\eqref{main_eq_quantum_Fisher_metric_001_001}, is beneficial for discussing non-full-rank density operators.
In this section, we first rewrite the quantum Fisher metric, Eq.~\eqref{main_eq_quantum_Fisher_metric_001_001}, in the bra-ket notation for both full-rank and non-full-rank cases.
We also demonstrate that the bra-ket notation of the SLD metric in Refs.~\cite{Liu_001, Liu_002, Safranek_001, Zanardi_001, Liu_003, Liu_004} can be easily obtained from our formula.
Furthermore, we show the condition for the quantum Fisher metric, Eq.~\eqref{main_eq_quantum_Fisher_metric_001_001}, to be well-defined and discuss the valid regime of $\alpha$ for the quantum Fisher metric induced by the rescaled sandwiched quantum R\'enyi divergence, Eq.~\eqref{main_eq_def_rescaled_sandwiched_quantum_Renyi_divergence_001_001} to be well-defined.
Due to the condition, we can easily understand the reason why the SLD metric has been often used compared to other metrics for non-full-rank density operators, including pure states, and that the SLD metric is not unique for them.

\subsection{Full-rank case}

We discuss the bra-ket expression of the quantum Fisher metric, Eq.~\eqref{main_eq_quantum_Fisher_metric_001_001}, in the case of full-rank density operators.
Note that, in the case of a full-rank density operator, we have $p_i \in \mathbb{R}_{> 0}$ for $i = 1, 2, \dots, N$ when the density operator is expressed in the form of Eq.~\eqref{main_eq_def_parameterized_quantum_state_001_001}.
In this case, the m-representation of $\partial_m$ at $\hat{\rho}_\theta$ reads, for $m = 1, 2, \dots, N_\theta$,
\begin{align}
	[\hat{\partial}_m]_{\hat{\rho}_\theta}^\mathrm{m} & = \partial_m \sum_{i=1}^N p_i | \psi_i \rangle \langle \psi_i |                                                                             \\
	                                                  & = \sum_{i=1}^N (\partial_m p_i) | \psi_i \rangle \langle \psi_i | + \sum_{i=1}^N p_i | \partial_m \psi_i \rangle \langle \psi_i | \nonumber \\
	                                                  & \quad + \sum_{i=1}^N p_i | \psi_i \rangle \langle \partial_m \psi_i |.
\end{align}
Then, the quantum Fisher metric, Eq.~\eqref{main_eq_quantum_Fisher_metric_001_001}, becomes, for $m, n = 1, 2, \dots, N_\theta$,
\begin{align}
	 & g_{\hat{\rho}_\theta, f (\cdot)} (\partial_m, \partial_n) \nonumber                                                                                                                                                                         \\
	 & \quad = \sum_{i, j = 1}^N \frac{1}{p_j f (p_i / p_j)} \langle \psi_j | [\hat{\partial}_m]_{\hat{\rho}_\theta}^\mathrm{m} | \psi_i \rangle \langle \psi_i | [\hat{\partial}_n]_{\hat{\rho}_\theta}^\mathrm{m} | \psi_j \rangle               \\
	 & \quad = \sum_{i = 1}^N \frac{(\partial_m p_i) (\partial_n p_i)}{p_i} \nonumber                                                                                                                                                              \\
	 & \qquad + \sum_{i, j = 1}^N \frac{(p_i - p_j)^2}{p_j f (p_i / p_j)} \Re [\langle \psi_j | \partial_m \psi_i \rangle \langle \partial_n \psi_i | \psi_j \rangle], \label{main_eq_quantum_Fisher_metric_f_ij_element_braket_full_rank_001_001}
\end{align}
where
\begin{align}
	 & \Re [\langle \psi_j | \partial_m \psi_i \rangle \langle \partial_n \psi_i | \psi_j \rangle] \nonumber                                                                                                                                                                                       \\
	 & \quad \coloneqq \frac{\langle \psi_j | \partial_m \psi_i \rangle \langle \partial_n \psi_i | \psi_j \rangle + \langle \psi_j | \partial_n \psi_i \rangle \langle \partial_m \psi_i | \psi_j \rangle}{2}. \label{main_eq_def_real_part_inner_product_partial_derivative_bra_and_ket_001_001}
\end{align}
From Eq.~\eqref{main_eq_def_real_part_inner_product_partial_derivative_bra_and_ket_001_001}, $\Re [\langle \psi_j | \partial_m \psi_i \rangle \langle \partial_n \psi_i | \psi_j \rangle]$ is symmetric with respect to $i \leftrightarrow j$.
Note that the following relation holds for $p_i, p_j \in \mathbb{R}_{> 0}$ from Eq.~\eqref{main_eq_condition_Petz_function_001_012}:
\begin{align}
	p_j f (p_i / p_j) = p_i f (p_j / p_i). \label{main_eq_condition_Petz_function_002_001}
\end{align}

In the case of $f_\mathrm{SLD} (\cdot)$, Eq.~\eqref{main_eq_def_f_SLD_001_001}, Eq.~\eqref{main_eq_quantum_Fisher_metric_f_ij_element_braket_full_rank_001_001} reads
\begin{align}
	 & g_{\hat{\rho}_\theta, f_\mathrm{SLD} (\cdot)} (\partial_m, \partial_n) \nonumber                                                                          \\
	 & \quad = \sum_{i = 1}^N \frac{(\partial_m p_i) (\partial_n p_i)}{p_i} \nonumber                                                                            \\
	 & \qquad + \sum_{i, j = 1}^N \frac{2 (p_i - p_j)^2}{p_i + p_j} \Re [\langle \psi_j | \partial_m \psi_i \rangle \langle \partial_n \psi_i | \psi_j \rangle].
\end{align}
This result is also written in Ref.~\cite{Zanardi_001}.

\subsection{Non-full-rank case} \label{main_sec_bra-ket_notation_non-full-rank-case_001_001}

Next, we derive the bra-ket expression of the quantum Fisher metric, Eq.~\eqref{main_eq_quantum_Fisher_metric_001_001}, in the case of non-full-rank density operators.
As will be clarified below, Petz functions need to satisfy the additional condition $f (0) > 0$ for non-full-rank density operators.

For $M < N$, Eq.~\eqref{main_eq_def_parameterized_quantum_state_001_001} becomes
\begin{align}
	\hat{\rho}_\theta & = \sum_{i=1}^M p_i | \psi_i \rangle \langle \psi_i |. \label{note_eq_density_matrix_non-full_rank_001_001}
\end{align}
Equation~\eqref{note_eq_density_matrix_non-full_rank_001_001} can be expressed as follows:
\begin{align}
	\hat{\rho}_\theta & = \sum_{i=1}^N p_i | \psi_i \rangle \langle \psi_i |, \label{note_eq_density_matrix_non-full_rank_001_002}
\end{align}
where $p_i = 0$ for $i = M+1, M+2, \dots, N$.
One may expect that we can obtain the quantum Fisher metric, Eq.~\eqref{main_eq_quantum_Fisher_metric_001_001} by plugging Eq.~\eqref{note_eq_density_matrix_non-full_rank_001_002} into Eq.~\eqref{main_eq_quantum_Fisher_metric_f_ij_element_braket_full_rank_001_001}.
Unfortunately, this is not the case since Eq.~\eqref{main_eq_quantum_Fisher_metric_f_ij_element_braket_full_rank_001_001} has zero in the denominator.

We explain the technique to circumvent this problem as follows.
In the case of Eq.~\eqref{note_eq_density_matrix_non-full_rank_001_001}, Eq.~\eqref{main_eq_quantum_Fisher_metric_f_ij_element_braket_full_rank_001_001} is transformed into
\begin{align}
	 & g_{\hat{\rho}_\theta, f (\cdot)} (\partial_m, \partial_n) \nonumber                                                                                                                                          \\
	 & \quad = \sum_{i = 1}^M \frac{(\partial_m p_i) (\partial_n p_i)}{p_i} \nonumber                                                                                                                               \\
	 & \qquad + \sum_{i, j = 1}^M \frac{(p_i - p_j)^2}{p_j f (p_i / p_j)} \Re [\langle \psi_j | \partial_m \psi_i \rangle \langle  \partial_n \psi_i |\psi_j \rangle] \nonumber                                     \\
	 & \qquad + \sum_{i = 1}^M \sum_{j = M + 1}^N \frac{2 p_i^2}{p_j f (p_i / p_j)} \Re [\langle \psi_j | \partial_m \psi_i \rangle \langle \partial_n \psi_i | \psi_j \rangle]                                     \\
	 & \quad = \sum_{i = 1}^M \frac{(\partial_m p_i) (\partial_n p_i)}{p_i} \nonumber                                                                                                                               \\
	 & \qquad + \sum_{i, j = 1}^M \frac{(p_i - p_j)^2}{p_j f (p_i / p_j)} \Re [\langle \psi_j | \partial_m \psi_i \rangle \langle  \partial_n \psi_i |\psi_j \rangle] \nonumber                                     \\
	 & \qquad + \sum_{i = 1}^M \sum_{j = M + 1}^N \frac{2 p_i}{f (0)} \Re [\langle \psi_j | \partial_m \psi_i \rangle \langle \partial_n \psi_i | \psi_j \rangle]                                                   \\
	 & \quad = \sum_{i = 1}^M \frac{(\partial_m p_i) (\partial_n p_i)}{p_i} \nonumber                                                                                                                               \\
	 & \qquad + \sum_{i, j = 1}^M \bigg( \frac{(p_i - p_j)^2}{p_j f (p_i / p_j)} - \frac{2 p_i}{f (0)} \bigg) \Re [\langle \psi_j | \partial_m \psi_i \rangle \langle \partial_n \psi_i | \psi_j \rangle] \nonumber \\
	 & \qquad + \sum_{i = 1}^M \frac{2 p_i}{f (0)} \Re [\langle \partial_n \psi_i | \partial_m \psi_i \rangle], \label{main_eq_quantum_Fisher_metric_f_ij_element_braket_non-full_rank_001_001}
\end{align}
where $p_j = 0$ for $j = M + 1, M + 2, \dots, N$.
Note that we have used the following relation:
\begin{align}
	\sum_{j = M + 1}^N | \psi_j \rangle \langle \psi_j | & = \hat{1} - \sum_{j = 1}^M | \psi_j \rangle \langle \psi_j |.
\end{align}
Furthermore, extending Eq.~\eqref{main_eq_condition_Petz_function_002_001}, we have defined the following rule:
\begin{align}
	0 f (p_i / 0) & \coloneqq \lim_{p_j \to 0+} p_j f (p_i / p_j) \\
	              & = \lim_{p_j \to 0+} p_i f (p_j / p_i)         \\
	              & = p_i f (0).
\end{align}
For pure states ($M = 1$), Eq.~\eqref{main_eq_quantum_Fisher_metric_f_ij_element_braket_non-full_rank_001_001} is computed as
\begin{align}
	g_{\hat{\rho}_\theta, f (\cdot)} (\partial_m, \partial_n) & = - \frac{2}{f (0)} \Re [\langle \psi_1 | \partial_m \psi_1 \rangle \langle \partial_n \psi_1 | \psi_1 \rangle] \nonumber                                     \\
	                                                          & \quad + \frac{2}{f (0)} \Re [\langle \partial_n \psi_1 | \partial_m \psi_1 \rangle]. \label{note_eq_quantum_Fisher_metric_f_ij_element_braket_rank_1_001_001}
\end{align}
Note that $\hat{\rho}_\theta = | \psi_1 \rangle \langle \psi_1 |$ for $M = 1$.
Equation~\eqref{note_eq_quantum_Fisher_metric_f_ij_element_braket_rank_1_001_001} implies that the choice of the Petz function $f (\cdot)$ varies only the overall factor for pure states ($M = 1$).
In the case of Eq.~\eqref{main_eq_def_f_alpha_001_001} with $\alpha \to 0+$, Eq.~\eqref{note_eq_quantum_Fisher_metric_f_ij_element_braket_rank_1_001_001} reads
\begin{align}
	\lim_{\alpha \to 0+} g_{\hat{\rho}_\theta, f_\alpha^\mathrm{sw} (\cdot)} (\partial_m, \partial_n) & = - 2 \Re [\langle \psi_1 | \partial_m \psi_1 \rangle \langle \partial_n \psi_1 | \psi_1 \rangle] \nonumber                                     \\
	                                                                                                  & \quad + 2 \Re [\langle \partial_n \psi_1 | \partial_m \psi_1 \rangle]. \label{note_eq_quantum_Fisher_metric_f_ij_element_braket_rank_1_002_001}
\end{align}

Finally, we consider the SLD metric.
In the case of $f_\mathrm{SLD} (\cdot)$, Eq.~\eqref{main_eq_def_f_SLD_001_001}, Eq.~\eqref{main_eq_quantum_Fisher_metric_f_ij_element_braket_non-full_rank_001_001} reads
\begin{align}
	 & g_{\hat{\rho}_\theta, f_\mathrm{SLD} (\cdot)} (\partial_m, \partial_n) \nonumber                                                                                                               \\
	 & \quad = \sum_{i = 1}^M \frac{(\partial_m p_i) (\partial_n p_i)}{p_i} \nonumber                                                                                                                 \\
	 & \qquad + \sum_{i, j = 1}^M \bigg( \frac{2 (p_j^2 - p_i^2) - 8 p_i p_j}{p_i + p_j} \bigg) \Re [\langle \psi_j | \partial_m \psi_i \rangle \langle \partial_n \psi_i | \psi_j \rangle] \nonumber \\
	 & \qquad + \sum_{i = 1}^M 4 p_i \Re [\langle \partial_n \psi_i | \partial_m \psi_i \rangle] \nonumber                                                                                            \\
	 & \quad = \sum_{i = 1}^M \frac{(\partial_m p_i) (\partial_n p_i)}{p_i} \nonumber                                                                                                                 \\
	 & \qquad - \sum_{i, j = 1}^M \bigg( \frac{8 p_i p_j}{p_i + p_j} \bigg) \Re [\langle \psi_j | \partial_m \psi_i \rangle \langle \partial_n \psi_i | \psi_j \rangle] \nonumber                     \\
	 & \qquad + \sum_{i = 1}^M 4 p_i \Re [\langle \partial_n \psi_i | \partial_m \psi_i \rangle]. \label{note_eq_quantum_Fisher_metric_SLD_ij_element_braket_non-full_rank_001_001}
\end{align}
In Ref.~\cite{Liu_003}, a similar expression to Eq.~\eqref{note_eq_quantum_Fisher_metric_SLD_ij_element_braket_non-full_rank_001_001} is obtained.
Furthermore, for pure states ($M = 1$), Eq.~\eqref{note_eq_quantum_Fisher_metric_SLD_ij_element_braket_non-full_rank_001_001} becomes
\begin{align}
	g_{\hat{\rho}_\theta, f_\mathrm{SLD} (\cdot)} (\partial_m, \partial_n) & = - 4 \Re [\langle \psi_1 | \partial_m \psi_1 \rangle \langle \partial_n \psi_1 | \psi_1 \rangle] \nonumber                                       \\
	                                                                       & \quad + 4 \Re [\langle \partial_n \psi_1 | \partial_m \psi_1 \rangle]. \label{note_eq_quantum_Fisher_metric_SLD_ij_element_braket_rank_1_001_001}
\end{align}
Equation~\eqref{note_eq_quantum_Fisher_metric_f_ij_element_braket_rank_1_002_001} is half as large as Eq.~\eqref{note_eq_quantum_Fisher_metric_SLD_ij_element_braket_rank_1_001_001}.
In the literature~\cite{Cheng_001}, Eq.~\eqref{note_eq_quantum_Fisher_metric_SLD_ij_element_braket_rank_1_001_001} is called the Fubini-Study metric or quantum geometric tensor.

\subsection{$\alpha$ such that $f_\alpha^\mathrm{sw} (0) > 0$}

Non-full-rank density operators have eigenvalues equal to zero; therefore, the condition $f_\alpha^\mathrm{sw} (0) > 0$ is necessary for the quantum Fisher metric, Eq.~\eqref{main_eq_quantum_Fisher_metric_001_001} to be well-defined.
Then we consider $\alpha$ such that $f_\alpha^\mathrm{sw} (0) > 0$.

\begin{figure}[t]
	\centering
	\includegraphics[scale=0.60]{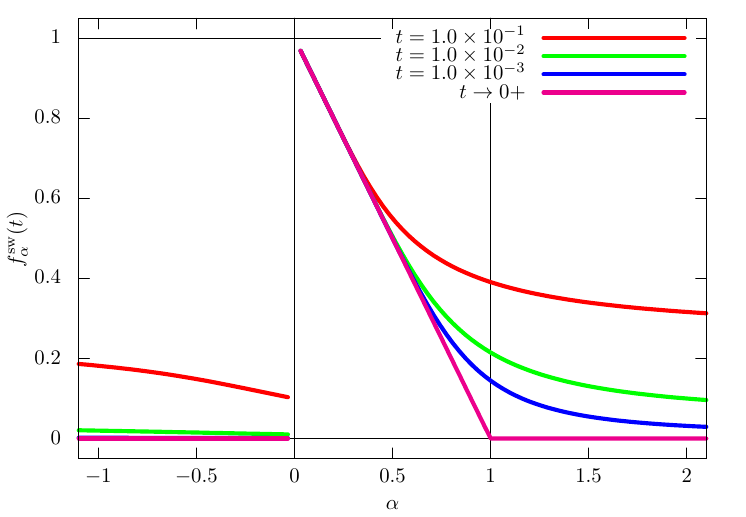}
	\caption{Value of $f_\alpha^\mathrm{sw} (t)$, Eq.~\eqref{main_eq_def_f_alpha_001_001}, for small $t$.}
	\label{main_fig_Petz_function_f(0)_001_001}
\end{figure}

From Eq.~\eqref{main_eq_def_f_alpha_001_001}, we get
\begin{align}
	\lim_{t \to 0+} f_\alpha^\mathrm{sw} (t) & =
	\begin{cases}
		1 - \alpha & (\alpha \in (0, 1)),                         \\
		0          & (\alpha \in (- \infty, 0) \cup [1, \infty)).
	\end{cases} \label{main_eq_f_alpha_(0)_001_001}
\end{align}
Eq.~\eqref{main_eq_f_alpha_(0)_001_001} implies that $\alpha \in (0, 1)$ must hold for $f_\alpha^\mathrm{sw} (t) > 0$ to be applied for pure states.
In Fig.~\ref{main_fig_Petz_function_f(0)_001_001}, we plot $f_\alpha^\mathrm{sw} (t)$ for small $t > 0$.

\section{Relationship with other metrics}

In the literature of quantum information theory~\cite{Nielsen_002, Wilde_003, Safranek_002, Cheng_003}, several different quantum metrics are known.
In this section, we address the relationship of the quantum Fisher metric with them.
We also mention the SR method~\cite{Becca_001}, which is a similar approach to NG.

\subsection{Fubini-Study distance, Bures angle, Bures distance, and fidelity}

In the realm of quantum information, several distances between quantum states have been proposed.
Here, we review some of them.
The first one is the Fubini-Study distance~\cite{Cheng_003}, which is a distance between pure states and was used to formulate QNG in Ref.~\cite{Stokes_001}:
\begin{align}
	D_\mathrm{FB} (| \psi \rangle, | \bar{\psi} \rangle) & \coloneqq \arccos \sqrt{\frac{\langle \psi | \bar{\psi} \rangle \langle{\bar{\psi} | \psi \rangle}}{\langle \bar{\psi} | \bar{\psi} \rangle \langle{\psi | \psi \rangle}}}. \label{main_eq_def_Fubini_Study_distance_001_001}
\end{align}
For mixed states, the Fubini-Study distance, Eq.~\eqref{main_eq_def_Fubini_Study_distance_001_001}, is generalized to the Bures angle, defined as follows:
\begin{align}
	A_\mathrm{Bures} (\hat{\rho}_\theta, \hat{\rho}_{\bar{\theta}}) & \coloneqq \arccos \Big( \sqrt{F (\hat{\rho}_\theta, \hat{\rho}_{\bar{\theta}})} \Big), \label{main_eq_def_Bures_angle_001_001}
\end{align}
where
\begin{align}
	F (\hat{\rho}_\theta, \hat{\rho}_{\bar{\theta}}) & \coloneqq \bigg( \mathrm{Tr} \bigg[ \sqrt{\sqrt{\hat{\rho}_\theta} \hat{\rho}_{\bar{\theta}} \sqrt{\hat{\rho}_\theta}} \bigg] \bigg)^2. \label{main_eq_def_Fidelity_quantum_states_001_001}
\end{align}
Equation~\eqref{main_eq_def_Fidelity_quantum_states_001_001} is called the fidelity of quantum states.
Note that the Fubini-Study distance, Eq.~\eqref{main_eq_def_Fubini_Study_distance_001_001}, and the Bures angle, Eq.~\eqref{main_eq_def_Bures_angle_001_001}, satisfy the following equality:
\begin{align}
	A_\mathrm{Bures} (| \psi \rangle \langle \psi |, | \bar{\psi} \rangle \langle \bar{\psi} |) & = D_\mathrm{FB} (| \psi \rangle, | \bar{\psi} \rangle).
\end{align}
As a similar quantity to the Bures angle, Eq.~\eqref{main_eq_def_Bures_angle_001_001}, the Bures distance is known:
\begin{align}
	D_\mathrm{Bures} (\hat{\rho}_\theta, \hat{\rho}_{\bar{\theta}}) & \coloneqq \sqrt{2 \Big( 1 - \sqrt{F (\hat{\rho}_\theta, \hat{\rho}_{\bar{\theta}})} \Big)}. \label{main_eq_def_Bures_distance_001_001}
\end{align}

It is known that Eqs.~\eqref{main_eq_def_Bures_angle_001_001} and \eqref{main_eq_def_Bures_distance_001_001} lead to the SLD metric, Eq.~\eqref{main_eq_SLD_metric_002_001}.
The reason is as follows.
First, we have
\begin{align}
	 & A_\mathrm{Bures} (\hat{\rho}_\theta, \hat{\rho}_{\theta + \Delta \theta}) \nonumber                                                                \\
	 & \quad = \arccos \Big( \sqrt{F (\hat{\rho}_\theta, \hat{\rho}_{\theta + \Delta \theta})} \Big)                                                      \\
	 & \quad = \arccos \Big( 1 - \Big( 1 - \sqrt{F (\hat{\rho}_\theta, \hat{\rho}_{\theta + \Delta \theta})} \Big) \Big)                                  \\
	 & \quad = \sqrt{2 \Big( 1 - \sqrt{F (\hat{\rho}_\theta, \hat{\rho}_{\theta + \Delta \theta})} \Big)} + \mathcal{O} (\| \Delta \theta \|^\frac{3}{2}) \\
	 & \quad = D_\mathrm{Bures} (\hat{\rho}_\theta, \hat{\rho}_{\theta + \Delta \theta}) + \mathcal{O} (\| \Delta \theta \|^\frac{3}{2}).
\end{align}
Note that we have used the following formula:
\begin{align}
	\arccos (1 - x) & = \sqrt{2 x} + \frac{(2 x)^\frac{3}{2}}{24} + \mathcal{O} (x^2).
\end{align}
Then, we have~\footnote{See Appendix~\ref{main_sec_derivation_Bures_metric_001_001} for the derivation of Eq.~\eqref{main_eq_relationship_Bures_distance_SLD_metric_001_001}.}
\begin{align}
	[D_\mathrm{Bures} (\hat{\rho}_\theta, \hat{\rho}_{\theta + \Delta \theta})]^2 & = \frac{1}{4} \mathrm{Tr} \bigg[ \hat{\rho}_\theta \bigg( \frac{\hat{L}_i \hat{L}_j + \hat{L}_j \hat{L}_i}{2} \bigg) \bigg] \Delta \theta^i \Delta \theta^j \nonumber \\
	                                                                              & \quad + O (\| \Delta \theta \|^3), \label{main_eq_relationship_Bures_distance_SLD_metric_001_001}
\end{align}
where $\hat{L}_i$ is an operator such that
\begin{align}
	\frac{\mathrm{d}}{\mathrm{d}\theta^i} \hat{\rho}_\theta & = \frac{1}{2} (\hat{L}_i \hat{\rho}_\theta + \hat{\rho}_\theta \hat{L}_i). \label{main_eq_differential_equation_SLD_L_i_001_001}
\end{align}
Note that Eqs.~\eqref{main_eq_differential_equation_SLD_L_X_001_001} and \eqref{main_eq_differential_equation_SLD_L_i_001_001} are identical when $X = \partial_i$.
Equation~\eqref{main_eq_relationship_Bures_distance_SLD_metric_001_001} implies that
\begin{align}
	g_\mathrm{Bures} (\partial_i, \partial_j) & = \frac{1}{4} g_\mathrm{SLD} (\partial_i, \partial_j), \label{main_eq_relation_Bures_metric_SLD_metric_001_001}
\end{align}
where $g_\mathrm{Bures} (\cdot, \cdot)$ is a metric such that
\begin{align}
	[D_\mathrm{Bures} (\hat{\rho}_\theta, \hat{\rho}_{\theta + \Delta \theta})]^2 & = g_\mathrm{Bures} (\partial_i, \partial_j) \Delta \theta^i \Delta \theta^j.
\end{align}

\subsection{Stochastic reconfiguration method}

Following Ref.~\cite{Becca_001}, we review the SR method, which aims to find the parameter of a parameterized wave function $| \psi_{\theta} \rangle$ that minimizes the expected value of a given quantum Hamiltonian $\hat{H}$.
Let us consider the following update rule for the parameter that characterizes a wave function:
\begin{align}
	\theta_{t+1} & = \theta_t + \Delta \theta_t (\delta_\tau). \label{main_eq_parameter_update_SR_001_001}
\end{align}
In the SR method, $\Delta \theta_t (\delta_\tau)$ in Eq.~\eqref{main_eq_parameter_update_SR_001_001} is computed as
\begin{align}
	\Delta \theta_t (\delta_\tau) & \coloneqq \argmin_{\Delta \theta_t \in \mathbb{R}^{N_\theta}} D_\mathrm{FB} (\mathrm{e}^{- 2 \delta_\tau \hat{H}} | \psi_{\theta_t} \rangle, | \psi_{\theta_t + \Delta \theta_t} \rangle) \\
	                                                & = - \delta_\tau S^{-1} \nabla_\theta \langle \hat{H} \rangle_{\theta_t}, \label{main_eq_Delta_theta_SR_001_001}
\end{align}
where
\begin{align}
	\langle \hat{H} \rangle_\theta & \coloneqq \langle \psi_\theta | \hat{H} | \psi_\theta \rangle.
\end{align}
The key point is that $S$ in Eq.~\eqref{main_eq_Delta_theta_SR_001_001} is proportional to the SLD metric because of the nature of the Fubini-Study distance, Eq.~\eqref{main_eq_def_Fubini_Study_distance_001_001}; that is, it follows from Eq.~\eqref{main_eq_relation_Bures_metric_SLD_metric_001_001}.

\section{Monotone and non-monotone quantum Fisher metrics} \label{main_sec_monotonicity_quantum_Fisher_metric_001_001}

Operator functions are functions that take an operator as an argument.
The key point about them is that some basic notions of functions that take a scalar value as an argument, such as monotonicity and convexity, do not hold directly for them.
More precisely, functions that satisfy monotonicity for scalar values may not be monotonic for operators, while the inverse is true in general.
Here, we review some important properties of operator functions~\cite{Bhatia_001, Horn_001, Hiai_002, Hiai_007}.

\subsection{Definition of operator-monotone functions}

We begin with the definition of operator-monotone functions.
Letting $\mathcal{H}$ and $\mathcal{B} (\mathcal{H})$ be the Hilbert space of interest and the set of bounded linear operators on $\mathcal{H}$, respectively, we define, for $\hat{A} \in \mathcal{B} (\mathcal{H})$,
\begin{align}
	\sigma (\hat{A}) & \coloneqq \{ \lambda \in \mathbb{C} | \lambda \hat{1} - \hat{A} \in \mathcal{S} \},
\end{align}
where $\mathcal{S}$ is the set of noninvertible operators in $\mathcal{B} (\mathcal{H})$.

Let us consider $f (\cdot): J \to \mathbb{R}$ where $J$ is an interval of $\mathbb{R}$.
If $f (\cdot)$ satisfies the following relation, then $f (\cdot)$ is called an operator-monotone function for self-adjoint operators $\hat{A}, \hat{B} \in \mathcal{B} (\mathcal{H})$ with $\sigma (\hat{A}), \sigma (\hat{B}) \subset J$~\cite{Bhatia_001, Hiai_007}:
\begin{align}
	\hat{A} \succeq \hat{B} & \Rightarrow f (\hat{A}) \succeq f (\hat{B}), \label{main_eq_def_operator_monotone_function_001_001}
\end{align}
where
\begin{align}
	\hat{A} \succeq \hat{B} & :\Leftrightarrow \text{$\forall | \psi \rangle \in \mathcal{H}$, $\langle \psi | (\hat{A} - \hat{B}) | \psi \rangle \ge 0$}.
\end{align}

\subsection{Monotone metrics}

Next, we turn our attention to monotone metrics.
A metric $g_{\hat{\rho}} (X, Y)$ is called a monotone metric if and only if it satisfies the following relation for an arbitrary CPTP map $\gamma (\cdot)$:
\begin{align}
	g_{\hat{\rho}} (X, X) & \ge g_{\gamma (\hat{\rho})} (\gamma_* (X), \gamma_* (X)), \label{main_eq_inequality_CPTP_dynamics_metric_X_X_001_001}
\end{align}
where $\gamma_* (\cdot)$ is the pushforward associated with $\gamma (\cdot)$.
Note that Eq.~\eqref{main_eq_def_f_alpha_001_001} is operator-monotone if and only if $\alpha \in (- \infty, -1] \cup [\frac{1}{2}, \infty)$ and Eq.~\eqref{main_eq_quantum_Fisher_metric_derivative_alpha_Renyi_divergence_001_001} is a monotone metric~\cite{Takahashi_001}.

The importance of Eq.~\eqref{main_eq_inequality_CPTP_dynamics_metric_X_X_001_001} can be easily understood.
In general, quantum dynamics can be expressed as a CPTP map~\cite{Wilde_003}.
Furthermore, it is naively expected that any quantum state reaches an equilibrium state, and thus the ``distance" between two distinct quantum states that obey the same dynamics gets closer to zero over time.
Thus, the monotonicity represented by Eq.~\eqref{main_eq_inequality_CPTP_dynamics_metric_X_X_001_001} is often assumed in physics.

\subsection{Partial order on operator-monotone functions}

We define the following partial order on $f (\cdot): \mathbb{R}_{\ge 0} \to \mathbb{R}_{\ge 0}$:
\begin{align}
	\text{$f (\cdot) \preceq \tilde{f} (\cdot)$} & :\Leftrightarrow \text{$\forall t \in \mathbb{R}_{\ge 0}$, $f (t) \le \tilde{f} (t)$}. \label{main_eq_def_order_operator_monotone_functions_001_001}
\end{align}
For monotone functions, the following theorem holds~\cite{Petz_001, Hiai_007}:
\begin{theorem} \label{main_theorem_maximum_minimum_elements_order_operator_monotone_functions_001_001}
	$f_\mathrm{SLD} (\cdot)$ and $f_\mathrm{rRLD} (\cdot)$ are the maximum and minimum elements with respect to the inequality on operator functions, Eq.~\eqref{main_eq_def_order_operator_monotone_functions_001_001}, under the condition of monotonicity, Eq.~\eqref{main_eq_def_operator_monotone_function_001_001}, with $f (1) = 1$ and $f (t) = t f (t^{-1})$.
	That is, the following relation holds for any $f (t)$ such that it satisfies the condition of monotonicity, Eq.~\eqref{main_eq_def_operator_monotone_function_001_001} with $f (1) = 1$ and $f (t) = t f (t^{-1})$:
	\begin{align}
		f_\mathrm{rRLD} (\cdot) \preceq f (\cdot) \preceq f_\mathrm{SLD} (\cdot). \label{main_eq_maximum_minimum_elements_order_operator_monotone_functions_001_001}
	\end{align}
\end{theorem}
The proof of Theorem~\ref{main_theorem_maximum_minimum_elements_order_operator_monotone_functions_001_001} is shown in Appendix~\ref{main_sec_theorem_maximum_minimum_elements_order_operator_monotone_functions_001_001}.

\subsection{Properties of Eq.~\eqref{main_eq_def_f_alpha_001_001}}

We here address some properties of Eq.~\eqref{main_eq_def_f_alpha_001_001}.
Substituting $\alpha = \frac{1}{\beta}$ into Eq.~\eqref{main_eq_def_f_alpha_001_001}, we have
\begin{align}
	f_\frac{1}{\beta}^\mathrm{sw} (t) & = \bigg( 1 - \frac{1}{\beta} \bigg) \frac{1 - t^\beta}{1 - t^{\beta - 1}}. \label{main_eq_def_f_beta_001_001}
\end{align}
We state two theorems on Eq.~\eqref{main_eq_def_f_beta_001_001} shown in Ref.~\cite{Takahashi_001}.
The first one is on the $\beta$-dependence of Eq.~\eqref{main_eq_def_f_beta_001_001}~\cite{Takahashi_001}:
\begin{theorem} \label{main_theorem_monotonicity_Petz_function_rescaled_sandwiched_quantum_Renyi_divergence_001_001}
	Eq.~\eqref{main_eq_def_f_beta_001_001} monotonically increases when $\beta$ increases.
\end{theorem}
Theorem~\ref{main_theorem_monotonicity_Petz_function_rescaled_sandwiched_quantum_Renyi_divergence_001_001} implies that $f_\alpha^\mathrm{sw} (\cdot)$, Eq.~\eqref{main_eq_def_f_alpha_001_001}, is monotonically decreasing with $\alpha$ increasing, but $\alpha = 0$ is singular.
Thus, $f_\alpha^\mathrm{sw} (\cdot)$, Eq.~\eqref{main_eq_def_f_alpha_001_001}, monotonically increases toward $\alpha \to 0+$.

The proof of Theorem~\ref{main_theorem_monotonicity_Petz_function_rescaled_sandwiched_quantum_Renyi_divergence_001_001} is as follows:
\begin{proof}
	We show the monotonicity of Eq.~\eqref{main_eq_def_f_beta_001_001} with respect to $\beta$.

	Taking the derivative of Eq.~\eqref{main_eq_def_f_beta_001_001},
	\begin{align}
		\ln f_\frac{1}{\beta}^\mathrm{sw} (t) & = \ln \bigg| \frac{1}{\beta} - 1 \bigg| + \ln | t^\beta - 1 | - \ln | 1 - t^{\beta - 1} |. \label{main_eq_log_f_beta_001_001}
	\end{align}

	The derivative of Eq.~\eqref{main_eq_log_f_beta_001_001} with respect to $\beta$ takes the following form:
	\begin{align}
		\frac{\partial}{\partial \beta} \ln f_\frac{1}{\beta}^\mathrm{sw} (t) & = \frac{1}{1 - \beta^{-1}} \frac{1}{\beta^2} + \frac{1}{1 - t^\beta} (- \ln t) t^\beta \nonumber \\
		                                                                      & \quad - \frac{1}{1 - t^{\beta - 1}} (- \ln t) t^{\beta - 1}                                      \\
		                                                                      & = \bigg( \frac{1}{1 - t^\beta} (- \ln t) t^\beta - \frac{1}{\beta} \bigg) \nonumber              \\
		                                                                      & \quad - \bigg( \frac{1}{1 - t^{\beta - 1}} (- \ln t) t^{\beta - 1} - \frac{1}{\beta - 1} \bigg)  \\
		                                                                      & = F_\beta (t) - F_{\beta - 1} (t), \label{main_eq_derivative_log_f_001_001}
	\end{align}
	where
	\begin{align}
		F_\beta (t) & \coloneqq \frac{\ln t}{1 - t^{- \beta}} - \frac{1}{\beta}. \label{main_eq_def_F_001_001}
	\end{align}
	Here, we have used $\frac{\mathrm{d}}{\mathrm{d}x} \ln |x| = 1 / x$.
	Next, we compute the derivative of Eq.~\eqref{main_eq_def_F_001_001}:
	\begin{align}
		\frac{\partial}{\partial \beta} F_\beta (t) & = \frac{(\ln t)^2}{(1 - t^{- \beta})^2} t^{- \beta} + \frac{1}{\beta^2} \\
		                                            & \ge 0.
	\end{align}
	Then, Eq.~\eqref{main_eq_def_F_001_001} monotonically increases with $\beta$, and therefore Eq.~\eqref{main_eq_derivative_log_f_001_001} is positive for $\beta \in \mathbb{R}$ and $t \in \mathbb{R}_{> 0}$.
	Thus, the proof is completed.
\end{proof}

\begin{figure}[t]
	\centering
	\includegraphics[scale=0.60]{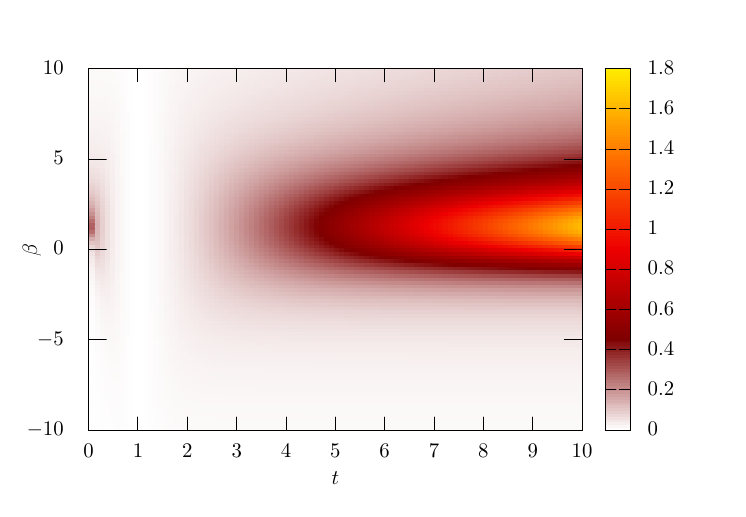}
	\caption{$\frac{\partial}{\partial \beta} f_\frac{1}{\beta}^\mathrm{sw} (t)$, Eq.~\eqref{main_eq_derivative_f_beta_001_001}.}
	\label{main_fig_gnuplot_operator_monotone_functions_003_001}
\end{figure}

We also confirm the monotonicity of Eq.~\eqref{main_eq_def_f_beta_001_001} numerically.
The derivative of Eq.~\eqref{main_eq_def_f_beta_001_001} with respect to $\beta$ reads
\begin{align}
	\frac{\partial}{\partial \beta} f_\frac{1}{\beta}^\mathrm{sw} (t) & = \frac{(1 - t^\beta) (1 - t^{\beta - 1}) + \beta (1 - \beta) t^{\beta - 1} (\ln t) (t - 1)}{\beta^2 (1 - t^{\beta - 1})^2}. \label{main_eq_derivative_f_beta_001_001}
\end{align}
In Fig.~\ref{main_fig_gnuplot_operator_monotone_functions_003_001}, we plot $\frac{\partial}{\partial \beta} f_\frac{1}{\beta}^\mathrm{sw} (t)$, Eq.~\eqref{main_eq_derivative_f_beta_001_001}.
This figure supports Theorem~\ref{main_theorem_monotonicity_Petz_function_rescaled_sandwiched_quantum_Renyi_divergence_001_001}.
As Fig.~\ref{main_fig_gnuplot_operator_monotone_functions_003_001} shows, $\frac{\partial}{\partial \beta} f_\frac{1}{\beta}^\mathrm{sw} (t) > 0$; thus $f_\alpha^\mathrm{sw} (t)$ monotonically decreases with $\alpha$ increased within the regimes of $\alpha > 0$ and $\alpha < 0$, respectively.
Note that, in Ref.~\cite{Takahashi_001}, Theorem~\ref{main_theorem_monotonicity_Petz_function_rescaled_sandwiched_quantum_Renyi_divergence_001_001} is implied though not explicitly written.

The second one is on the operator-monotonicity of Eq.~\eqref{main_eq_def_f_beta_001_001}~\cite{Takahashi_001}:
\begin{theorem} \label{main_theorem_monotonicity_Petz_function_rescaled_sandwiched_quantum_Renyi_divergence_002_001}
	Eq.~\eqref{main_eq_def_f_beta_001_001} is operator-monotone if and only if $\beta \in [-1, 2]$.
\end{theorem}
Theorem~\ref{main_theorem_monotonicity_Petz_function_rescaled_sandwiched_quantum_Renyi_divergence_002_001} means that $f_\alpha^\mathrm{sw} (\cdot)$, Eq.~\eqref{main_eq_def_f_alpha_001_001}, is operator-monotone for $\alpha \in (\infty, -1] \cup [1/2, \infty)$.
See Ref.~\cite{Takahashi_001} for the proof of Theorem~\ref{main_theorem_monotonicity_Petz_function_rescaled_sandwiched_quantum_Renyi_divergence_002_001}.

\subsection{Corollaries of Theorem~\ref{main_theorem_monotonicity_Petz_function_rescaled_sandwiched_quantum_Renyi_divergence_001_001}}

Here we state corollaries of Theorem~\ref{main_theorem_monotonicity_Petz_function_rescaled_sandwiched_quantum_Renyi_divergence_001_001}.
It is known that Eq.~\eqref{main_eq_def_f_alpha_001_001} leads to a monotone function for $\alpha \in (- \infty, -1] \cup [\frac{1}{2}, \infty)$~\cite{Takahashi_001}.
For $\alpha \in (- \infty, -1] \cup [\frac{1}{2}, \infty)$, the following relation holds for Eq.~\eqref{main_eq_def_f_alpha_001_001}~\cite{Takahashi_001}:
\begin{align}
	f_\frac{1}{2}^\mathrm{sw} (\cdot) \succeq f_\alpha^\mathrm{sw} (\cdot) \succeq f_{-1}^\mathrm{sw} (\cdot). \label{main_eq_relation_Petz_functions_monotone_001_001}
\end{align}
The equality in Eq.~\eqref{main_eq_relation_Petz_functions_monotone_001_001} holds for $t = 1$.
As explained later, Eq.~\eqref{main_eq_relation_Petz_functions_monotone_001_001} implies that the SLD metric realizes the fastest convergence of QNG among the monotone metrics.

Due to Theorem~\ref{main_theorem_maximum_minimum_elements_order_operator_monotone_functions_001_001}, the Petz function, Eq.~\eqref{main_eq_def_f_alpha_001_001}, for $\alpha \in (-1, 0) \cup (0, 1/2)$ yields non-monotone quantum Fisher metrics.
Note that Eq.~\eqref{main_eq_def_f_alpha_001_001} is singular at $\alpha = 0$.
For $\alpha_1 \in (0, 1/2)$, $\alpha_2 \in (-\infty, -1] \cup [1/2, \infty)$, and $\alpha_3 \in (-1, 0)$, the Petz function, Eq.~\eqref{main_eq_def_f_alpha_001_001}, satisfies the following relation:
\begin{align}
	f_{\alpha_1}^\mathrm{sw} (\cdot) \succeq f_{\alpha_2}^\mathrm{sw} (\cdot) \succeq f_{\alpha_3}^\mathrm{sw} (\cdot). \label{main_eq_relation_Petz_functions_non-monotone_001_001}
\end{align}
The equality in Eq.~\eqref{main_eq_relation_Petz_functions_non-monotone_001_001} holds for $t = 1$.
For $\alpha_1, \alpha_2 \in (0, 1/2), (-1, 0), (-\infty, -1], [1/2, \infty)$, the Petz function, Eq.~\eqref{main_eq_def_f_alpha_001_001}, satisfies
\begin{align}
	\alpha_1 \le \alpha_2 & \Leftrightarrow f_{\alpha_1}^\mathrm{sw} (\cdot) \succeq f_{\alpha_2}^\mathrm{sw} (\cdot). \label{main_eq_relation_Petz_functions_non-monotone_001_002}
\end{align}
The equality in Eq.~\eqref{main_eq_relation_Petz_functions_non-monotone_001_002} holds for $t = 1$.
Note that the following equality holds for the Petz function, Eq.~\eqref{main_eq_def_f_alpha_001_001}:
\begin{align}
	\lim_{\alpha \to \infty} f_\alpha^\mathrm{sw} (\cdot) & = \lim_{\alpha \to -\infty} f_\alpha^\mathrm{sw} (\cdot). \label{main_eq_Petz_function_limit_001_001}
\end{align}
More importantly, we have the following relation for the Petz function, Eq.~\eqref{main_eq_def_f_alpha_001_001} for $\alpha \ne 0$:
\begin{align}
	\lim_{\bar{\alpha} \to 0+} f_{\bar{\alpha}}^\mathrm{sw} (\cdot) \succeq f_\alpha^\mathrm{sw} (\cdot). \label{main_eq_Petz_function_limit_002_001}
\end{align}
From the viewpoint of the design problem for $\alpha$ in Eqs.~\eqref{main_eq_def_rescaled_sandwiched_quantum_Renyi_divergence_001_001} and \eqref{main_eq_def_f_alpha_001_001}, Eq.~\eqref{main_eq_Petz_function_limit_002_001} plays an important role.
Figure~\ref{main_fig_gnuplot_operator_monotone_functions_002_001} demonstrates Eqs.~\eqref{main_eq_relation_Petz_functions_non-monotone_001_001}, \eqref{main_eq_relation_Petz_functions_non-monotone_001_002}, \eqref{main_eq_Petz_function_limit_001_001}, and \eqref{main_eq_Petz_function_limit_002_001}.

\section{Fisher metrics and convergence speed of QNG} \label{main_sec_speed_QNG_001_001}

Convergence speed is one of the most important features of optimization algorithms that characterize their efficiency.
In this section, we explain the relation between quantum Fisher metrics and the convergence speed of QNG, following Ref.~\cite{Sasaki_001}.

\subsection{Petz functions and quantum Fisher metrics}

In general, the following theorem holds:
\begin{theorem} \label{main_theorem_relation_Petz_function_quantum_Fisher_metrics_001_001}
	Let us assume that $f (\cdot)$ and $g_{\hat{\rho}_\theta, f (\cdot)} (\cdot, \cdot)$ are related via Eq.~\eqref{main_eq_quantum_Fisher_metric_001_001}.
	Then, we have
	\begin{align}
		f (\cdot) \preceq \tilde{f} (\cdot) & \Leftrightarrow g_{\hat{\rho}_\theta, f (\cdot)} (\cdot, \cdot) \succeq g_{\hat{\rho}_\theta, \tilde{f} (\cdot)} (\cdot, \cdot), \label{main_eq_relation_Petz_function_quantum_Fisher_metric_001_001}
	\end{align}
	where
	\begin{align}
		\text{$g (\cdot, \cdot) \succeq \tilde{g} (\cdot, \cdot)$} & :\Leftrightarrow \text{$\forall X$, $g (X, X) \ge \tilde{g} (X, X)$}.
	\end{align}
\end{theorem}

The proof of Theorem~\ref{main_theorem_relation_Petz_function_quantum_Fisher_metrics_001_001} is as follows:
\begin{proof}
	We show Eq.~\eqref{main_eq_relation_Petz_function_quantum_Fisher_metric_001_001}.
	From the quantum Fisher metric, Eq.~\eqref{main_eq_quantum_Fisher_metric_001_002}, we have
	\begin{align}
		g_{\hat{\rho}_\theta, f (\cdot)} (X, X) & = \sum_{i, j = 1}^N \frac{1}{p_j f (p_i / p_j)} \langle \psi_j | \hat{X}_{\hat{\rho}_\theta}^\mathrm{m} | \psi_i \rangle \langle \psi_i | \hat{X}_{\hat{\rho}_\theta}^\mathrm{m} | \psi_j \rangle \\
		                                        & = \sum_{i, j = 1}^N \frac{1}{p_j f (p_i / p_j)} | \langle \psi_j | \hat{X}_{\hat{\rho}_\theta}^\mathrm{m} | \psi_i \rangle |^2.
	\end{align}
	Since $f (\cdot) \preceq \tilde{f} (\cdot)$, we have
	\begin{align}
		g_{\hat{\rho}_\theta, f (\cdot)} (X, X) & = \sum_{i, j = 1}^N \frac{1}{p_j f (p_i / p_j)} | \langle \psi_j | \hat{X}_{\hat{\rho}_\theta}^\mathrm{m} | \psi_i \rangle |^2           \\
		                                        & \ge \sum_{i, j = 1}^N \frac{1}{p_j \tilde{f} (p_i / p_j)} | \langle \psi_j | \hat{X}_{\hat{\rho}_\theta}^\mathrm{m} | \psi_i \rangle |^2 \\
		                                        & = g_{\hat{\rho}_\theta, \tilde{f} (\cdot)} (X, X).
	\end{align}
	Thus, Eq.~\eqref{main_eq_relation_Petz_function_quantum_Fisher_metric_001_001} is proved.
\end{proof}

\subsection{Diagonal approximation of the quantum Fisher metrics}

	The diagonal approximation of the Fisher metric is often employed when NG is applied to practical cases, and it is quite natural to consider it for QNG.
	Here, we deal with the problem of whether Theorem~\ref{main_theorem_relation_Petz_function_quantum_Fisher_metrics_001_001} holds or not when the diagonal approximation is used.

We define the following diagonal metric:
\begin{align}
	\bar{g}_{\hat{\rho}_\theta, f (\cdot)} (\partial_i, \partial_j) & \coloneqq
	\begin{cases}
		g_{\hat{\rho}_\theta, f (\cdot)} (\partial_i, \partial_j) & (i = j),   \\
		0                                                         & (i \ne j).
	\end{cases} \label{main_eq_def_tilde_g_001_001}
\end{align}
In partial usages, Eq.~\eqref{main_eq_def_tilde_g_001_001} is of great importance and is often used instead of Eq.~\eqref{main_eq_quantum_Fisher_metric_001_001} because the computation of the inverse matrix of Eq.~\eqref{main_eq_quantum_Fisher_metric_001_001} is numerically costly.

Similarly to Theorem~\ref{main_theorem_relation_Petz_function_quantum_Fisher_metrics_001_001}, the following theorem holds:
\begin{theorem} \label{main_theorem_relation_Petz_function_quantum_Fisher_metrics_002_001}
	The following relation holds for Eq.~\eqref{main_eq_def_tilde_g_001_001}:
	\begin{align}
		f (t) \preceq \tilde{f} (\cdot) & \Leftrightarrow \bar{g}_{\hat{\rho}_\theta, f (\cdot)} (\cdot, \cdot) \succeq \bar{g}_{\hat{\rho}_\theta, \tilde{f} (\cdot)} (\cdot, \cdot). \label{main_eq_relation_Petz_function_quantum_Fisher_metric_002_001}
	\end{align}
\end{theorem}
The proof of Theorem~\ref{main_theorem_relation_Petz_function_quantum_Fisher_metrics_002_001} is almost the same as that of Theorem~\ref{main_theorem_relation_Petz_function_quantum_Fisher_metrics_001_001}.

\subsection{Dependence of the convergence speed of QNG on the quantum Fisher metrics}

We discuss the dependence of the convergence speed of QNG on the quantum Fisher metrics.
Metrics $g (\cdot, \cdot)$ and $\tilde{g} (\cdot, \cdot)$ satisfy the following relation:
\begin{align}
	g (\cdot, \cdot) \succeq \tilde{g} (\cdot, \cdot) & \Leftrightarrow g^{-1} (\cdot, \cdot) \preceq \tilde{g}^{-1} (\cdot, \cdot).
\end{align}
In the case of Eq.~\eqref{main_eq_update_theta_001_001}, we have
\begin{align}
	L (\theta_{\tau+1}) - L (\theta_\tau) & = L (\theta_\tau + \Delta \theta) - L (\theta_\tau)                                                                      \\
	                                      & = \nabla_\theta L (\theta_\tau)^\intercal \Delta \theta                                                                  \\
	                                      & = - \sqrt{2 \epsilon \nabla_\theta L (\theta_\tau)^\intercal G_\alpha^{-1} (\theta_\tau) \nabla_\theta L (\theta_\tau)}.
\end{align}
Similarly, in the case of Eq.~\eqref{main_eq_update_theta_001_002}, we also have
\begin{align}
	L (\theta_{\tau+1}) - L (\theta_\tau) & = - \eta \nabla_\theta L (\theta_\tau)^\intercal G_\alpha^{-1} (\theta_\tau) \nabla_\theta L (\theta_\tau).
\end{align}
Thus, for both of Eqs.~\eqref{main_eq_update_theta_001_001} and \eqref{main_eq_update_theta_001_002}, $G_\alpha (\theta_\tau)$ decreases the value of $L (\theta_\tau)$ faster than $G_{\alpha'} (\theta_\tau)$ when $G_\alpha (\theta_\tau) \preceq G_{\alpha'} (\theta_\tau)$, that is, $G_\alpha^{-1} (\theta_\tau) \succeq G_{\alpha'}^{-1} (\theta_\tau)$.

\section{Designing geometries via Petz functions} \label{main_sec_design_of_geometry_via_Petz_functions_001_001}

The relationship between the quantum Fisher metric and the convergence speed of QNG implies that we may design an efficient algorithm by constructing a Petz function without explicitly considering quantum divergences.
This section explores a naive method to engineer a Petz function.
Furthermore, we perform numerical simulations of the newly designed Petz function.

Let us consider the linear combination of two Petz functions:
\begin{align}
	f_\alpha^\mathrm{lin} (t) & \coloneqq (1 - \alpha) f_1 (t) + \alpha f_2 (t), \label{main_eq_def_combined_Petz_function_001_001}
\end{align}
where $f_1 (\cdot)$ and $f_2 (\cdot)$ satisfy Eq.~\eqref{main_eq_condition_Petz_function_001_001}.
Then, Eq.~\eqref{main_eq_def_combined_Petz_function_001_001} also satisfies Eq.~\eqref{main_eq_condition_Petz_function_001_001}.

\begin{theorem} \label{main_theorem_operator_monotonicity_f_alpha_001_001}
	When $f_1 (\cdot)$ and $f_2 (\cdot)$ in Eq.~\eqref{main_eq_def_combined_Petz_function_001_001} are operator-monotone, then $f_\alpha^\mathrm{lin} (\cdot)$ in Eq.~\eqref{main_eq_def_combined_Petz_function_001_001} is also operator-monotone for $\alpha \in [0, 1]$.
\end{theorem}
The proof of Theorem~\ref{main_theorem_operator_monotonicity_f_alpha_001_001} is as follows:
\begin{proof}
	From Eq.~\eqref{main_eq_def_combined_Petz_function_001_001}, we have
	\begin{align}
		 & f_\alpha^\mathrm{lin} (t) - f_\alpha^\mathrm{lin} (t') \nonumber                            \\
		 & \quad = [(1 - \alpha) f_1 (t) + \alpha f_2 (t)] - [(1 - \alpha) f_1 (t') + \alpha f_2 (t')] \\
		 & \quad = (1 - \alpha) [f_1 (t) - f_1 (t')] + \alpha [f_2 (t) - f_2 (t')].
	\end{align}
	Then, when $\hat{A} \succeq \hat{B}$, we get
	\begin{align}
		 & f_\alpha^\mathrm{lin} (\hat{A}) - f_\alpha^\mathrm{lin} (\hat{B}) \nonumber                   \\
		 & \quad = (1 - \alpha) [f_1 (\hat{A}) - f_1 (\hat{B})] + \alpha [f_2 (\hat{A}) - f_2 (\hat{B})] \\
		 & \quad \ge 0.
	\end{align}
	Thus, the claim Theorem~\ref{main_theorem_operator_monotonicity_f_alpha_001_001} holds.
\end{proof}

\begin{figure}[t]
	\centering
	\includegraphics[scale=0.60]{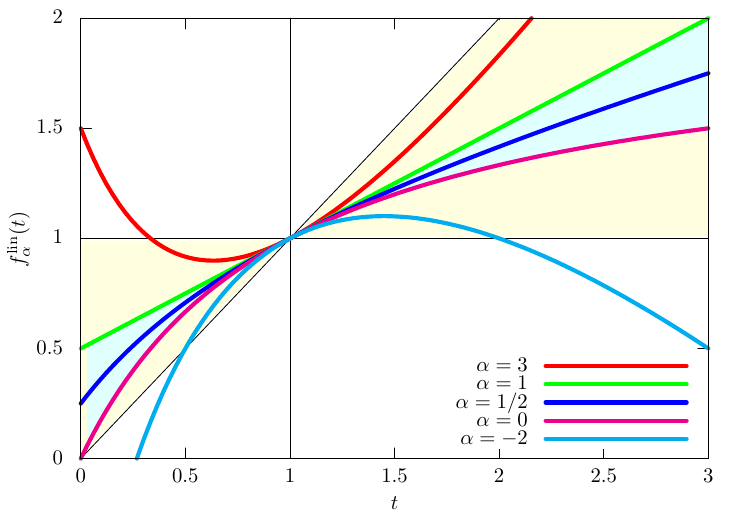}
	\caption{$f_\alpha^\mathrm{lin} (t)$, Eq.~\eqref{main_eq_def_combined_Petz_function_001_001}, in the case of $f_1 (t) = f_\mathrm{rRLD} (t)$ and $f_2 (t) = f_\mathrm{SLD} (t)$. The regimes of the monotone Petz functions and the Petz functions associated with the rescaled sandwiched quantum R\'enyi divergence are highlighted by light cyan and light yellow, respectively.}
	\label{main_fig_gnuplot_operator_monotone_functions_004_001}
\end{figure}

In Fig.~\ref{main_fig_gnuplot_operator_monotone_functions_004_001}, $f_\alpha^\mathrm{lin} (t)$, Eq.~\eqref{main_eq_def_combined_Petz_function_001_001}, in the case of $f_1 (t) = f_\mathrm{rRLD} (t)$ and $f_2 (t) = f_\mathrm{SLD} (t)$ is shown.
As shown in Eq.~\eqref{main_eq_f0+_001_001}, $f_\alpha^\mathrm{lin} (\cdot)$, Eq.~\eqref{main_eq_def_combined_Petz_function_001_001}, with $\alpha > 2$ has a larger value than $f_\alpha^\mathrm{sw} (t)$, Eq.~\eqref{main_eq_def_f_alpha_001_001}, for at least small $t$.
In general, $f_\alpha^\mathrm{lin} (\cdot)$, Eq.~\eqref{main_eq_def_combined_Petz_function_001_001}, and $f_{\alpha'}^\mathrm{sw} (t)$, Eq.~\eqref{main_eq_def_f_alpha_001_001}, do not satisfy the partial order defined in Eq.~\eqref{main_eq_def_order_operator_monotone_functions_001_001}.

\section{Numerical simulations} \label{main_sec_numerical_simulations_001_001}

In this section, we present numerical simulations to support the main claim of this paper.
We also provide additional numerical simulations to confirm that the observations made here also hold for other models in Appendix~\ref{main_sec_additional_numerical_simulations_001_001}.

\subsection{Setup}

Using $\theta \coloneqq [\theta^1, \theta^2, \theta^3]^\intercal$, we define the following parameterized quantum state:
\begin{align}
	\hat{\rho}_\theta & \coloneqq \hat{R}_3 (\theta) \hat{\rho}_\mathrm{ini} \hat{R}_3^\dagger (\theta),
\end{align}
where
\begin{align}
	\hat{R}_3 (\theta) & \coloneqq \hat{R}_z (\theta^3) \hat{R}_y (\theta^2) \hat{R}_z (\theta^1)                                                 \\
	                   & =
	\begin{bmatrix}
		\mathrm{e}^{- \mathrm{i} \frac{\theta^1 + \theta^3}{2}} \cos \theta^2 & - \mathrm{e}^{\mathrm{i} \frac{\theta^1 - \theta^3}{2}} \sin \theta^2 \\
		\mathrm{e}^{- \mathrm{i} \frac{\theta^1 - \theta^3}{2}} \sin \theta^2 & \mathrm{e}^{- \mathrm{i} \frac{\theta^1 + \theta^3}{2}} \cos \theta^2
	\end{bmatrix}, \\
	\hat{R}_z (\phi)   & \coloneqq \exp \bigg( - \mathrm{i} \frac{\phi}{2} \hat{\sigma}_z \bigg),                                                 \\
	\hat{R}_y (\phi)   & \coloneqq \exp \bigg( - \mathrm{i} \frac{\phi}{2} \hat{\sigma}_y \bigg).
\end{align}
Furthermore, we define
\begin{align}
	\hat{\rho}_\mathrm{ini} & \coloneqq \frac{1}{2} (\hat{1} + x \hat{\sigma}_x + y \hat{\sigma}_y + z \hat{\sigma}_z),
\end{align}
where $\hat{1}$ is the $2 \times 2$ identity operator and $x^2 + y^2 + z^2 < 1$.

Then, we consider the following minimization problem with respect to $\theta$:
\begin{align}
	\min_{\theta \in \mathbb{R}^{N_\theta}} L (\theta),
\end{align}
where
\begin{align}
	L (\theta) & \coloneqq \mathrm{Tr} [(\hat{\rho}_\theta - \hat{\rho}_{\theta_*})^\dagger (\hat{\rho}_\theta - \hat{\rho}_{\theta_*})].
\end{align}

To stabilize the numerical calculations, we conduct the following operations on the density operator and metric:
\begin{align}
	\hat{\rho}_\theta & \leftarrow (1 - \delta) \hat{\rho}_\theta + \delta \frac{\hat{1}}{N}, \\
	G_\alpha (\theta) & \leftarrow (1 - \xi) G_\alpha (\theta) + \xi I,
\end{align}
where $I$ is the identity matrix whose size is identical to that of $G_\alpha (\theta)$, and $\delta$ and $\xi$ are tiny positive numbers.

\subsection{Results}

\begin{figure}[t]
	\centering
	\includegraphics[scale=0.60]{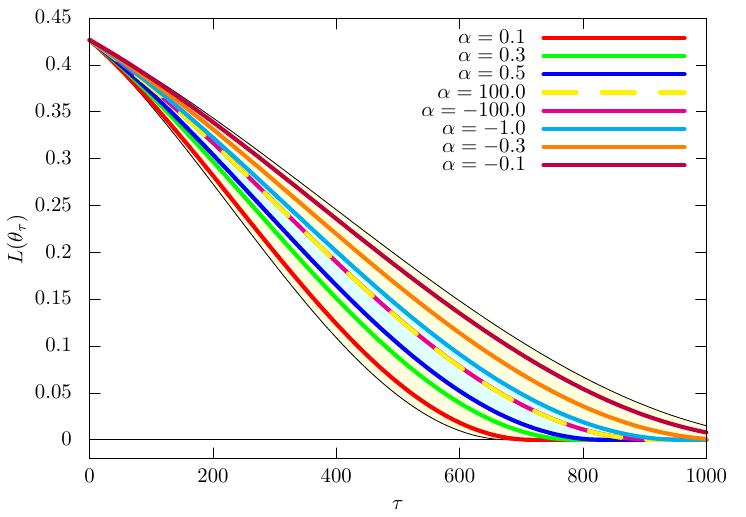}
	\includegraphics[scale=0.60]{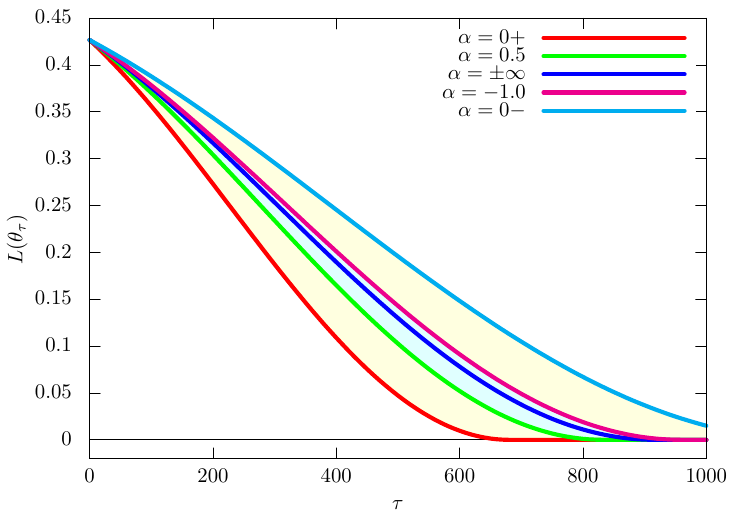}
	\caption{Cost functions in the case of $f_\alpha^\mathrm{sw} (t)$, Eq.~\eqref{main_eq_def_f_alpha_001_001}, for several $\alpha$ in the case of Eq.~\eqref{main_eq_update_theta_001_001}. We set $\epsilon = 1.0 \times 10^{-6}$, $\xi = 1.0 \times 10^{-3}$, and $\delta = 1.0 \times 10^{-3}$. The regimes of the monotone metrics and the rescaled sandwiched quantum R\'enyi divergence are highlighted by light cyan and light yellow, respectively.}
	\label{main_fig_performance_QNG_sandwiched_rot-states_001_001}
\end{figure}

\begin{figure}[t]
	\centering
	\includegraphics[scale=0.60]{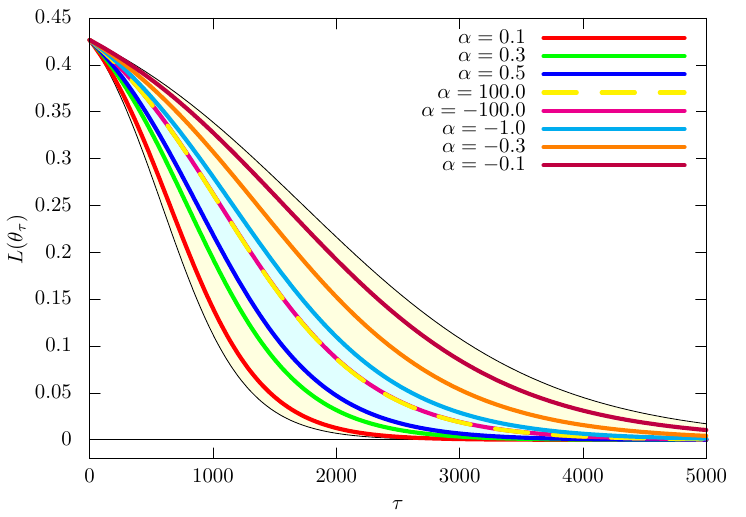}
	\includegraphics[scale=0.60]{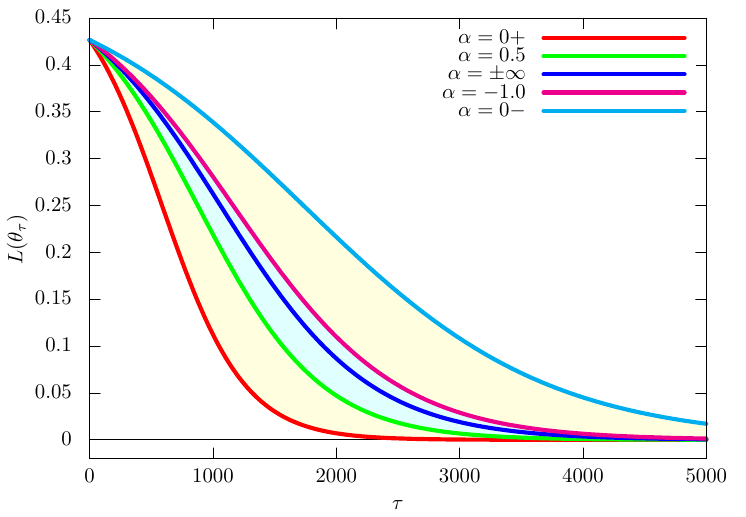}
	\caption{Cost functions in the case of $f_\alpha^\mathrm{sw} (t)$, Eq.~\eqref{main_eq_def_f_alpha_001_001}, for several $\alpha$ in the case of Eq.~\eqref{main_eq_update_theta_001_002}. We set $\eta = 1.0 \times 10^{-3}$, $\xi = 1.0 \times 10^{-3}$, and $\delta = 1.0 \times 10^{-3}$. The regimes of the monotone metrics and the rescaled sandwiched quantum R\'enyi divergence are highlighted by light cyan and light yellow, respectively.}
	\label{main_fig_performance_QNG_sandwiched_rot-states_001_002}
\end{figure}

In the following calculations, we set $[x, y, z] = [0.5, 0.0, 0.0]$, $\theta_0 = [\pi / 2, \pi / 2, \pi / 4]^\intercal$, and $\theta_* = [0.0, 0.0, 0.0]^\intercal$.
First, we consider the Petz function associated with the rescaled sandwiched quantum R\'enyi divergence, Eq.~\eqref{main_eq_def_f_alpha_001_001}.
In Fig.~\ref{main_fig_performance_QNG_sandwiched_rot-states_001_001}, we plot the time evolution of the cost function for the case of Eq.~\eqref{main_eq_update_theta_001_001}.
In Fig.~\ref{main_fig_performance_QNG_sandwiched_rot-states_001_002}, we show the time evolution of the cost function for the case of Eq.~\eqref{main_eq_update_theta_001_002}.
In both cases, QNG with $\alpha \in (0.0, 0.5)$ outperforms SLD-based QNG proposed in Refs.~\cite{Stokes_001, Koczor_001}.
Furthermore, these figures show that the gaps between different values of $\alpha$ become larger in the case of Eq.~\eqref{main_eq_update_theta_001_002} than in Eq.~\eqref{main_eq_update_theta_001_001}.

\begin{figure}[t]
	\centering
	\includegraphics[scale=0.60]{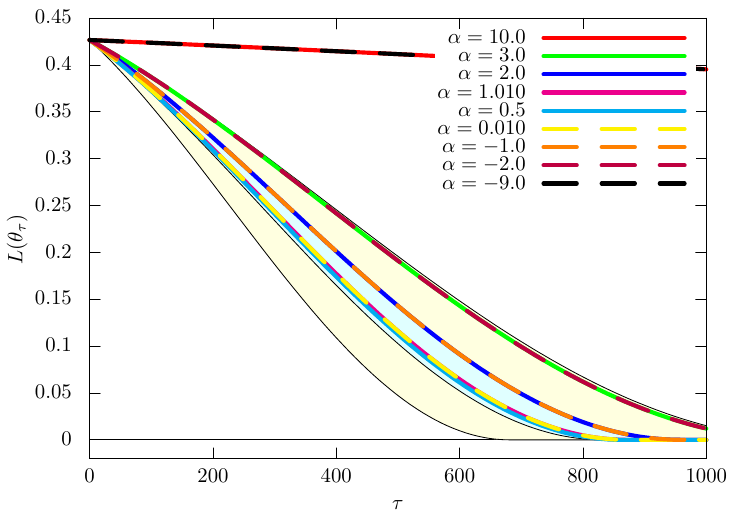}
	\caption{Cost functions in the case of $f_\alpha^\mathrm{st} (t)$, Eq.~\eqref{main_eq_def_tilde_f_alpha_001_001}, for several $\alpha$ in the case of Eq.~\eqref{main_eq_update_theta_001_001}. We set $\epsilon = 1.0 \times 10^{-6}$, $\xi = 1.0 \times 10^{-3}$, and $\delta = 1.0 \times 10^{-3}$. The regimes of the monotone metrics and the rescaled sandwiched quantum R\'enyi divergence are highlighted by light cyan and light yellow, respectively.}
	\label{main_fig_performance_QNG_standard_rot-states_001_001}
\end{figure}

\begin{figure}[t]
	\centering
	\includegraphics[scale=0.60]{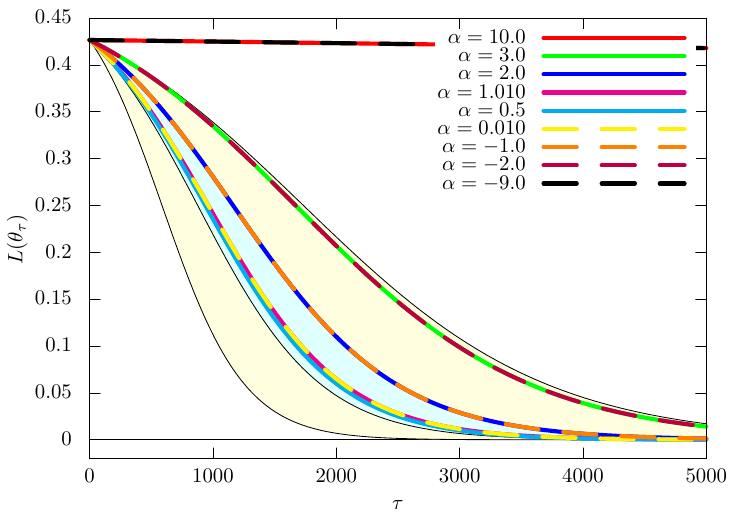}
	\caption{Cost functions in the case of $f_\alpha^\mathrm{st} (t)$, Eq.~\eqref{main_eq_def_tilde_f_alpha_001_001}, for several $\alpha$ in the case of Eq.~\eqref{main_eq_update_theta_001_002}. We set $\eta = 1.0 \times 10^{-3}$, $\xi = 1.0 \times 10^{-3}$, and $\delta = 1.0 \times 10^{-3}$. The regimes of the monotone metrics and the rescaled sandwiched quantum R\'enyi divergence are highlighted by light cyan and light yellow, respectively.}
	\label{main_fig_performance_QNG_standard_rot-states_001_002}
\end{figure}

Next, we conduct the calculations in the case of the Petz function associated with the rescaled standard quantum R\'enyi divergence, Eq.~\eqref{main_eq_def_tilde_f_alpha_001_001}.
In Fig.~\ref{main_fig_performance_QNG_standard_rot-states_001_001}, we plot the time evolution of the cost function for the case of Eq.~\eqref{main_eq_update_theta_001_001}.
In Fig.~\ref{main_fig_performance_QNG_standard_rot-states_001_002}, we show the time evolution of the cost function for the case of Eq.~\eqref{main_eq_update_theta_001_002}.
As expected, the convergence speed of QNG with the Petz function associated with the rescaled standard quantum R\'enyi divergence, Eq.~\eqref{main_eq_def_tilde_f_alpha_001_001}, becomes slower than that with the Petz function associated with the rescaled sandwiched quantum R\'enyi divergence, Eq.~\eqref{main_eq_def_f_alpha_001_001}

\begin{figure}[t]
	\centering
	\includegraphics[scale=0.60]{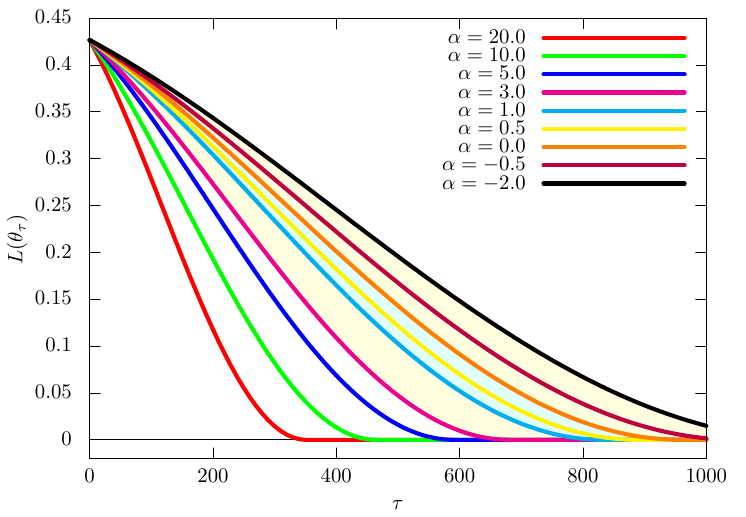}
	\caption{Cost functions in the case of $f_\alpha^\mathrm{lin} (t)$, Eq.~\eqref{main_eq_def_combined_Petz_function_001_001}, for several $\alpha$ in the case of Eq.~\eqref{main_eq_update_theta_001_001}. We set $\epsilon = 1.0 \times 10^{-6}$, $\xi = 1.0 \times 10^{-3}$, and $\delta = 1.0 \times 10^{-3}$. The regimes of the monotone metrics and the rescaled sandwiched quantum R\'enyi divergence are highlighted by light cyan and light yellow, respectively.}
	\label{main_fig_performance_QNG_linear_rot-states_001_001}
\end{figure}

\begin{figure}[t]
	\centering
	\includegraphics[scale=0.60]{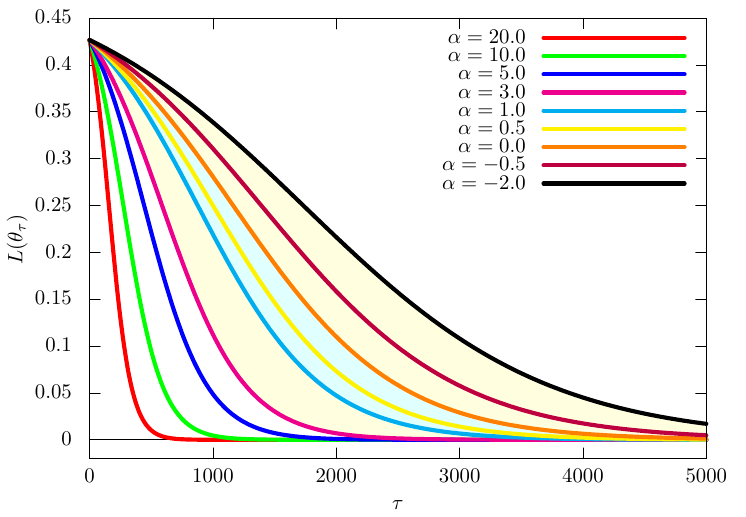}
	\caption{Cost functions in the case of $f_\alpha^\mathrm{lin} (t)$, Eq.~\eqref{main_eq_def_combined_Petz_function_001_001}, for several $\alpha$ in the case of Eq.~\eqref{main_eq_update_theta_001_002}. We set $\eta = 1.0 \times 10^{-3}$, $\xi = 1.0 \times 10^{-3}$, and $\delta = 1.0 \times 10^{-3}$. The regimes of the monotone metrics and the rescaled sandwiched quantum R\'enyi divergence are highlighted by light cyan and light yellow, respectively.}
	\label{main_fig_performance_QNG_linear_rot-states_001_002}
\end{figure}

Finally, we investigate Eq.~\eqref{main_eq_def_combined_Petz_function_001_001} for the Petz function that defines the quantum Fisher metric.
In Fig.~\ref{main_fig_performance_QNG_linear_rot-states_001_001}, we plot the time evolution of the cost function for the case of Eq.~\eqref{main_eq_update_theta_001_001}.
In Fig.~\ref{main_fig_performance_QNG_linear_rot-states_001_002}, we show the time evolution of the cost function for the case of Eq.~\eqref{main_eq_update_theta_001_002}.
For large $\alpha$, this Petz function accelerates the convergence speed of QNG.
This fact implies that designing a ``good" Petz function is important for the performance of QNG.

\section{Discussions} \label{main_sec_discussions_001_001}

We here discuss possible issues regarding QNG.

\subsection{Why QNG outperforms Newton's method for finding second-order phase transitions}

In computational physics, Monte Carlo methods play an important role, especially for strongly correlated materials and lattice QCD, and the VMC method, which aims to find the ground state of a given Hamiltonian by assuming a parameterized trial function, is one of the most popular methods because it does not suffer from the negative sign problem~\cite{Scherer_001, Tahara_001}.
To stabilize and accelerate the VMC method, the SR algorithm was proposed~\cite{Sorella_001, Sorella_002, Sorella_003, Becca_001, Mazzola_001, Park_001, Xie_001}.
Interestingly, SLD-based QNG and the SR algorithm were developed in different fields, but it is known that these are mathematically equivalent~\cite{Becca_001}.

In SLD-based QNG, the second derivative of Eq.~\eqref{main_eq_def_rescaled_sandwiched_quantum_Renyi_divergence_001_001} with $\alpha = \frac{1}{2}$ is used instead of that of Eq.~\eqref{main_eq_def_quantum_KL_divergence_001_001}.
At critical points where second-order phase transitions occur, the second derivative of thermodynamic quantities diverges, and then the second derivative of Eq.~\eqref{main_eq_def_quantum_KL_divergence_001_001} is expected to be unstable.
Furthermore, as discussed above, SLD-based QNG is faster and more stable than BKM-based QNG.
Thus, SLD-based QNG and the SR algorithm are beneficial for optimization problems such as the VMC method.
In the case of Newton's method, the second derivative of the free energy is used, but this is also expected to be unstable at critical points.

\subsection{Constraints of QNG}

QNG for full-rank density operators is mainly investigated in this paper; so, the reader may think that it cannot be directly applied to optimization problems of pure states, such as the VMC method.
However, the assumption that density operators are of full rank is not a physical problem for several reasons.
The first reason is that creating pure states is extremely difficult in experiments~\cite{Taranto_001}.
The second one is that the cost to realize projective measurements is not finite~\cite{Guryanova_001}.
The third one is that quantum systems are also governed by the third law of thermodynamics~\cite{Mohammady_001}.
More interestingly, the QNG framework generalized in this paper can be applied to pure states by limiting ourselves to the Petz functions such that $f (0) > 0$, as discussed in Sec.~\ref{main_sec_bra-ket_notation_non-full-rank-case_001_001}.

\section{Conclusions} \label{main_sec_conclusions_001_001}

In this paper, we have generalized QNG based on QIG.
More specifically, QNG in the literature typically refers to the SLD metric, but the SLD metric is not the unique metric that corresponds to the classical Fisher metric in the classical limit.
Furthermore, in QNG, the e-representation, which is specified by the Petz function, corresponds to a specific type of the logarithmic derivative of density operators and the choice of the Petz function identifies the form of the quantum Fisher metric.
In this paper, we have investigated how QNG depends on the choice of quantum divergences and quantum Fisher metrics and proposed a new approach to designing a geometry that may improve optimization efficiency.
\begin{acknowledgments}
	H.M. was supported by JSPS KAKENHI Grant Number JP25H01499.
	H.M. thanks Yusuke Nomura and Toi Sasaki for fruitful discussions.
\end{acknowledgments}

\appendix

\section{Derivations of some equations}

To make this paper as self-contained as possible, we provide the derivations of some equations in the main text~\cite{Takahashi_001}.

\subsection{Derivation of Eq.~\eqref{main_eq_Delta_theta_shift_KL_001_001}} \label{main_appendix_Fisher_metric_KL_001_001}

The detailed derivation of Eq.~\eqref{main_eq_Delta_theta_shift_KL_001_001} is as follows:
\begin{align}
	 & D_\mathrm{KL} (p_{\theta + \Delta \theta} (\cdot) \| p_\theta (\cdot)) \nonumber                                                                                                                      \\
	 & \quad = \sum_{x=1}^N p_{\theta + \Delta \theta} (x) \ln \frac{p_{\theta + \Delta \theta} (x)}{p_\theta (x)}                                                                                           \\
	 & \quad \approx \sum_{x=1}^N (p_\theta (x) + \nabla_\theta p_\theta (x) \cdot \Delta \theta) \nonumber                                                                                                  \\
	 & \qquad \times \ln \bigg( 1 + \frac{1}{p_\theta (x)} \nabla_\theta p_\theta (x) \cdot \Delta \theta \bigg)                                                                                             \\
	 & \quad \approx \sum_{x=1}^N (p_\theta (x) + \nabla_\theta p_\theta (x) \cdot \Delta \theta) \nonumber                                                                                                  \\
	 & \qquad \times \bigg[ \frac{1}{p_\theta (x)} \nabla_\theta p_\theta (x) \cdot \Delta \theta - \frac{1}{2} \bigg( \frac{1}{p_\theta (x)} \nabla_\theta p_\theta (x) \cdot \Delta \theta \bigg)^2 \bigg] \\
	 & \quad \approx \sum_{x=1}^N \nabla_\theta p_\theta (x) \cdot \Delta \theta \nonumber                                                                                                                   \\
	 & \qquad + \frac{1}{2} \sum_{x=1}^N p_\theta (x) \bigg( \frac{1}{p_\theta (x)} \nabla_\theta p_\theta (x) \cdot \Delta \theta \bigg)^2                                                                  \\
	 & \quad = \frac{1}{2} \sum_{x=1}^N p_\theta (x) \bigg( \sum_{i, j = 1}^{N_\theta} (\partial_i \ln p_\theta (x)) (\partial_j \ln p_\theta (x)) \bigg) \Delta \theta^i \Delta \theta^j                    \\
	 & \quad = \frac{1}{2} \sum_{i, j = 1}^{N_\theta} g_{p_\theta (\cdot)} (\partial_i, \partial_j) \Delta \theta^i \Delta \theta^j.
\end{align}

\subsection{Derivation of Eq.~\eqref{main_eq_Delta_theta_shift_rescaled_classical_Renyi_001_001}} \label{main_appendix_Fisher_metric_rescaled_classical_Renyi_001_001}

\begin{widetext}
	The detailed derivation of Eq.~\eqref{main_eq_Delta_theta_shift_rescaled_classical_Renyi_001_001} is as follows:
	\begin{align}
		 & R_\alpha (p_{\theta + \Delta \theta} (\cdot) \| p_\theta (\cdot)) \nonumber                                                                                                                                                                                                                                         \\
		 & \quad = \frac{1}{\alpha (\alpha - 1)} \ln \sum_{x=1}^N p_{\theta + \Delta \theta}^\alpha (x) p_\theta^{1 - \alpha} (x)                                                                                                                                                                                              \\
		 & \quad \approx \frac{1}{\alpha (\alpha - 1)} \ln \sum_{x=1}^N (p_\theta (x) + \nabla_\theta p_\theta (x) \cdot \Delta \theta)^\alpha p_\theta^{1 - \alpha} (x)                                                                                                                                                       \\
		 & \quad = \frac{1}{\alpha (\alpha - 1)} \ln \sum_{x=1}^N \bigg( \sum_{n=0}^\alpha \binom{\alpha}{n} p_\theta^{\alpha - n} (x) (\nabla_\theta p_\theta (x) \cdot \Delta \theta)^n \bigg) p_\theta^{1 - \alpha} (x)                                                                                                     \\
		 & \quad \approx \frac{1}{\alpha (\alpha - 1)} \ln \sum_{x=1}^N \bigg( \sum_{n=0}^2 \binom{\alpha}{n} p_\theta^{\alpha - n} (x) (\nabla_\theta p_\theta (x) \cdot \Delta \theta)^n \bigg) p_\theta^{1 - \alpha} (x)                                                                                                    \\
		 & \quad = \frac{1}{\alpha (\alpha - 1)} \ln \sum_{x=1}^N \bigg( p_\theta^\alpha (x) + \alpha p_\theta^{\alpha - 1} (x) (\nabla_\theta p_\theta (x) \cdot \Delta \theta) + \frac{\alpha (\alpha - 1)}{2} p_\theta^{\alpha - 2} (x) (\nabla_\theta p_\theta (x) \cdot \Delta \theta)^2 \bigg) p_\theta^{1 - \alpha} (x) \\
		 & \quad = \frac{1}{\alpha (\alpha - 1)} \ln \sum_{x=1}^N \bigg( p_\theta (x) + \alpha \nabla_\theta p_\theta (x) \cdot \Delta \theta + \frac{\alpha (\alpha - 1)}{2} p_\theta^{-1} (x) (\nabla_\theta p_\theta (x) \cdot \Delta \theta)^2 \bigg)                                                                      \\
		 & \quad = \frac{1}{\alpha (\alpha - 1)} \ln \bigg[ \sum_{x=1}^N p_\theta (x) + \alpha \nabla_\theta \bigg( \sum_{x=1}^N p_\theta (x) \bigg) \cdot \Delta \theta + \frac{\alpha (\alpha - 1)}{2} \sum_{x=1}^N p_\theta (x) \bigg( \frac{\nabla_\theta p_\theta (x) \cdot \Delta \theta}{p_\theta (x)} \bigg)^2 \bigg]  \\
		 & \quad = \frac{1}{\alpha (\alpha - 1)} \ln \bigg[ 1 + \frac{\alpha (\alpha - 1)}{2} \sum_{x=1}^N p_\theta (x) \bigg( \frac{\nabla_\theta p_\theta (x) \cdot \Delta \theta}{p_\theta (x)} \bigg)^2 \bigg]                                                                                                             \\
		 & \quad \approx \frac{1}{2} \sum_{x=1}^N p_\theta (x) \bigg( \frac{\nabla_\theta p_\theta (x) \cdot \Delta \theta}{p_\theta (x)} \bigg)^2                                                                                                                                                                             \\
		 & \quad = \frac{1}{2} \sum_{x=1}^N p_\theta (x) (\nabla_\theta \ln p_\theta (x) \cdot \Delta \theta)^2                                                                                                                                                                                                                \\
		 & \quad = \frac{1}{2} \sum_{x=1}^N p_\theta (x) \bigg( \sum_{i=1}^{N_\theta} \partial_i \ln p_\theta (x) \Delta \theta^i \bigg) \bigg( \sum_{j=1}^{N_\theta} \partial_j \ln p_\theta (x) \Delta \theta^j \bigg)                                                                                                       \\
		 & \quad = \frac{1}{2} \sum_{i, j = 1}^{N_\theta} \bigg( \sum_{x=1}^N p_\theta (x) (\partial_i \ln p_\theta (x)) (\partial_j \ln p_\theta (x)) \bigg) \Delta \theta^i \Delta \theta^j                                                                                                                                  \\
		 & \quad = \frac{1}{2} \sum_{i, j = 1}^{N_\theta} g_{p_\theta (\cdot)} (\partial_i, \partial_j) \Delta \theta^i \Delta \theta^j.
	\end{align}
\end{widetext}

\subsection{Derivation of Eq.~\eqref{main_eq_derivative_rescaled_classical_Renyi_divergence_discrete_distribution_001_001}} \label{main_sec_Fisher_metric_rescaled_Renyi_divergence_001_001}

The first-order derivative of Eq.~\eqref{main_eq_def_rescaled_classical_Renyi_divergence_discrete_distribution_001_001} is given by
\begin{align}
	\bar{\partial}_j R_\alpha (p_{\bar{\theta}} (\cdot) \| p_\theta (\cdot)) |_{\bar{\theta} = \theta} & = \bar{\partial}_j \frac{1}{\alpha (\alpha - 1)} \ln \sum_{i = 1}^N \bar{p}_i^\alpha p_i^{1 - \alpha}                                                                                                                                                                 \\
	                                                                                                   & = \frac{1}{\alpha - 1} \frac{\bar{p}_j^{\alpha - 1} p_j^{1 - \alpha} - \bar{p}_N^{\alpha - 1} p_N^{1 - \alpha}}{\sum_{i = 1}^N \bar{p}_i^\alpha p_i^{1 - \alpha}}. \label{main_eq_first_derivative_rescaled_classical_Renyi_divergence_discrete_distribution_001_001}
\end{align}
From Eq.~\eqref{main_eq_first_derivative_rescaled_classical_Renyi_divergence_discrete_distribution_001_001}, the second-order derivative of Eq.~\eqref{main_eq_def_rescaled_classical_Renyi_divergence_discrete_distribution_001_001} in the limit as $\bar{\theta} \to \theta$ is computed as
\begin{widetext}
	\begin{align}
		\bar{\partial}_i (\bar{\partial}_j R_\alpha (p_{\bar{\theta}} (\cdot) \| p_\theta (\cdot)) |_{\bar{\theta} = \theta}) & = \bar{\partial}_i \bigg( \bar{\partial}_j \frac{1}{\alpha (\alpha - 1)} \ln \sum_{i = 1}^N \bar{p}_i^\alpha p_i^{1 - \alpha} \bigg)                                                                                                                                                                                                                                                                                        \\
		                                                                                                                      & = \bar{\partial}_i \bigg( \frac{1}{\alpha - 1} \frac{\bar{p}_j^{\alpha - 1} p_j^{1 - \alpha} - \bar{p}_N^{\alpha - 1} p_N^{1 - \alpha}}{\sum_{i = 1}^N \bar{p}_i^\alpha p_i^{1 - \alpha}} \bigg)                                                                                                                                                                                                                            \\
		                                                                                                                      & = \frac{\delta_{i, j} \bar{p}_j^{\alpha - 2} p_j^{1 - \alpha} + \bar{p}_N^{\alpha - 2} p_N^{1 - \alpha}}{\sum_{i = 1}^N \bar{p}_i^\alpha p_i^{1 - \alpha}} + \frac{\alpha}{\alpha - 1} \frac{(\bar{p}_i^{\alpha - 1} p_i^{1 - \alpha} - \bar{p}_N^{\alpha - 1} p_N^{1 - \alpha}) (\bar{p}_j^{\alpha - 1} p_j^{1 - \alpha} - \bar{p}_N^{\alpha - 1} p_N^{1 - \alpha})}{(\sum_{i = 1}^N \bar{p}_i^\alpha p_i^{1 - \alpha})^2} \\
		                                                                                                                      & \xrightarrow{\bar{\theta} \to \theta} \delta_{i, j} p_i^{-1} + p_N^{-1}.
	\end{align}
	Thus, we obtain Eq.~\eqref{main_eq_derivative_rescaled_classical_Renyi_divergence_discrete_distribution_001_001}.
\end{widetext}

\subsection{Quantum $F$-divergence and its Petz function} \label{main_sec_quantum_F_divergence_001_001}

Introducing $F (\cdot)$ such that $F (1) = 0$, we define the quantum $F$-divergence as
\begin{align}
	D_{F (\cdot)} (\hat{\rho}_{\bar{\theta}} \| \hat{\rho}_\theta) & \coloneqq \mathrm{Tr} [ \hat{\rho}_\theta F (\Delta_{\hat{\rho}_{\bar{\theta}}, \hat{\rho}_\theta}) \hat{I}], \label{main_eq_def_quantum_f_divergence_001_001}
\end{align}
where $\Delta_{\hat{\rho}_{\bar{\theta}}, \hat{\rho}_\theta} \hat{\sigma} \coloneqq \hat{\rho}_{\bar{\theta}} \hat{\sigma} \hat{\rho}_\theta^{-1}$.
Then, $F (\cdot)$ in the quantum $F$-divergence, Eq.~\eqref{main_eq_def_quantum_f_divergence_001_001}, and its Petz function $f (\cdot)$ satisfy the following relationship:
\begin{align}
	F (t) + t F (1/ t) & = \frac{(1 - t)^2}{f (t)}. \label{main_eq_Petz_function_quantum_f_divergence_001_001}
\end{align}

\begin{widetext}
	We derive Eq.~\eqref{main_eq_Petz_function_quantum_f_divergence_001_001} as follows.
	To simplify the discussion, we consider a parameterized density operator whose eigenvectors are parameterized:
	\begin{align}
		\hat{\rho}_\theta & \coloneqq \sum_{i = 1}^N p_i | \psi_i (\theta) \rangle \langle \psi_i (\theta) |. \label{main_eq_def_parameterized_density_operator_prove_Petz_function_alpha_divergence_001_001}
	\end{align}
	To emphasize the fact that only the eigenvectors depend on $\theta$, we write them as $\{ | \psi_i (\theta) \rangle \}_{i=1}^N$.
	Note that $\{ p_i \}_{i=1}^N$ does not depend on $\theta$.

	From Eq.~\eqref{main_eq_quantum_e_representation_001_001}, the e-representation of $\partial_k$ with respect to $f (\cdot)$ and Eq.~\eqref{main_eq_def_parameterized_density_operator_prove_Petz_function_alpha_divergence_001_001} is given by
	\begin{align}
		[\hat{\partial}_k]_{\hat{\rho}_\theta, f (\cdot)}^\mathrm{e} & = F^{-1} (\Delta_{\hat{\rho}_\theta}) ([\partial_k \hat{\rho}_\theta] \hat{\rho}_\theta^{-1}). \label{main_eq_quantum_e_representation_proof_Petz_function_001_001}
	\end{align}
	From Eq.~\eqref{main_eq_quantum_m_representation_001_001}, the m-representation of $\partial_k$ with respect to Eq.~\eqref{main_eq_def_parameterized_density_operator_prove_Petz_function_alpha_divergence_001_001} is given by
	\begin{align}
		[\hat{\partial_k}]_{\hat{\rho}_\theta}^\mathrm{m} & = \partial_k \hat{\rho}_\theta                                                                                                                          \\
		                                                  & = \sum_{i = 1}^N p_i (| \partial_k \psi_i (\theta) \rangle \langle \psi_i (\theta) | + | \psi_i (\theta) \rangle \langle \partial_k \psi_i (\theta) |).
	\end{align}
	Thus, we have
	\begin{align}
		\langle \psi_i (\theta) | (\partial_k \hat{\rho}_\theta) | \psi_j (\theta) \rangle & = p_j \langle \psi_i (\theta) | \partial_k \psi_j (\theta) \rangle + p_i \langle \partial_k \psi_i (\theta) | \psi_j (\theta) \rangle. \label{main_eq_derivative_density_operator_proof_Petz_function_001_001}
	\end{align}

	Using $\Delta_{\hat{\rho}_\theta} \hat{A} \coloneqq \hat{\rho}_\theta \hat{A} \hat{\rho}_\theta^{-1}$, we have
	\begin{align}
		\Delta_{\hat{\rho}_\theta}^n ([\hat{\partial}_k]_{\hat{\rho}_\theta, f (\cdot)}^\mathrm{e} \hat{\rho}_\theta) & = \hat{\rho}_\theta^n [\hat{\partial}_k]_{\hat{\rho}_\theta, f (\cdot)}^\mathrm{e} \hat{\rho}_\theta^{1 - n}                                                                                                                  \\
		                                                                                                              & = \sum_{i, j = 1}^N p_i^n p_j^{n-1} | \psi_i (\theta) \rangle \langle \psi_i (\theta) | [\hat{\partial}_k]_{\hat{\rho}_\theta, f (\cdot)}^\mathrm{e} | \psi_j (\theta) \rangle \langle \psi_j (\theta) |                      \\
		                                                                                                              & = \sum_{i, j = 1}^N p_j \bigg( \frac{p_i}{p_j} \bigg)^n \langle \psi_i (\theta) | [\hat{\partial}_k]_{\hat{\rho}_\theta, f (\cdot)}^\mathrm{e} | \psi_j (\theta) \rangle | \psi_i (\theta) \rangle \langle \psi_j (\theta) |.
	\end{align}
	Then, we have
	\begin{align}
		f (\Delta_{\hat{\rho}_\theta}) ([\hat{\partial}_k]_{\hat{\rho}_\theta, f (\cdot)}^\mathrm{e}) \hat{\rho}_\theta & = \sum_{i, j = 1}^N p_j F \bigg( \frac{p_i}{p_j} \bigg) \langle \psi_i (\theta) | [\hat{\partial}_k]_{\hat{\rho}_\theta, f (\cdot)}^\mathrm{e} | \psi_j (\theta) \rangle | \psi_i (\theta) \rangle \langle \psi_j (\theta) |.
	\end{align}
	Sandwiching this with $| \psi_i (\theta) \rangle$ and $| \psi_j (\theta) \rangle$, we get
	\begin{align}
		\langle \psi_i (\theta) | f (\Delta_{\hat{\rho}_\theta}) ([\hat{\partial}_k]_{\hat{\rho}_\theta, f (\cdot)}^\mathrm{e}) \hat{\rho}_\theta | \psi_j (\theta) \rangle & = p_j F \bigg( \frac{p_i}{p_j} \bigg) \langle \psi_i (\theta) | [\hat{\partial}_k]_{\hat{\rho}_\theta, f (\cdot)}^\mathrm{e} | \psi_j (\theta) \rangle. \label{main_eq_f_Xe_rho_proof_Petz_function_001_001}
	\end{align}
	From Eq.~\eqref{main_eq_quantum_e_representation_proof_Petz_function_001_001},
	\begin{align}
		\langle \psi_i (\theta) | (\partial_k \hat{\rho}_\theta) | \psi_j (\theta) \rangle & = \langle \psi_i (\theta) | f (\Delta_{\hat{\rho}_\theta}) ([\hat{\partial}_k]_{\hat{\rho}_\theta, f (\cdot)}^\mathrm{e}) \hat{\rho}_\theta | \psi_j (\theta) \rangle. \label{main_eq_quantum_e_representation_proof_Petz_function_001_002}
	\end{align}
	Substituting Eqs.~\eqref{main_eq_derivative_density_operator_proof_Petz_function_001_001} and \eqref{main_eq_f_Xe_rho_proof_Petz_function_001_001} into Eq.~\eqref{main_eq_quantum_e_representation_proof_Petz_function_001_002}, we get
	\begin{align}
		\langle \psi_i (\theta) | [\hat{\partial}_k]_{\hat{\rho}_\theta, f (\cdot)}^\mathrm{e} | \psi_j (\theta) \rangle & = \frac{p_j \langle \psi_i (\theta) | \partial_k \psi_j (\theta) \rangle + p_i \langle \partial_k \psi_i (\theta) | \psi_j (\theta) \rangle}{p_j f (p_i / p_j)}. \label{main_eq_quantum_e_representation_proof_Petz_function_001_003}
	\end{align}

	To simplify calculations, we define
	\begin{align}
		r_{k; i, j} & \coloneqq \langle \partial_k \psi_i (\theta) | \psi_j (\theta) \rangle. \label{main_eq_def_r_kij_proof_Petz_function_001_001}
	\end{align}
	Using Eq.~\eqref{main_eq_def_r_kij_proof_Petz_function_001_001}, we have
	\begin{align}
		\langle \psi_i (\theta) | \partial_k \psi_j (\theta) \rangle & = - r_{k; i, j}.
	\end{align}
	We also get
	\begin{align}
		\langle \psi_i (\theta) | [\hat{\partial}_k]_{\hat{\rho}_\theta}^\mathrm{m} | \psi_j (\theta) \rangle & = \langle \psi_i (\theta) | (\partial_k \hat{\rho}_\theta) | \psi_j (\theta) \rangle            \\
		                                                                                                      & = (p_i - p_j) r_{k; i, j}. \label{main_eq_quantum_m_representation_proof_Petz_function_001_001}
	\end{align}
	Then, Eq.~\eqref{main_eq_quantum_e_representation_proof_Petz_function_001_003} can be written as
	\begin{align}
		\langle \psi_i (\theta) | [\hat{\partial}_k]_{\hat{\rho}_\theta, f (\cdot)}^\mathrm{e} | \psi_j (\theta) \rangle & = \frac{p_i - p_j}{p_j f (p_i / p_j)} r_{k; i, j}. \label{main_eq_quantum_e_representation_proof_Petz_function_001_004}
	\end{align}
	Exchanging $i \leftrightarrow j$ in Eq.~\eqref{main_eq_quantum_e_representation_proof_Petz_function_001_004}, we get
	\begin{align}
		\langle \psi_j (\theta) | [\hat{\partial}_k]_{\hat{\rho}_\theta, f (\cdot)}^\mathrm{e} | \psi_i (\theta) \rangle & = \frac{p_j - p_i}{p_i f (p_j / p_i)} r_{k; j, i}. \label{main_eq_quantum_e_representation_proof_Petz_function_001_005}
	\end{align}
	Using Eqs.~\eqref{main_eq_quantum_m_representation_proof_Petz_function_001_001} and \eqref{main_eq_quantum_e_representation_proof_Petz_function_001_005}, the quantum Fisher metric, Eq.~\eqref{main_eq_quantum_Fisher_metric_001_001}, is given by
	\begin{align}
		g_{\hat{\rho}_\theta, f (\cdot)} (\partial_k, \partial_l) & = \sum_{i, j = 1}^N \langle \psi_i (\theta) | [\hat{\partial}_k]_{\hat{\rho}_\theta}^\mathrm{m} | \psi_j (\theta) \rangle \langle \psi_j (\theta) | [\hat{\partial}_l]_{\hat{\rho}_\theta, f (\cdot)}^\mathrm{e} | \psi_i (\theta) \rangle \\
		                                                          & = - \sum_{i, j = 1}^N \frac{(p_i - p_j)^2}{p_i f (p_j / p_i)} r_{k; i, j} r_{l; j, i}. \label{main_eq_quantum_Fisher_metric_proof_Petz_function_001_001}
	\end{align}

	The quantum $F$-divergence, Eq.~\eqref{main_eq_def_quantum_f_divergence_001_001}, is given by
	\begin{align}
		D_{F} (\hat{\rho}_{\bar{\theta}} \| \hat{\rho}_\theta) & = \mathrm{Tr} \bigg[ \bigg( \sum_{i = 1}^N p_i | \psi_i (\theta) \rangle \langle \psi_i (\theta) | \bigg) f \Big( \Delta_{\sum_{i = 1}^N p_i | \psi_i (\bar{\theta}) \rangle \langle \psi_i (\bar{\theta}) |, \sum_{i = 1}^N p_i | \psi_i (\theta) \rangle \langle \psi_i (\theta) |} \Big) \hat{1} \bigg] \\
		                                                       & = \sum_{i, j = 1}^N p_i F \bigg( \frac{p_j}{p_i} \bigg) | \langle \psi_j (\bar{\theta}) | \psi_i (\theta) \rangle |^2, \label{main_eq_f_divergence_proof_Petz_function_001_001}
	\end{align}
	where, similarly to Eq.~\eqref{main_eq_def_parameterized_density_operator_prove_Petz_function_alpha_divergence_001_001}, we have defined $\hat{\rho}_{\bar{\theta}}$ as
	\begin{align}
		\hat{\rho}_{\bar{\theta}} & \coloneqq \sum_{i = 1}^N p_i | \psi_i (\bar{\theta}) \rangle \langle \psi_i (\bar{\theta}) |.
	\end{align}
	Applying $\bar{\partial}_k \coloneqq \frac{\partial}{\partial \bar{\theta}^k}$ and $\partial_l \coloneqq \frac{\partial}{\partial \theta^l}$ to Eq.~\eqref{main_eq_f_divergence_proof_Petz_function_001_001}, we have
	\begin{align}
		\bar{\partial}_k \partial_l D (\hat{\rho}_{\bar{\theta}} \| \hat{\rho}_\theta) & = \sum_{i, j = 1}^N p_i F \bigg( \frac{p_j}{p_i} \bigg) \bar{\partial}_i \partial_j | \langle \psi_j (\bar{\theta}) | \psi_i (\theta) \rangle |^2. \label{main_eq_derivative_f_divergence_proof_Petz_function_001_001}
	\end{align}
	Extracting the part concerning the derivative in Eq.~\eqref{main_eq_derivative_f_divergence_proof_Petz_function_001_001}, we get
	\begin{align}
		\bar{\partial}_k \partial_l | \langle \psi_j (\bar{\theta}) | \psi_i (\theta) \rangle |^2 & = \bar{\partial}_k \partial_l \langle \psi_j (\bar{\theta}) | \psi_i (\theta) \rangle \langle \psi_i (\theta) | \psi_j (\bar{\theta}) \rangle                                                                                                                                                         \\
		                                                                                          & = \bar{\partial}_k (\langle \psi_j (\bar{\theta}) | \partial_l \psi_i (\theta) \rangle \langle \psi_i (\theta) | \psi_j (\bar{\theta}) \rangle + \langle \psi_j (\bar{\theta}) | \psi_i (\theta) \rangle \langle \partial_l \psi_i (\theta) | \psi_j (\bar{\theta}) \rangle)                          \\
		                                                                                          & = \langle \bar{\partial}_k \psi_j (\bar{\theta}) | \partial_l \psi_i (\theta) \rangle \langle \psi_i (\theta) | \psi_j (\bar{\theta}) \rangle + \langle \psi_j (\bar{\theta}) | \partial_l \psi_i (\theta) \rangle \langle \psi_i (\theta) | \bar{\partial}_k \psi_j (\bar{\theta}) \rangle \nonumber \\
		                                                                                          & \quad + \langle \bar{\partial}_k \psi_j (\bar{\theta}) | \psi_i (\theta) \rangle \langle \partial_l \psi_i (\theta) | \psi_j (\bar{\theta}) \rangle + \langle \psi_j (\bar{\theta}) | \psi_i (\theta) \rangle \langle \partial_l \psi_i (\theta) | \bar{\partial}_k \psi_j (\bar{\theta}) \rangle.
	\end{align}
	Then, we have
	\begin{align}
		\lim_{\bar{\theta} \to \theta} \bar{\partial}_k \partial_l D (\hat{\rho}_{\bar{\theta}} \| \hat{\rho}_\theta) & = \sum_{i, j = 1}^N p_i F \bigg( \frac{p_j}{p_i} \bigg) (\langle \bar{\partial}_k \psi_j (\bar{\theta}) | \partial_l \psi_i (\theta) \rangle \langle \psi_i (\theta) | \psi_j (\bar{\theta}) \rangle + \langle \psi_j (\bar{\theta}) | \partial_l \psi_i (\theta) \rangle \langle \psi_i (\theta) | \bar{\partial}_k \psi_j (\bar{\theta}) \rangle \nonumber \\
		                                                                                                              & \qquad \qquad \qquad + \langle \bar{\partial}_k \psi_j (\bar{\theta}) | \psi_i (\theta) \rangle \langle \partial_l \psi_i (\theta) | \psi_j (\bar{\theta}) \rangle + \langle \psi_j (\bar{\theta}) | \psi_i (\theta) \rangle \langle \partial_l \psi_i (\theta) | \bar{\partial}_k \psi_j (\bar{\theta}) \rangle)                                            \\
		                                                                                                              & = \sum_{i, j = 1}^N p_i F \bigg( \frac{p_j}{p_i} \bigg) (\langle \bar{\partial}_k \psi_j (\bar{\theta}) | \partial_l \psi_i (\theta) \rangle \delta_{i, j} + r_{k; j, i} r_{l; i, j} + r_{k; i, j} r_{l; j, i} + \delta_{i, j} \langle \partial_l \psi_i (\theta) | \bar{\partial}_k \psi_j (\bar{\theta}) \rangle)                                          \\
		                                                                                                              & = \sum_{i, j = 1}^N \bigg[ p_i F \bigg( \frac{p_j}{p_i} \bigg) + p_j F \bigg( \frac{p_i}{p_j} \bigg) \bigg] r_{k; i, j} r_{l; j, i}. \label{main_eq_derivative_f_divergence_proof_Petz_function_001_002}
	\end{align}
	From Eqs.~\eqref{main_eq_quantum_Fisher_metric_proof_Petz_function_001_001} and \eqref{main_eq_derivative_f_divergence_proof_Petz_function_001_002}, we get
	\begin{align}
		g_{\hat{\rho}_\theta, f (\cdot)} (\partial_k, \partial_l) = - \lim_{\bar{\theta} \to \theta} \bar{\partial}_k \partial_l D (\hat{\rho}_{\bar{\theta}} \| \hat{\rho}_\theta) & \Leftrightarrow \text{$\forall p_i, p_j > 0$, $\frac{(p_i - p_j)^2}{p_i f (p_j / p_i)} = p_i F \bigg( \frac{p_j}{p_i} \bigg) + p_j F \bigg( \frac{p_i}{p_j} \bigg)$} \\
		                                                                                                                                                                            & \Leftrightarrow \text{$\forall t > 0$, $\frac{(1 - t)^2}{f (t)} = F (t) + t F (1 / t)$}.
	\end{align}
	Here, we have used the following relation:
	\begin{align}
		\bar{X} Y D (\hat{\rho}_{\bar{\theta}} \| \hat{\rho}_\theta) & = X \bar{Y} D (\hat{\rho}_{\bar{\theta}} \| \hat{\rho}_\theta)          \\
		                                                             & = - \bar{X} \bar{Y} D (\hat{\rho}_{\bar{\theta}} \| \hat{\rho}_\theta).
	\end{align}
	Thus, we obtain Eq.~\eqref{main_eq_Petz_function_quantum_f_divergence_001_001}.
\end{widetext}

\subsection{Derivation of Eq.~\eqref{main_eq_def_tilde_f_alpha_001_001}} \label{main_sec_derivatin_tilde_f_001_001}

We define the standard quantum $\alpha$-divergence as
\begin{align}
	D_{\alpha} (\hat{\rho}_{\bar{\theta}} \| \hat{\rho}_\theta) & \coloneqq \frac{4}{1 - \alpha^2} \Big( 1 - \mathrm{Tr} \Big[ \hat{\rho}_{\bar{\theta}}^\frac{1 + \alpha}{2} \hat{\rho}_\theta^\frac{1 - \alpha}{2} \Big] \Big). \label{main_eq_def_standard_quantum_alpha_divergence_001_001}
\end{align}
The quantum $F$-divergence, Eq.~\eqref{main_eq_def_quantum_f_divergence_001_001}, with the following $F_\alpha (\cdot)$ is equivalent to the standard quantum $\alpha$-divergence, Eq.~\eqref{main_eq_def_standard_quantum_alpha_divergence_001_001}:
\begin{align}
	F_\alpha (t) & \coloneqq
	\begin{cases}
		\frac{4}{1 - \alpha^2} \Big( 1 - t^\frac{1 + \alpha}{2} \Big) & (\alpha \ne 1, -1), \\
		t \ln t                                                       & (\alpha = 1),       \\
		- \ln t                                                       & (\alpha = -1).
	\end{cases} \label{main_eq_f_function_standard_quahtum_alpha_divergence_001_001}
\end{align}
Substituting Eq.~\eqref{main_eq_f_function_standard_quahtum_alpha_divergence_001_001} into Eq.~\eqref{main_eq_Petz_function_quantum_f_divergence_001_001}, we get
\begin{align}
	f (t) & = \frac{(1 - t)^2}{F (t) + t F (1 / t)}                                                                                                     \\
	      & = \frac{(1 - t)^2}{\frac{4}{1 - \alpha^2} ( 1 - t^\frac{1 + \alpha}{2} ) + t (\frac{4}{1 - \alpha^2} ( 1 - (1 / t)^\frac{1 + \alpha}{2} ))} \\
	      & = \frac{(1 - \alpha^2) (t - 1)^2}{4 (1 + t - t^\frac{1 + \alpha}{2} - t^\frac{1 - \alpha}{2})}.
\end{align}

The rescaled standard quantum R\'enyi divergence, Eq.~\eqref{main_eq_def_rescaled_standard_quantum_Renyi_divergence_001_001}, and the standard quantum $\alpha$-divergence, Eq.~\eqref{main_eq_def_standard_quantum_alpha_divergence_001_001}, satisfy the following relationship:
\begin{align}
	R_\alpha^\mathrm{st} (\hat{\rho}_{\bar{\theta}} \| \hat{\rho}_\theta) & = \frac{1}{\alpha (\alpha - 1)} \ln (1 + \alpha (\alpha - 1) D_{2 \alpha - 1} (\hat{\rho}_{\bar{\theta}} \| \hat{\rho}_\theta)).
\end{align}
Then the metric induced by the rescaled standard quantum R\'enyi divergence, Eq.~\eqref{main_eq_def_rescaled_standard_quantum_Renyi_divergence_001_001}, is computed using the standard quantum $\alpha$-divergence, Eq.~\eqref{main_eq_def_standard_quantum_alpha_divergence_001_001}, as follows:
\begin{align}
	\bar{X} \bar{Y} R_\alpha^\mathrm{st} (\hat{\rho}_{\bar{\theta}} \| \hat{\rho}_\theta) |_{\bar{\theta} = \theta} & = \bar{X} \bar{Y} D_{2 \alpha - 1} (\hat{\rho}_{\bar{\theta}} \| \hat{\rho}_\theta) |_{\bar{\theta} = \theta}.
\end{align}
Thus, we obtain Eq.~\eqref{main_eq_def_tilde_f_alpha_001_001}.

\subsection{Preliminary to the derivation of Eq.~\eqref{main_eq_def_f_alpha_001_001}} \label{main_sec_dervation_f_alpha_preliminary_001_001}

Here, we derive the derivative of a polynomial function with respect to a matrix, which will be used for deriving Eq.~\eqref{main_eq_def_f_alpha_001_001} later.
The formal definition of the derivative of a matrix function is called the G\^ateaux derivative.
The definition of the G\^ateaux derivative of $f (\cdot)$ at $\hat{A}$ in the direction $\hat{B}$ is given by
\begin{align}
	\mathrm{D} f (\cdot) [\hat{A}, \hat{B}] & \coloneqq \lim_{t \to 0} \frac{f (\hat{A} + t \hat{B}) - f (\hat{A})}{t}. \label{main_eq_def_Gateaux_derivative_001_001}
\end{align}
\begin{widetext}
	Then, the chain rule of the G\^ateuax derivative, Eq.~\eqref{main_eq_def_Gateaux_derivative_001_001}, is given by
	\begin{align}
		\mathrm{D} g \circ f (\cdot) [\hat{A}, \hat{B}] & = \lim_{t \to 0} \frac{1}{t} (g \circ f (\hat{A} + t \hat{B}) - g \circ f (\hat{A}))                                                             \\
		                                                & = \lim_{t \to 0} \frac{1}{t} (g (f (\hat{A} + t \hat{B})) - g (f (\hat{A})))                                                                     \\
		                                                & = \lim_{t \to 0} \frac{1}{t} (g (f (\hat{A}) + t \mathrm{D} f (\cdot) [\hat{A}, \hat{B}]) - g (f (\hat{A})))                                     \\
		                                                & = \lim_{t \to 0} \frac{1}{t} (g (f (\hat{A})) + t \mathrm{D} g (\cdot) [f (\hat{A}), \mathrm{D} f (\cdot) [\hat{A}, \hat{B}]] - g (f (\hat{A}))) \\
		                                                & = \mathrm{D} g (\cdot) [f (\hat{A}), \mathrm{D} f (\cdot) [\hat{A}, \hat{B}]]. \label{main_eq_chain_rule_Gateaux_derivative_001_001}
	\end{align}
\end{widetext}

We define the following function:
\begin{align}
	f_\lambda (x) & \coloneqq x^\lambda. \label{main_eq_def_f_lambda_polynomial_function_001_001}
\end{align}
The G\^ateaux derivative of Eq.~\eqref{main_eq_def_f_lambda_polynomial_function_001_001} is
\begin{align}
	 & \mathrm{D} f_\lambda (\cdot) [\hat{A}, \hat{B}] \nonumber                                                                                                                                                                                                  \\
	 & \quad = \lambda \int_0^1 \mathrm{d}t \, \int_0^\infty \mathrm{d}s \, \hat{A}^{t \lambda} (s \hat{1} + \hat{A})^{-1} \hat{B} (s \hat{1} + \hat{A})^{-1} \hat{A}^{(1 - t) \lambda}. \label{main_eq_equality_derivative_operator_polynomial_function_001_001}
\end{align}

First, we define
\begin{subequations} \label{main_eq_def_g_lambda_logarithmic_function_h_exponential_function_001_001}
	\begin{align}
		g_\lambda (x) & \coloneqq \lambda \ln x, \label{main_eq_def_g_lambda_logarithmic_function_001_001} \\
		h (x)         & \coloneqq \mathrm{e}^x. \label{main_eq_def_h_exponential_function_001_001}
	\end{align}
\end{subequations}
Then, Eq.~\eqref{main_eq_def_f_lambda_polynomial_function_001_001} can be expressed as the composite function of $g_\lambda (\cdot)$ and $h (\cdot)$ defined in Eq.~\eqref{main_eq_def_g_lambda_logarithmic_function_h_exponential_function_001_001}:
\begin{align}
	f_\lambda (x) & = h (g_\lambda (x)).
\end{align}

\begin{widetext}
	In general, we have
	\begin{align}
		\frac{\mathrm{d}}{\mathrm{d}t} (\mathrm{e}^{- t \hat{A}} \mathrm{e}^{t (\hat{A} + u \hat{B})}) & = - \hat{A} \mathrm{e}^{- t \hat{A}} \mathrm{e}^{t (\hat{A} + u \hat{B})} + \mathrm{e}^{- t \hat{A}} (\hat{A} + u \hat{B}) \mathrm{e}^{t (\hat{A} + u \hat{B})}                                                               \\
		                                                                                               & = - \hat{A} \mathrm{e}^{- t \hat{A}} \mathrm{e}^{t (\hat{A} + u \hat{B})} + \mathrm{e}^{- t \hat{A}} \hat{A} \mathrm{e}^{t (\hat{A} + u \hat{B})} + \mathrm{e}^{- t \hat{A}} (u \hat{B}) \mathrm{e}^{t (\hat{A} + u \hat{B})} \\
		                                                                                               & = u \mathrm{e}^{- t \hat{A}} \hat{B} \mathrm{e}^{t (\hat{A} + u \hat{B})}.
	\end{align}
\end{widetext}
Then, we get
\begin{align}
	\mathrm{e}^{- \hat{A}} \mathrm{e}^{\hat{A} + u \hat{B}} - \hat{1} & = u \int_0^1 \mathrm{d}t \, \mathrm{e}^{- t \hat{A}} \hat{B} \mathrm{e}^{t (\hat{A} + u \hat{B})}.
\end{align}
By multiplying $\mathrm{e}^{\hat{A}}$ from the left, we have
\begin{align}
	\mathrm{e}^{\hat{A} + u \hat{B}} - \mathrm{e}^{\hat{A}} & = u \int_0^1 \mathrm{d}t \, \mathrm{e}^{(1 - t) \hat{A}} \hat{B} \mathrm{e}^{t (\hat{A} + u \hat{B})}.
\end{align}
Then, the G\^ateaux derivative of Eq.~\eqref{main_eq_def_h_exponential_function_001_001} is computed as
\begin{align}
	\mathrm{D} h (\cdot) [\hat{A}, \hat{B}] & = \lim_{u \to 0} \frac{1}{u} (\mathrm{e}^{\hat{A} + u \hat{B}} - \mathrm{e}^{\hat{A}})                                                                       \\
	                                        & = \lim_{u \to 0} \frac{1}{u} \bigg( u \int_0^1 \mathrm{d}t \, \mathrm{e}^{(1 - t) \hat{A}} \hat{B} \mathrm{e}^{t (\hat{A} + u \hat{B})} \bigg)               \\
	                                        & = \int_0^1 \mathrm{d}t \, \mathrm{e}^{(1 - t) \hat{A}} \hat{B} \mathrm{e}^{t \hat{A}}. \label{main_eq_Gateaux_derivative_D_exponential_function_A_B_001_001}
\end{align}

In general, we have
\begin{align}
	 & \int_0^\infty \mathrm{d}s \, ((s + 1)^{-1} - (s + x)^{-1}) \nonumber                               \\
	 & \quad = \lim_{\alpha \to \infty} \int_0^\alpha \mathrm{d}s \, ((s + 1)^{-1} - (s + x)^{-1})        \\
	 & \quad = \lim_{\alpha \to \infty} [\ln (s + 1) - \ln (s + x)]_0^\alpha                              \\
	 & \quad = \lim_{\alpha \to \infty} \bigg( \ln \frac{\alpha + 1}{1} - \ln \frac{\alpha + x}{x} \bigg) \\
	 & \quad = \lim_{\alpha \to \infty} \bigg( \ln x - \ln \frac{\alpha + x}{\alpha + 1} \bigg)           \\
	 & \quad = \ln x - \lim_{\alpha \to \infty} \ln \frac{\alpha + x}{\alpha + 1}                         \\
	 & \quad = \ln x.
\end{align}
Then, we obtain
\begin{align}
	\ln \hat{X} & = \int_0^\infty \mathrm{d}s \, ( (s + 1)^{-1} \hat{1} - (s \hat{1} + \hat{X})^{-1}).
\end{align}

\begin{widetext}
	Then, the G\^ateaux derivative of Eq.~\eqref{main_eq_def_g_lambda_logarithmic_function_001_001} is computed as
	\begin{align}
		\mathrm{D} g_\lambda (\cdot) [\hat{A}, \hat{B}] & = \lambda \lim_{u \to 0} \frac{1}{u} (\ln (\hat{A} + u \hat{B}) - \ln \hat{A}) \nonumber                                                                                                                                           \\
		                                                & = \lambda \lim_{u \to 0} \frac{1}{u} \bigg( \int_0^\infty \mathrm{d}s \, ((s + 1)^{-1} \hat{1} - (s \hat{1} + \hat{A} + u \hat{B})^{-1}) - \int_0^\infty \mathrm{d}s \, ((s + 1)^{-1} \hat{1} - (s \hat{1} + \hat{A})^{-1}) \bigg) \\
		                                                & = \lambda \lim_{u \to 0} \frac{1}{u} \int_0^\infty \mathrm{d}s \, [((s + 1)^{-1} \hat{1} - (s \hat{1} + \hat{A} + u \hat{B})^{-1}) - ((s + 1)^{-1} \hat{1} - (s \hat{1} + \hat{A})^{-1})]                                          \\
		                                                & = \lambda \lim_{u \to 0} \frac{1}{u} \int_0^\infty \mathrm{d}s \, (- (s \hat{1} + \hat{A} + u \hat{B})^{-1} + (s \hat{1} + \hat{A})^{-1})                                                                                          \\
		                                                & = \lambda \int_0^\infty \mathrm{d}s \, (s \hat{1} + \hat{A})^{-1} \hat{B} (s \hat{1} + \hat{A})^{-1}, \label{main_eq_Gateaux_derivative_D_logarithmic_function_A_B_001_001}
	\end{align}
	where we have used
	\begin{align}
		(s \hat{1} + \hat{Q})^{-1} - (s \hat{1} + \hat{P})^{-1} & = (s \hat{1} + \hat{P})^{-1} (\hat{P} - \hat{Q}) (s \hat{1} + \hat{Q})^{-1}.
	\end{align}

	The chain rule of the G\^ateaux derivative, Eq.~\eqref{main_eq_chain_rule_Gateaux_derivative_001_001}, with Eqs.~\eqref{main_eq_Gateaux_derivative_D_exponential_function_A_B_001_001} and \eqref{main_eq_Gateaux_derivative_D_logarithmic_function_A_B_001_001} leads to the G\^ateaux derivative of Eq.~\eqref{main_eq_def_f_lambda_polynomial_function_001_001}:
	\begin{align}
		\mathrm{D} f_\lambda (\cdot) [\hat{A}, \hat{B}] & = \mathrm{D} h \circ g_\lambda (\cdot) [\hat{A}, \hat{B}]                                                                                                                                                              \\
		                                                & = \mathrm{D} h (\cdot) [g_\lambda (\hat{A}), \mathrm{D} g_\lambda (\cdot) [\hat{A}, \hat{B}]]                                                                                                                          \\
		                                                & = \mathrm{D} h (\cdot) \bigg[ g_\lambda (\hat{A}), \lambda \int_0^\infty \mathrm{d}s \, (s \hat{1} + \hat{A})^{-1} \hat{B} (s \hat{1} + \hat{A})^{-1} \bigg]                                                           \\
		                                                & = \int_0^1 \mathrm{d}t \, \mathrm{e}^{(1 - t) \lambda \ln \hat{A}} \bigg( \lambda \int_0^\infty \mathrm{d}s \, (s \hat{1} + \hat{A})^{-1} \hat{B} (s \hat{1} + \hat{A})^{-1} \bigg) \mathrm{e}^{t \lambda \ln \hat{A}} \\
		                                                & = \lambda \int_0^1 \mathrm{d}t \, \int_0^\infty \mathrm{d}s \, \hat{A}^{(1 - t) \lambda} (s \hat{1} + \hat{A})^{-1} \hat{B} (s \hat{1} + \hat{A})^{-1} \hat{A}^{t \lambda}.
	\end{align}
\end{widetext}
Reassigning $t \to 1 - t$, we obtain Eq.~\eqref{main_eq_equality_derivative_operator_polynomial_function_001_001}.

\subsection{Derivation of Eq.~\eqref{main_eq_def_f_alpha_001_001}} \label{main_sec_dervation_f_alpha_main_001_001}

We show the derivation of Eq.~\eqref{main_eq_def_f_alpha_001_001}~\cite{Takahashi_001}.
The first-order derivative of Eq.~\eqref{main_eq_def_rescaled_sandwiched_quantum_Renyi_divergence_001_001} is computed as
\begin{align}
	\bar{X} R_\alpha^\mathrm{sw} (\hat{\rho}_{\bar{\theta}} \| \hat{\rho}_\theta) & = \frac{1}{\alpha (\alpha - 1)} \bar{X} \ln \mathrm{Tr} [\hat{A}_{\theta, \bar{\theta}}^\alpha]                                                                                        \\
	                                                                              & = \frac{1}{\alpha (\alpha - 1)} \frac{\bar{X} \mathrm{Tr} [\hat{A}_{\theta, \bar{\theta}}^\alpha]}{\mathrm{Tr} [\hat{A}_{\theta, \bar{\theta}}^\alpha]}                                \\
	                                                                              & = \frac{1}{\alpha - 1} \frac{\mathrm{Tr} [\hat{A}_{\theta, \bar{\theta}}^{\alpha - 1} (\bar{X} \hat{A}_{\theta, \bar{\theta}})]}{\mathrm{Tr} [\hat{A}_{\theta, \bar{\theta}}^\alpha]}  \\
	                                                                              & = \frac{1}{\alpha - 1} \frac{\mathrm{Tr} [\hat{A}_{\theta, \bar{\theta}}^{\alpha - 1} \hat{B}_{\theta, \bar{\theta}}^{\bar{X}}]}{\mathrm{Tr} [\hat{A}_{\theta, \bar{\theta}}^\alpha]},
\end{align}
where
\begin{align}
	\hat{A}_{\theta, \bar{\theta}}           & \coloneqq \hat{\rho}_\theta^\frac{1 - \alpha}{2 \alpha} \hat{\rho}_{\bar{\theta}} \hat{\rho}_\theta^\frac{1 - \alpha}{2 \alpha}, \label{main_eq_def_hat_A_proof_Petz_function_sandwiched_Renyi_divergence_001_001} \\
	\hat{B}_{\theta, \bar{\theta}}^{\bar{X}} & \coloneqq \bar{X} \hat{A}_{\theta, \bar{\theta}}                                                                                                                                                                   \\
	                                         & = \hat{\rho}_\theta^\frac{1 - \alpha}{2 \alpha} [\bar{X} \hat{\rho}_{\bar{\theta}}] \hat{\rho}_\theta^\frac{1 - \alpha}{2 \alpha}.
\end{align}
Then, we have
\begin{align}
	\bar{X} R_\alpha^\mathrm{sw} (\hat{\rho}_{\bar{\theta}} \| \hat{\rho}_\theta) |_{\bar{\theta} = \theta} & = \frac{1}{\alpha - 1} \frac{\mathrm{Tr} [\hat{A}_{\theta, \theta}^{\alpha - 1} \hat{B}_{\theta, \theta}^X]}{\mathrm{Tr} [\hat{A}_{\theta, \theta}^\alpha]}     \\
	                                                                                                        & = \frac{1}{\alpha - 1} \frac{\mathrm{Tr} \Big[ \hat{\rho}_\theta^\frac{\alpha - 1}{\alpha} \hat{B}_{\theta, \theta}^X \Big]}{\mathrm{Tr} [ \hat{\rho}_\theta ]} \\
	                                                                                                        & = \frac{1}{\alpha - 1} \mathrm{Tr} [\bar{X} \hat{\rho}_{\bar{\theta}} |_{\bar{\theta} = \theta}]                                                                \\
	                                                                                                        & = \frac{1}{\alpha - 1} \bar{X} \mathrm{Tr} [\hat{\rho}_{\bar{\theta}}] |_{\bar{\theta} = \theta}                                                                \\
	                                                                                                        & = 0, \label{main_eq_equality_first_derivative_divergence_001_001}
\end{align}
where
\begin{align}
	\hat{A}_{\theta, \theta}   & \coloneqq \lim_{\bar{\theta} \to \theta} \hat{A}_{\theta, \bar{\theta}}                                              \\
	                           & = \hat{\rho}_\theta^\frac{1}{\alpha},                                                                                \\
	\hat{B}_{\theta, \theta}^X & \coloneqq \lim_{\bar{\theta} \to \theta} \hat{B}_{\theta, \bar{\theta}}^{\bar{X}}                                    \\
	                           & = \hat{\rho}_\theta^\frac{1 - \alpha}{2 \alpha} [X \hat{\rho}_\theta] \hat{\rho}_\theta^\frac{1 - \alpha}{2 \alpha}.
\end{align}

The second-order derivative of Eq.~\eqref{main_eq_def_rescaled_sandwiched_quantum_Renyi_divergence_001_001} is computed as
\begin{align}
	 & \bar{X} \bar{Y} R_\alpha^\mathrm{sw} (\hat{\rho}_{\bar{\theta}} \| \hat{\rho}_\theta) \nonumber                                                                                                                                                                                                                                                                                                                  \\
	 & \quad = \frac{1}{\alpha - 1} \bar{X} \frac{\mathrm{Tr} [\hat{A}_{\theta, \bar{\theta}}^{\alpha - 1} \hat{B}_{\theta, \bar{\theta}}^{\bar{Y}}]}{\mathrm{Tr} [\hat{A}_{\theta, \bar{\theta}}^\alpha]}                                                                                                                                                                                                              \\
	 & \quad = \frac{1}{\alpha - 1} \frac{\bar{X} \mathrm{Tr} [ \hat{A}_{\theta, \bar{\theta}}^{\alpha - 1} \hat{B}_{\theta, \bar{\theta}}^{\bar{Y}}]}{\mathrm{Tr} [\hat{A}_{\theta, \bar{\theta}}^\alpha]} \nonumber                                                                                                                                                                                                   \\
	 & \qquad - \frac{\alpha}{\alpha - 1} \frac{\mathrm{Tr} [\hat{A}_{\theta, \bar{\theta}}^{\alpha - 1} \hat{B}_{\theta, \bar{\theta}}^{\bar{X}}] \mathrm{Tr} [\hat{A}_{\theta, \bar{\theta}}^{\alpha - 1} \hat{B}_{\theta, \bar{\theta}}^{\bar{Y}}]}{\mathrm{Tr} [\hat{A}_{\theta, \bar{\theta}}^\alpha]^2}                                                                                                           \\
	 & \quad = \frac{1}{\alpha - 1} \frac{\mathrm{Tr} [ (\bar{X} \hat{A}_{\theta, \bar{\theta}}^{\alpha - 1}) \hat{B}_{\theta, \bar{\theta}}^{\bar{Y}} ]}{\mathrm{Tr} [\hat{A}_{\theta, \bar{\theta}}^\alpha]} + \frac{1}{\alpha - 1} \frac{\mathrm{Tr} [\hat{A}_{\theta, \bar{\theta}}^{\alpha - 1} \hat{C}_{\theta, \bar{\theta}}^{\bar{X}, \bar{Y}}]}{\mathrm{Tr} [\hat{A}_{\theta, \bar{\theta}}^\alpha]} \nonumber \\
	 & \qquad - \frac{\alpha}{\alpha - 1} \frac{\mathrm{Tr} [\hat{A}_{\theta, \bar{\theta}}^{\alpha - 1} \hat{B}_{\theta, \bar{\theta}}^{\bar{X}}] \mathrm{Tr} [\hat{A}_{\theta, \bar{\theta}}^{\alpha - 1} \hat{B}_{\theta, \bar{\theta}}^{\bar{Y}}]}{\mathrm{Tr} [\hat{A}_{\theta, \bar{\theta}}^\alpha]^2}, \label{main_eq_equality_second_derivative_divergence_001_001}
\end{align}
where
\begin{align}
	\hat{C}_{\theta, \bar{\theta}}^{\bar{X}, \bar{Y}} & \coloneqq \hat{\rho}_\theta^\frac{1 - \alpha}{2 \alpha} [\bar{X} \bar{Y} \hat{\rho}_{\bar{\theta}}] \hat{\rho}_\theta^\frac{1 - \alpha}{2 \alpha}. \label{main_eq_def_hat_C_proof_Petz_function_sandwiched_Renyi_divergence_001_001}
\end{align}
Note that, from Eqs.~\eqref{main_eq_def_hat_A_proof_Petz_function_sandwiched_Renyi_divergence_001_001} and \eqref{main_eq_def_hat_C_proof_Petz_function_sandwiched_Renyi_divergence_001_001}, we have
\begin{align}
	\mathrm{Tr} [\hat{A}_{\theta, \bar{\theta}}^{\alpha - 1} \hat{C}_{\theta, \bar{\theta}}^{\bar{X}, \bar{Y}}] |_{\bar{\theta} = \theta} & = \mathrm{Tr} [\bar{X} \bar{Y} \hat{\rho}_{\bar{\theta}}] |_{\bar{\theta} = \theta} \\
	                                                                                                                                      & = \bar{X} \bar{Y} \mathrm{Tr} [\hat{\rho}_{\bar{\theta}}] |_{\bar{\theta} = \theta} \\
	                                                                                                                                      & = 0. \label{main_eq_equality_XY_divergence_001}
\end{align}

Using Eqs.~\eqref{main_eq_equality_first_derivative_divergence_001_001} and \eqref{main_eq_equality_XY_divergence_001}, Eq.~\eqref{main_eq_equality_second_derivative_divergence_001_001} reads
\begin{align}
	\bar{X} \bar{Y} R_\alpha^\mathrm{sw} (\hat{\rho}_{\bar{\theta}} \| \hat{\rho}_\theta) |_{\bar{\theta} = \theta} & = \frac{1}{\alpha - 1} \mathrm{Tr} [(\bar{X} \hat{A}_{\theta, \bar{\theta}}^{\alpha - 1}) \hat{B}_{\theta, \bar{\theta}}^{\bar{Y}}] |_{\bar{\theta} = \theta}, \label{main_eq_equality_second_derivative_divergence_001_002}
\end{align}
where we have used the following equality:
\begin{align}
	\mathrm{Tr} [\hat{A}_{\theta, \bar{\theta}}^\alpha] |_{\bar{\theta} = \theta} & = 1.
\end{align}

\begin{widetext}
	Equation~\eqref{main_eq_equality_derivative_operator_polynomial_function_001_001} leads to
	\begin{align}
		\bar{X} \hat{A}_{\theta, \bar{\theta}}^{\alpha - 1} & = \mathrm{D} f_{\alpha - 1} (\cdot) [\hat{A}_{\theta, \bar{\theta}}, \bar{X} \hat{A}_{\theta, \bar{\theta}}]                                                                                                                                                                                                            \\
		                                                    & = \mathrm{D} f_{\alpha - 1} (\cdot) [\hat{A}_{\theta, \bar{\theta}}, \hat{B}_{\theta, \bar{\theta}}^{\bar{X}}]                                                                                                                                                                                                          \\
		                                                    & = (\alpha - 1) \int_0^1 \mathrm{d}t \, \int_0^\infty \mathrm{d}s \, \hat{A}_{\theta, \bar{\theta}}^{(\alpha - 1) t} (s \hat{1} + \hat{A}_{\theta, \bar{\theta}})^{-1} \hat{B}_{\theta, \bar{\theta}}^{\bar{X}} (s \hat{1} + \hat{A}_{\theta, \bar{\theta}})^{-1} \hat{A}_{\theta, \bar{\theta}}^{(\alpha - 1) (1 - t)}.
	\end{align}
	From Eqs.~\eqref{main_eq_quantum_Fisher_metric_derivative_alpha_Renyi_divergence_001_001} and \eqref{main_eq_equality_second_derivative_divergence_001_002}, we have
	\begin{align}
		g_{\hat{\rho}_\theta, f_\alpha (\cdot)} (X, Y) & = \bar{X} \bar{Y} R_\alpha^\mathrm{sw} (\hat{\rho}_{\bar{\theta}} \| \hat{\rho}_\theta) |_{\bar{\theta} = \theta}                                                                                                                                                                                                                                                                                                                                                                                                                                                                  \\
		                                               & = \frac{1}{\alpha - 1} \mathrm{Tr} [(\bar{X} \hat{A}_{\theta, \bar{\theta}}^{\alpha - 1}) \hat{B}_{\theta, \bar{\theta}}^{\bar{Y}}] |_{\bar{\theta} = \theta}                                                                                                                                                                                                                                                                                                                                                                                                                      \\
		                                               & = \frac{1}{\alpha - 1} \mathrm{Tr} [(\bar{X} \hat{A}_{\theta, \theta}^{\alpha - 1}) \hat{B}_{\theta, \theta}^Y]                                                                                                                                                                                                                                                                                                                                                                                                                                                                    \\
		                                               & = \mathrm{Tr} \bigg[ \bigg( \int_0^1 \mathrm{d}t \, \int_0^\infty \mathrm{d}s \, \hat{A}_{\theta, \theta}^{(\alpha - 1) t} (s \hat{1} + \hat{A}_{\theta, \theta})^{-1} \hat{B}_{\theta, \theta}^X (s \hat{1} + \hat{A}_{\theta, \theta})^{-1} \hat{A}_{\theta, \theta}^{(\alpha - 1) (1 - t)} \bigg) \hat{B}_{\theta, \theta}^Y \bigg]                                                                                                                                                                                                                                             \\
		                                               & = \mathrm{Tr} \bigg[ \bigg( \int_0^1 \mathrm{d}t \, \int_0^\infty \mathrm{d}s \, \hat{\rho}_\theta^{\frac{\alpha - 1}{\alpha} t} (s \hat{1} + \hat{\rho}_\theta^\frac{1}{\alpha})^{-1} \hat{\rho}_\theta^\frac{1 - \alpha}{2 \alpha} [X \hat{\rho}_\theta] \hat{\rho}_\theta^\frac{1 - \alpha}{2 \alpha} (s \hat{1} + \hat{\rho}_\theta^\frac{1}{\alpha})^{-1} \hat{\rho}_\theta^{\frac{\alpha - 1}{\alpha} (1 - t)} \bigg) \hat{\rho}_\theta^\frac{1 - \alpha}{2 \alpha} [Y \hat{\rho}_\theta] \hat{\rho}_\theta^\frac{1 - \alpha}{2 \alpha} \bigg]                               \\
		                                               & = \mathrm{Tr} \bigg[ \hat{\rho}_\theta^\frac{1 - \alpha}{\alpha} \bigg( \int_0^1 \mathrm{d}t \, \int_0^\infty \mathrm{d}s \, \hat{\rho}_\theta^{\frac{\alpha - 1}{\alpha} t} (s \hat{1} + \hat{\rho}_\theta^\frac{1}{\alpha})^{-1} \hat{X}_{\hat{\rho}_\theta}^\mathrm{m} (s \hat{1} + \hat{\rho}_\theta^\frac{1}{\alpha})^{-1} \hat{\rho}_\theta^{\frac{\alpha - 1}{\alpha} (1 - t)} \bigg) \hat{\rho}_\theta^\frac{1 - \alpha}{\alpha} \hat{Y}_{\hat{\rho}_\theta}^\mathrm{m} \bigg]. \label{main_eq_quantum_Fisher_metric_rescaled_sandwiched_quantum_Renyi_divergence_001_001}
	\end{align}
	From Eqs.~\eqref{main_eq_quantum_Fisher_metric_001_001} and \eqref{main_eq_quantum_Fisher_metric_rescaled_sandwiched_quantum_Renyi_divergence_001_001}, we get
	\begin{align}
		\hat{X}_{\hat{\rho}_\theta, f_\alpha (\cdot)}^\mathrm{e} & = \hat{\rho}_\theta^\frac{1 - \alpha}{\alpha} \bigg( \int_0^1 \mathrm{d}t \, \int_0^\infty \mathrm{d}s \, \hat{\rho}_\theta^{\frac{\alpha - 1}{\alpha} t} (s \hat{1} + \hat{\rho}_\theta^\frac{1}{\alpha})^{-1} \hat{X}_{\hat{\rho}_\theta}^\mathrm{m} (s \hat{1} + \hat{\rho}_\theta^\frac{1}{\alpha})^{-1} \hat{\rho}_\theta^{\frac{\alpha - 1}{\alpha} (1 - t)} \bigg) \hat{\rho}_\theta^\frac{1 - \alpha}{\alpha}. \label{main_eq_e-representation_X_f_alpha_001_001}
	\end{align}
	Using Eqs.~\eqref{main_eq_def_f_alpha_001_001} and \eqref{main_eq_e-representation_X_f_alpha_001_001}, Eq.~\eqref{main_eq_quantum_e_representation_001_002} for this case can be written as
	\begin{align}
		\langle \psi_i | \hat{X}_{\hat{\rho}_\theta, f_\alpha (\cdot)}^\mathrm{e} | \psi_j \rangle & = \langle \psi_i | \hat{\rho}_\theta^\frac{1 - \alpha}{\alpha} \bigg( \int_0^1 \mathrm{d}t \, \int_0^\infty \mathrm{d}s \, \hat{\rho}_\theta^{\frac{\alpha - 1}{\alpha} t} (s \hat{1} + \hat{\rho}_\theta^\frac{1}{\alpha})^{-1} \hat{X}_{\hat{\rho}_\theta}^\mathrm{m} (s \hat{1} + \hat{\rho}_\theta^\frac{1}{\alpha})^{-1} \hat{\rho}_\theta^{\frac{\alpha - 1}{\alpha} (1 - t)} \bigg) \hat{\rho}_\theta^\frac{1 - \alpha}{\alpha} | \psi_j \rangle \\
		                                                                                           & = p_i^\frac{1 - \alpha}{\alpha} \bigg( \int_0^1 \mathrm{d}t \, \int_0^\infty \mathrm{d}s \, p_i^{\frac{\alpha - 1}{\alpha} t} (s \hat{1} + p_i^\frac{1}{\alpha})^{-1} \langle \psi_i | \hat{X}_{\hat{\rho}_\theta}^\mathrm{m} | \psi_j \rangle (s \hat{1} + p_j^\frac{1}{\alpha})^{-1} p_j^{\frac{\alpha - 1}{\alpha} (1 - t)} \bigg) p_j^\frac{1 - \alpha}{\alpha}                                                                                     \\
		                                                                                           & = \langle \psi_i | \hat{X}_{\hat{\rho}_\theta}^\mathrm{m} | \psi_j \rangle p_i^\frac{1 - \alpha}{\alpha} p_j^\frac{1 - \alpha}{\alpha} \bigg( \int_0^1 \mathrm{d}t \, p_i^{\frac{\alpha - 1}{\alpha} t} p_j^{\frac{\alpha - 1}{\alpha} (1 - t)} \int_0^\infty \mathrm{d}s \, (s \hat{1} + p_i^\frac{1}{\alpha})^{-1} (s \hat{1} + p_j^\frac{1}{\alpha})^{-1} \bigg)                                                                                     \\
		                                                                                           & = \langle \psi_i | \hat{X}_{\hat{\rho}_\theta}^\mathrm{m} | \psi_j \rangle p_i^\frac{1 - \alpha}{\alpha} p_j^\frac{1 - \alpha}{\alpha} \frac{p_i^\frac{\alpha - 1}{\alpha} - p_j^\frac{\alpha - 1}{\alpha}}{\ln p_i^\frac{\alpha - 1}{\alpha} - \ln p_j^\frac{\alpha - 1}{\alpha}} \frac{\ln p_i^\frac{1}{\alpha} - \ln p_j^\frac{1}{\alpha}}{p_i^\frac{1}{\alpha} - p_j^\frac{1}{\alpha}}                                                              \\
		                                                                                           & = \langle \psi_i | \hat{X}_{\hat{\rho}_\theta}^\mathrm{m} | \psi_j \rangle \frac{1}{1 - \alpha} \frac{p_i^\frac{1 - \alpha}{\alpha} - p_j^\frac{1 - \alpha}{\alpha}}{p_i^\frac{1}{\alpha} - p_j^\frac{1}{\alpha}}                                                                                                                                                                                                                                       \\
		                                                                                           & = \langle \psi_i | \hat{X}_{\hat{\rho}_\theta}^\mathrm{m} | \psi_j \rangle \frac{1}{1 - \alpha} \frac{p_j^\frac{1 - \alpha}{\alpha} ((p_i / p_j)^\frac{1 - \alpha}{\alpha} - 1)}{p_j^\frac{1}{\alpha} ((p_i / p_j)^\frac{1}{\alpha} - 1)}                                                                                                                                                                                                               \\
		                                                                                           & = \langle \psi_i | \hat{X}_{\hat{\rho}_\theta}^\mathrm{m} | \psi_j \rangle \frac{1}{1 - \alpha} \frac{(1 - (p_i / p_j)^\frac{1 - \alpha}{\alpha})}{p_j (1 - (p_i / p_j)^\frac{1}{\alpha})}                                                                                                                                                                                                                                                              \\
		                                                                                           & = \frac{\langle \psi_i | \hat{X}_{\hat{\rho}_\theta}^\mathrm{m} | \psi_j \rangle}{p_j f_\alpha (p_i / p_j)},
	\end{align}
	where we have used the following formulas:
	\begin{align}
		\int_0^1 \mathrm{d}t \, x^t y^{1 - t}                  & = \frac{x - y}{\ln x - \ln y}, \\
		\int_0^\infty \mathrm{d}s \, \frac{1}{(s + x) (s + y)} & = \frac{\ln x - \ln y}{x - y}.
	\end{align}
	Thus, Eq.~\eqref{main_eq_def_f_alpha_001_001} is proved.
\end{widetext}

\subsection{Derivation of Eq.~\eqref{main_eq_f1_BKM_001_001}} \label{main_sec_derivation_f1_BKM_001_001}

We set $h \coloneqq \frac{1 - \alpha}{\alpha}$.
Note that $\alpha = \frac{1}{1 + h}$ and $h \to 0$ when $\alpha \to 1$.
Then, we have
\begin{align}
	f_\alpha (t) & = (1 - \alpha) \frac{1 - t^\frac{1}{\alpha}}{1 - t^\frac{1 - \alpha}{\alpha}}                       \\
	             & = \frac{\frac{1 - \alpha}{\alpha}}{t^\frac{1 - \alpha}{\alpha} - 1} \alpha (1 - t^\frac{1}{\alpha}) \\
	             & = \frac{h}{t^h - 1} \frac{1}{1 + h} (1 - t^{h + 1})                                                 \\
	             & \xrightarrow{h \to 0} \frac{1 - t}{\ln t}.
\end{align}
Here, we have used the following equality:
\begin{align}
	\lim_{h \to 0} \frac{h}{t^h - 1} & = \lim_{h \to 0} \frac{h}{\mathrm{e}^{h \ln t} - 1}     \\
	                                 & = \lim_{h \to 0} \frac{1}{(\ln t) \mathrm{e}^{h \ln t}} \\
	                                 & = \frac{1}{\ln t},
\end{align}
where l'Hopital's rule was used.
Then we get Eq.~\eqref{main_eq_f1_BKM_001_001}.

\subsection{Derivation of Eq.~\eqref{main_eq_def_f_pm_infty_001_001}} \label{main_sec_derivation_f_pm_infty_001_001}

From Eq.~\eqref{main_eq_def_f_alpha_001_001}, we have
\begin{align}
	f_\alpha^\mathrm{sw} (t) & = (1 - \alpha) \frac{1 - t^\frac{1}{\alpha}}{1 - t^\frac{1 - \alpha}{\alpha}}                                                  \\
	                         & = \frac{t}{t - t^\frac{1}{\alpha}} (\alpha - 1) (t^\frac{1}{\alpha} - 1)                                                       \\
	                         & = \frac{t}{t - t^\frac{1}{\alpha}} \alpha (t^\frac{1}{\alpha} - 1) - \frac{t}{t - t^\frac{1}{\alpha}} (t^\frac{1}{\alpha} - 1) \\
	                         & \xrightarrow{\alpha \to \infty} \frac{t}{t - 1} \ln t.
\end{align}
Thus, we obtain Eq.~\eqref{main_eq_def_f_pm_infty_001_001}.

\subsection{Derivation of Eq.~\eqref{main_eq_f0pm_001_001}} \label{main_sec_derivation_f_0pm_001_001}

We here derive Eq.~\eqref{main_eq_f0pm_001_001}.
First, we consider the case of $\alpha \to 0+$.
From Eq.~\eqref{main_eq_def_f_alpha_001_001}, we get
\begin{align}
	\lim_{\alpha \to 0+} f_\alpha^\mathrm{sw} (t) & = \lim_{\alpha \to 0+} (1 - \alpha) \frac{1 - t^\frac{1}{\alpha}}{1 - t^\frac{1 - \alpha}{\alpha}} \\
	                                              & =
	\begin{cases}
		\lim_{\alpha \to 0+} \frac{t^\frac{1}{\alpha}}{t^{\frac{1}{\alpha} - 1}} & (t > 1), \\
		1                                                                        & (t = 1), \\
		\lim_{\alpha \to 0+} 1                                                   & (t < 1).
	\end{cases}
\end{align}
Then Eq.~\eqref{main_eq_f0+_001_001} is shown.
Similarly, we have
\begin{align}
	\lim_{\alpha \to 0+} f_\alpha^\mathrm{sw} (t) & = \lim_{\alpha \to 0-} (1 - \alpha) \frac{1 - t^\frac{1}{\alpha}}{1 - t^\frac{1 - \alpha}{\alpha}} \\
	                                              & =
	\begin{cases}
		\lim_{\alpha \to 0+} 1                                                   & (t > 1), \\
		1                                                                        & (t = 1), \\
		\lim_{\alpha \to 0+} \frac{t^\frac{1}{\alpha}}{t^{\frac{1}{\alpha} - 1}} & (t < 1).
	\end{cases}
\end{align}
Then Eq.~\eqref{main_eq_f0-_001_001} holds.

\subsection{Derivation of Eq.~\eqref{main_eq_relationship_Bures_distance_SLD_metric_001_001}} \label{main_sec_derivation_Bures_metric_001_001}

First, we define
\begin{align}
	\hat{X} & \coloneqq \hat{\rho}_\theta^\frac{1}{2} \hat{\rho}_{\theta + \Delta \theta} \hat{\rho}_\theta^\frac{1}{2}                                                                                                                                                                                        \\
	        & = \hat{\rho}_\theta^2 + \hat{\rho}_\theta^\frac{1}{2} (\partial_i \hat{\rho}_\theta) \hat{\rho}_\theta^\frac{1}{2} \Delta \theta^i + \frac{1}{2} \hat{\rho}_\theta^\frac{1}{2} (\partial_i \partial_j \hat{\rho}_\theta) \hat{\rho}_\theta^\frac{1}{2} \Delta \theta^i \Delta \theta_j \nonumber \\
	        & \quad + \mathcal{O} (\| \Delta \theta \|^3), \label{main_eq_def_X_Bures_metric_001_001}
\end{align}
where
\begin{align}
	\hat{\rho}_{\theta + \Delta \theta} = \hat{\rho}_\theta + \partial_i \hat{\rho}_\theta \Delta \theta^i + \frac{1}{2} \partial_i \partial_j \hat{\rho}_\theta \Delta \theta^i \Delta \theta^j + \mathcal{O} (\| \Delta \theta \|^3).
\end{align}
Next, we define $\hat{B}_i$ and $\hat{C}_{i, j}$ such that
\begin{align}
	\hat{X}^\frac{1}{2} & = \sqrt{\hat{\rho}_\theta^\frac{1}{2} \hat{\rho}_{\theta + \Delta \theta} \hat{\rho}_\theta^\frac{1}{2}}                                                                                \\
	                    & = \hat{\rho}_\theta + \hat{B}_i \Delta \theta^i + \hat{C}_{i, j} \Delta \theta^i \Delta \theta^j + \mathcal{O} (\| \Delta \theta \|^3). \label{main_eq_def_Bi_Cij_Bures_metric_001_001}
\end{align}
From Eq.~\eqref{main_eq_def_Bi_Cij_Bures_metric_001_001}, we have
\begin{align}
	\hat{X} & = \hat{\rho}_\theta^2 + \hat{\rho}_\theta \hat{B}_i \Delta \theta^i + \hat{B}_i \hat{\rho}_\theta \Delta \theta^i + \frac{1}{2} (\hat{B}_i \hat{B}_j + \hat{B}_j \hat{B}_i) \Delta \theta^i \Delta \theta^i \nonumber     \\
	        & \quad + \hat{\rho}_\theta \hat{C}_{i, j} \Delta \theta^i \Delta \theta^j + \hat{C}_{i, j} \hat{\rho}_\theta \Delta \theta^i \Delta \theta^j + \mathcal{O} (\| \Delta \theta \|^3). \label{main_eq_X_Bures_metric_001_001}
\end{align}
From Eqs.~\eqref{main_eq_def_X_Bures_metric_001_001} and \eqref{main_eq_X_Bures_metric_001_001}, we get
\begin{align}
	\hat{\rho}_\theta^\frac{1}{2} (\partial_i \hat{\rho}_\theta) \hat{\rho}_\theta^\frac{1}{2}                        & = \hat{\rho}_\theta \hat{B}_i + \hat{B}_i \hat{\rho}_\theta, \label{main_eq_condition_1_Bures_metric_001_001}                                                                     \\
	\frac{1}{2} \hat{\rho}_\theta^\frac{1}{2} (\partial_i \partial_j \hat{\rho}_\theta) \hat{\rho}_\theta^\frac{1}{2} & = \frac{1}{2} (\hat{B}_i \hat{B}_j + \hat{B}_j \hat{B}_i) + \hat{\rho}_\theta \hat{C}_{i, j} + \hat{C}_{i, j} \hat{\rho}_\theta. \label{main_eq_condition_2_Bures_metric_001_001}
\end{align}
From Eqs.~\eqref{main_eq_differential_equation_SLD_L_i_001_001} and \eqref{main_eq_condition_1_Bures_metric_001_001}, we have
\begin{align}
	\hat{B}_i & = \frac{1}{2} \hat{\rho}_\theta^\frac{1}{2} \hat{L}_i \hat{\rho}_\theta^\frac{1}{2}. \label{main_eq_relationship_B_L_001_001}
\end{align}
In general, the following equality holds:
\begin{align}
	\partial_i \partial_j \hat{\rho}_\theta & = \frac{1}{2} (\partial_i \partial_j + \partial_j \partial_i) \hat{\rho}_\theta. \label{main_eq_second_order_derivative_Bures_metric_001_001}
\end{align}
\begin{widetext}
	From Eqs.~\eqref{main_eq_differential_equation_SLD_L_i_001_001}, \eqref{main_eq_relationship_B_L_001_001}, and \eqref{main_eq_second_order_derivative_Bures_metric_001_001}, Eq.~\eqref{main_eq_condition_2_Bures_metric_001_001} is rewritten as
	\begin{align}
		\hat{\rho}_\theta \hat{C}_{i, j} + \hat{C}_{i, j} \hat{\rho}_\theta & = \frac{1}{2} \hat{\rho}_\theta^\frac{1}{2} (\partial_i \partial_j \hat{\rho}_\theta) \hat{\rho}_\theta^\frac{1}{2} - \frac{1}{2} (\hat{B}_i \hat{B}_j + \hat{B}_j \hat{B}_i)                                                                                                                                                                                                                                                                                                                                                                           \\
		                                                                    & = \frac{1}{8} \hat{\rho}_\theta^\frac{1}{2} (\partial_i [\hat{L}_j \hat{\rho}_\theta + \hat{\rho}_\theta \hat{L}_j] + \partial_j [\hat{L}_i \hat{\rho}_\theta + \hat{\rho}_\theta \hat{L}_i]) \hat{\rho}_\theta^\frac{1}{2} - \frac{1}{8} \hat{\rho}_\theta^\frac{1}{2} \hat{L}_i \hat{\rho}_\theta \hat{L}_j \hat{\rho}_\theta^\frac{1}{2} - \frac{1}{8} \hat{\rho}_\theta^\frac{1}{2} \hat{L}_j \hat{\rho}_\theta \hat{L}_i \hat{\rho}_\theta^\frac{1}{2}                                                                                             \\
		                                                                    & = \frac{1}{8} \hat{\rho}_\theta^\frac{1}{2} (\partial_i \hat{L}_j) \hat{\rho}_\theta \hat{\rho}_\theta^\frac{1}{2} + \frac{1}{8} \hat{\rho}_\theta^\frac{1}{2} \hat{L}_j (\partial_i \hat{\rho}_\theta) \hat{\rho}_\theta^\frac{1}{2} + \frac{1}{8} \hat{\rho}_\theta^\frac{1}{2} (\partial_i \hat{\rho}_\theta) \hat{L}_j \hat{\rho}_\theta^\frac{1}{2} + \frac{1}{8} \hat{\rho}_\theta^\frac{1}{2} \hat{\rho}_\theta (\partial_i \hat{L}_j) \hat{\rho}_\theta^\frac{1}{2} \nonumber                                                                   \\
		                                                                    & \quad + \frac{1}{8} \hat{\rho}_\theta^\frac{1}{2} (\partial_j \hat{L}_i) \hat{\rho}_\theta \hat{\rho}_\theta^\frac{1}{2} + \frac{1}{8} \hat{\rho}_\theta^\frac{1}{2} \hat{L}_i (\partial_j \hat{\rho}_\theta) \hat{\rho}_\theta^\frac{1}{2} + \frac{1}{8} \hat{\rho}_\theta^\frac{1}{2} (\partial_j \hat{\rho}_\theta) \hat{L}_i \hat{\rho}_\theta^\frac{1}{2} + \frac{1}{8} \hat{\rho}_\theta^\frac{1}{2} \hat{\rho}_\theta (\partial_j \hat{L}_i) \hat{\rho}_\theta^\frac{1}{2} \nonumber                                                             \\
		                                                                    & \quad - \frac{1}{8} \hat{\rho}_\theta^\frac{1}{2} \hat{L}_i \hat{\rho}_\theta \hat{L}_j \hat{\rho}_\theta^\frac{1}{2} - \frac{1}{8} \hat{\rho}_\theta^\frac{1}{2} \hat{L}_j \hat{\rho}_\theta \hat{L}_i \hat{\rho}_\theta^\frac{1}{2}                                                                                                                                                                                                                                                                                                                   \\
		                                                                    & = \frac{1}{8} \hat{\rho}_\theta^\frac{1}{2} (\partial_i \hat{L}_j) \hat{\rho}_\theta \hat{\rho}_\theta^\frac{1}{2} + \frac{1}{16} \hat{\rho}_\theta^\frac{1}{2} \hat{L}_j (\hat{\rho}_\theta \hat{L}_i + \hat{L}_i \hat{\rho}_\theta) \hat{\rho}_\theta^\frac{1}{2} + \frac{1}{16} \hat{\rho}_\theta^\frac{1}{2} (\hat{\rho}_\theta \hat{L}_i + \hat{L}_i \hat{\rho}_\theta) \hat{L}_j \hat{\rho}_\theta^\frac{1}{2} + \frac{1}{8} \hat{\rho}_\theta^\frac{1}{2} \hat{\rho}_\theta (\partial_i \hat{L}_j) \hat{\rho}_\theta^\frac{1}{2} \nonumber       \\
		                                                                    & \quad + \frac{1}{8} \hat{\rho}_\theta^\frac{1}{2} (\partial_j \hat{L}_i) \hat{\rho}_\theta \hat{\rho}_\theta^\frac{1}{2} + \frac{1}{16} \hat{\rho}_\theta^\frac{1}{2} \hat{L}_i (\hat{\rho}_\theta \hat{L}_j + \hat{L}_j \hat{\rho}_\theta) \hat{\rho}_\theta^\frac{1}{2} + \frac{1}{16} \hat{\rho}_\theta^\frac{1}{2} (\hat{\rho}_\theta \hat{L}_j + \hat{L}_j \hat{\rho}_\theta) \hat{L}_i \hat{\rho}_\theta^\frac{1}{2} + \frac{1}{8} \hat{\rho}_\theta^\frac{1}{2} \hat{\rho}_\theta (\partial_j \hat{L}_i) \hat{\rho}_\theta^\frac{1}{2} \nonumber \\
		                                                                    & \quad - \frac{1}{8} \hat{\rho}_\theta^\frac{1}{2} \hat{L}_i \hat{\rho}_\theta \hat{L}_j \hat{\rho}_\theta^\frac{1}{2} - \frac{1}{8} \hat{\rho}_\theta^\frac{1}{2} \hat{L}_j \hat{\rho}_\theta \hat{L}_i \hat{\rho}_\theta^\frac{1}{2}                                                                                                                                                                                                                                                                                                                   \\
		                                                                    & = \frac{1}{8} \hat{\rho}_\theta^\frac{1}{2} (\partial_i \hat{L}_j) \hat{\rho}_\theta \hat{\rho}_\theta^\frac{1}{2} + \frac{1}{16} \hat{\rho}_\theta^\frac{1}{2} \hat{L}_j \hat{\rho}_\theta \hat{L}_i \hat{\rho}_\theta^\frac{1}{2} + \frac{1}{16} \hat{\rho}_\theta^\frac{1}{2} \hat{L}_j \hat{L}_i \hat{\rho}_\theta \hat{\rho}_\theta^\frac{1}{2} \nonumber                                                                                                                                                                                          \\
		                                                                    & \quad + \frac{1}{16} \hat{\rho}_\theta^\frac{1}{2} \hat{\rho}_\theta \hat{L}_i \hat{L}_j \hat{\rho}_\theta^\frac{1}{2} + \frac{1}{16} \hat{\rho}_\theta^\frac{1}{2} \hat{L}_i \hat{\rho}_\theta \hat{L}_j \hat{\rho}_\theta^\frac{1}{2} + \frac{1}{8} \hat{\rho}_\theta^\frac{1}{2} \hat{\rho}_\theta (\partial_i \hat{L}_j) \hat{\rho}_\theta^\frac{1}{2} \nonumber                                                                                                                                                                                    \\
		                                                                    & \quad + \frac{1}{8} \hat{\rho}_\theta^\frac{1}{2} (\partial_j \hat{L}_i) \hat{\rho}_\theta \hat{\rho}_\theta^\frac{1}{2} + \frac{1}{16} \hat{\rho}_\theta^\frac{1}{2} \hat{L}_i \hat{\rho}_\theta \hat{L}_j \hat{\rho}_\theta^\frac{1}{2} + \frac{1}{16} \hat{\rho}_\theta^\frac{1}{2} \hat{L}_i \hat{L}_j \hat{\rho}_\theta \hat{\rho}_\theta^\frac{1}{2} \nonumber                                                                                                                                                                                    \\
		                                                                    & \quad + \frac{1}{16} \hat{\rho}_\theta^\frac{1}{2} \hat{\rho}_\theta \hat{L}_j \hat{L}_i \hat{\rho}_\theta^\frac{1}{2} + \frac{1}{16} \hat{\rho}_\theta^\frac{1}{2} \hat{L}_j \hat{\rho}_\theta \hat{L}_i \hat{\rho}_\theta^\frac{1}{2} + \frac{1}{8} \hat{\rho}_\theta^\frac{1}{2} \hat{\rho}_\theta (\partial_j \hat{L}_i) \hat{\rho}_\theta^\frac{1}{2} \nonumber                                                                                                                                                                                    \\
		                                                                    & \quad - \frac{1}{8} \hat{\rho}_\theta^\frac{1}{2} \hat{L}_i \hat{\rho}_\theta \hat{L}_j \hat{\rho}_\theta^\frac{1}{2} - \frac{1}{8} \hat{\rho}_\theta^\frac{1}{2} \hat{L}_j \hat{\rho}_\theta \hat{L}_i \hat{\rho}_\theta^\frac{1}{2}                                                                                                                                                                                                                                                                                                                   \\
		                                                                    & = \frac{1}{8} \hat{\rho}_\theta^\frac{1}{2} (\partial_i \hat{L}_j) \hat{\rho}_\theta \hat{\rho}_\theta^\frac{1}{2}  + \frac{1}{16} \hat{\rho}_\theta^\frac{1}{2} \hat{L}_j \hat{L}_i \hat{\rho}_\theta \hat{\rho}_\theta^\frac{1}{2} + \frac{1}{16} \hat{\rho}_\theta^\frac{1}{2} \hat{\rho}_\theta \hat{L}_i \hat{L}_j \hat{\rho}_\theta^\frac{1}{2} + \frac{1}{8} \hat{\rho}_\theta^\frac{1}{2} \hat{\rho}_\theta (\partial_i \hat{L}_j) \hat{\rho}_\theta^\frac{1}{2} \nonumber                                                                      \\
		                                                                    & \quad + \frac{1}{8} \hat{\rho}_\theta^\frac{1}{2} (\partial_j \hat{L}_i) \hat{\rho}_\theta \hat{\rho}_\theta^\frac{1}{2}  + \frac{1}{16} \hat{\rho}_\theta^\frac{1}{2} \hat{L}_i \hat{L}_j \hat{\rho}_\theta \hat{\rho}_\theta^\frac{1}{2} + \frac{1}{16} \hat{\rho}_\theta^\frac{1}{2} \hat{\rho}_\theta \hat{L}_j \hat{L}_i \hat{\rho}_\theta^\frac{1}{2}  + \frac{1}{8} \hat{\rho}_\theta^\frac{1}{2} \hat{\rho}_\theta (\partial_j \hat{L}_i) \hat{\rho}_\theta^\frac{1}{2}. \label{main_eq_condition_2_Bures_metric_001_002}
	\end{align}
\end{widetext}
From Eq.~\eqref{main_eq_condition_2_Bures_metric_001_002}, we get
\begin{align}
	\hat{C}_{i, j} & = \frac{1}{16} \hat{\rho}_\theta^\frac{1}{2} (2 \partial_i \hat{L}_j + 2 \partial_j \hat{L}_i + \hat{L}_i \hat{L}_j + \hat{L}_j \hat{L}_i) \hat{\rho}_\theta^\frac{1}{2}.
\end{align}
The square of the Bures distance, Eq.~\eqref{main_eq_def_Bures_distance_001_001}, is computed as
\begin{align}
	[D_\mathrm{Bures} (\hat{\rho}_\theta, \hat{\rho}_{\theta + \Delta \theta})]^2 & = 2 - 2 \sqrt{F (\hat{\rho}_\theta, \hat{\rho}_{\theta + \Delta \theta})}                                                                                  \\
	                                                                              & = 2 - 2 \mathrm{Tr} [\hat{X}^\frac{1}{2}]                                                                                                                  \\
	                                                                              & = - 2 \mathrm{Tr} [\hat{B}_i] \Delta \theta^i - 2 \mathrm{Tr} [\hat{C}_{i, j}] \Delta \theta^i \Delta \theta^j. \label{main_eq_def_Bures_distance_002_001}
\end{align}
From Eq.~\eqref{main_eq_relationship_B_L_001_001}, we have
\begin{align}
	\mathrm{Tr} [\hat{B}_i] & = \mathrm{Tr} \bigg[ \frac{1}{2} \hat{\rho}_\theta^\frac{1}{2} \hat{L}_i \hat{\rho}_\theta^\frac{1}{2} \bigg] \\
	                        & = \frac{1}{2} \mathrm{Tr} \bigg[ \frac{\hat{\rho}_\theta \hat{L}_i + \hat{L}_i \hat{\rho}_\theta}{2} \bigg]   \\
	                        & = \frac{1}{2} \mathrm{Tr} [\partial_i \hat{\rho}_\theta]                                                      \\
	                        & = \frac{1}{2} \partial_i \mathrm{Tr} [\hat{\rho}_\theta]                                                      \\
	                        & = 0. \label{main_eq_equality_B_Bures_distance_001_001}
\end{align}
\begin{widetext}
	Furthermore, we get
	\begin{align}
		\partial_i \partial_j \mathrm{Tr} [\hat{\rho}_\theta] & = \partial_i \mathrm{Tr} [\partial_j \hat{\rho}_\theta]                                                                                                                                                                                                                                    \\
		                                                      & = \partial_i \mathrm{Tr} \bigg[ \frac{\hat{\rho}_\theta \hat{L}_j + \hat{L}_j \hat{\rho}_\theta}{2} \bigg]                                                                                                                                                                                 \\
		                                                      & = \frac{1}{2} \mathrm{Tr} [(\partial_i \hat{\rho}_\theta) \hat{L}_j + \hat{\rho}_\theta (\partial_i \hat{L}_j) + (\partial_i \hat{L}_j) \hat{\rho}_\theta + \hat{L}_j (\partial_i \hat{\rho}_\theta)]                                                                                      \\
		                                                      & = \frac{1}{2} \mathrm{Tr} \bigg[ \frac{\hat{\rho}_\theta \hat{L}_i + \hat{L}_i \hat{\rho}_\theta}{2} \hat{L}_j + \hat{\rho}_\theta (\partial_i \hat{L}_j)+ (\partial_i \hat{L}_j) \hat{\rho}_\theta + \hat{L}_j \frac{\hat{\rho}_\theta \hat{L}_i + \hat{L}_i \hat{\rho}_\theta}{2} \bigg] \\
		                                                      & = \mathrm{Tr} [\hat{\rho}_\theta (\partial_i \hat{L}_j + \partial_j \hat{L}_i)] + \mathrm{Tr} [\hat{\rho}_\theta (\hat{L}_i \hat{L}_j + \hat{L}_j \hat{L}_i)]. \label{main_eq_equality_second_derivative_trace_rho_001_001}
	\end{align}
\end{widetext}
From Eq.~\eqref{main_eq_equality_second_derivative_trace_rho_001_001} and $\partial_i \partial_j \mathrm{Tr} [\hat{\rho}_\theta] = 0$, the following equality holds:
\begin{align}
	\mathrm{Tr} [\hat{\rho}_\theta (\partial_i \hat{L}_j + \partial_j \hat{L}_i)] & = - \mathrm{Tr} [\hat{\rho}_\theta (\hat{L}_i \hat{L}_j + \hat{L}_j \hat{L}_i)]. \label{main_eq_equality_second_derivative_trace_rho_001_002}
\end{align}
From Eq.~\eqref{main_eq_equality_second_derivative_trace_rho_001_002}, we obtain
\begin{align}
	\mathrm{Tr} [\hat{C}_{i, j}] & = \frac{1}{16} \mathrm{Tr} [\hat{\rho}_\theta^\frac{1}{2} (2 \partial_i \hat{L}_j + 2 \partial_j \hat{L}_i + \hat{L}_i \hat{L}_j + \hat{L}_j \hat{L}_i) \hat{\rho}_\theta^\frac{1}{2}] \\
	                             & = \frac{1}{8} \mathrm{Tr} [\hat{\rho}_\theta (\partial_i \hat{L}_j + \partial_j \hat{L}_i)] + \frac{1}{16} \mathrm{Tr} [\hat{\rho}_\theta (\hat{L}_i \hat{L}_j + \hat{L}_j \hat{L}_i)] \\
	                             & = - \frac{1}{8} \mathrm{Tr} \bigg[ \hat{\rho}_\theta \bigg( \frac{\hat{L}_i \hat{L}_j + \hat{L}_j \hat{L}_i}{2} \bigg) \bigg]. \label{main_eq_equality_C_Bures_distance_001_001}
\end{align}
Substituting Eqs.~\eqref{main_eq_equality_B_Bures_distance_001_001} and \eqref{main_eq_equality_C_Bures_distance_001_001} into Eq.~\eqref{main_eq_def_Bures_distance_002_001}, we get
\begin{align}
	[D_\mathrm{Bures} (\hat{\rho}_\theta, \hat{\rho}_{\theta + \Delta \theta})]^2 & = - 2 \mathrm{Tr} [\hat{C}_{i, j}] \Delta \theta^i \Delta \theta^j                                                                                                    \\
	                                                                              & = \frac{1}{4} \mathrm{Tr} \bigg[ \hat{\rho}_\theta \bigg( \frac{\hat{L}_i \hat{L}_j + \hat{L}_j \hat{L}_i}{2} \bigg) \bigg] \Delta \theta^i \Delta \theta^j \nonumber \\
	                                                                              & \quad + + \mathcal{O} (\| \Delta \theta \|^3).
\end{align}
Thus, we obtain Eq.~\eqref{main_eq_relationship_Bures_distance_SLD_metric_001_001}.

\subsection{Proof of Theorem~\ref{main_theorem_maximum_minimum_elements_order_operator_monotone_functions_001_001}} \label{main_sec_theorem_maximum_minimum_elements_order_operator_monotone_functions_001_001}

To prove Theorem~\ref{main_theorem_maximum_minimum_elements_order_operator_monotone_functions_001_001}, the following lemma is useful:
\begin{lemma} \label{main_lemma_operator_monotonicity_t/f(t)_001_001}
	If $f (t)$ is operator-monotone, then $t / f (t)$ is also operator-monotone.
\end{lemma}
See Ref.~\cite{Hiai_007} for the proof of Lemma~\ref{main_lemma_operator_monotonicity_t/f(t)_001_001}.

The proof of Theorem~\ref{main_theorem_maximum_minimum_elements_order_operator_monotone_functions_001_001} is as follows:
\begin{proof}

	Let us consider a operator-monotone function $f (\cdot)$ such that $f (1) = 1$, and $f (t) = t f (t^{-1})$.
	With simple calculations, we have
	\begin{align}
		 & \frac{\mathrm{d}}{\mathrm{d}t} t f (t^{-1}) = f (t^{-1}) - \frac{1}{t} f' (t^{-1})             \\
		 & \Rightarrow \frac{\mathrm{d}}{\mathrm{d}t} t f (t^{-1}) \bigg|_{t = 1} = f (1) - f' (1)        \\
		 & \Rightarrow f' (1) = f (1) - f' (1). \label{main_eq_Petz_function_derivative_relation_001_001}
	\end{align}
	Combining Eqs.~\eqref{main_eq_condition_Petz_function_001_011} and \eqref{main_eq_Petz_function_derivative_relation_001_001}, we get
	\begin{align}
		f' (1) & = \frac{1}{2}.
	\end{align}
	Because of the concavity of operator-monotone functions, $f_\mathrm{SLD} (\cdot)$ is the maximum element among the operator-monotone functions such that $f (1) = 1$ and $f' (1) = 1/2$:
	\begin{align}
		f (\cdot) & \preceq f_\mathrm{SLD} (\cdot). \label{main_eq_maximum_minimum_elements_order_operator_monotone_functions_002_001}
	\end{align}

	To conduct the same argument for the minimum element of the operator-monotone functions such that $f (1) = 1$ and $f' (1) = 1/2$, we define
	\begin{align}
		\tilde{f} (t) & \coloneqq \frac{1}{f (t^{-1})}. \label{main_eq_def_g_maximum_minimum_elements_order_operator_monotone_functions_001_001}
	\end{align}
	Note that $\tilde{f} (t)$ in Eq.~\eqref{main_eq_def_g_maximum_minimum_elements_order_operator_monotone_functions_001_001} is operator-monotone due to $\tilde{f} (t) = t / f (t)$ and Lemma~\ref{main_lemma_operator_monotonicity_t/f(t)_001_001}.
	Because of Eq.~\eqref{main_eq_def_g_maximum_minimum_elements_order_operator_monotone_functions_001_001}, we have
	\begin{align}
		\tilde{f} (1) & = 1.
	\end{align}
	Similarly to Eq.~\eqref{main_eq_Petz_function_derivative_relation_001_001}, we get
	\begin{align}
		\frac{\mathrm{d}}{\mathrm{d}t} \tilde{f} (t) & = t^{-2} \frac{f' (t^{-1})}{f^2 (t^{-1})},
	\end{align}
	and
	\begin{align}
		\frac{\mathrm{d}}{\mathrm{d}t} \tilde{f} (t) \bigg|_{t = 1} & = f' (1)       \\
		                                                            & = \frac{1}{2}.
	\end{align}
	Eq.~\eqref{main_eq_def_g_maximum_minimum_elements_order_operator_monotone_functions_001_001} leads to
	\begin{align}
		\text{$\forall t \in \mathbb{R}_{\ge 0}$, $\tilde{f} (t) \le f_\mathrm{SLD} (t)$}.
	\end{align}
	Then, we have
	\begin{align}
		\text{$\forall t \in \mathbb{R}_{\ge 0}$, $f (t) \ge 1 / f_\mathrm{SLD} (t^{-1})$}.
	\end{align}
	Since $f_\mathrm{rRLD} (t) = 1 / f_\mathrm{SLD} (t^{-1})$, we have, when $f (\cdot)$ is an operator-monotone function such that $f (1) = 1$ and $f' (1) = 1/2$,
	\begin{align}
		f (\cdot) & \succeq f_\mathrm{rRLD} (\cdot).  \label{main_eq_maximum_minimum_elements_order_operator_monotone_functions_002_002}
	\end{align}
	Combining Eqs.~\eqref{main_eq_maximum_minimum_elements_order_operator_monotone_functions_002_001} and \eqref{main_eq_maximum_minimum_elements_order_operator_monotone_functions_002_002}, we get Eq.~\eqref{main_eq_maximum_minimum_elements_order_operator_monotone_functions_001_001}.
\end{proof}

\section{Additional numerical simulations} \label{main_sec_additional_numerical_simulations_001_001}

In Sec.~\ref{main_sec_numerical_simulations_001_001}, we have shown numerical simulations.
To support the result, we perform additional numerical simulations.

\subsection{Setup I}

As a parameterized quantum state, we assume
\begin{align}
	\hat{\rho}_\theta & \coloneqq \hat{U} (\theta) \hat{\rho}_\mathrm{ini} \hat{U}^\dagger (\theta),
\end{align}
where
\begin{align}
	\hat{\rho}_\mathrm{ini} & \coloneqq \hat{\rho}_1 \otimes \hat{\rho}_2,
\end{align}
and
\begin{align}
	\hat{U} (\theta) & \coloneqq
	\begin{tikzcd}[row sep = 0.2cm, column sep = 0.2cm]
		& \gate{\hat{R}_3 (\theta^1)} & \ctrl{1}  & \targ \qw & \qw \\
		& \gate{\hat{R}_3 (\theta^2)} & \targ \qw & \ctrl{-1} & \qw
	\end{tikzcd},
\end{align}
where $\theta \coloneqq [\theta^1, \theta^2]^\intercal$, and $\theta^i = [\theta^{i, 1}, \theta^{i, 2}, \theta^{i, 3}]^\intercal$ for $i = 1, 2$.
Then, we consider the following cost function:
\begin{align}
	\min_{\theta \in \mathbb{R}^{N_\theta}} L (\theta),
\end{align}
where
\begin{align}
	L (\theta) & \coloneqq \mathrm{Tr} [\hat{\rho}_\theta (\hat{\sigma}_1^z + 0.1 \hat{\sigma}_1^x \hat{\sigma}_2^x)].
\end{align}

\subsection{Results I}

\begin{figure}[t]
	\centering
	\includegraphics[scale=0.60]{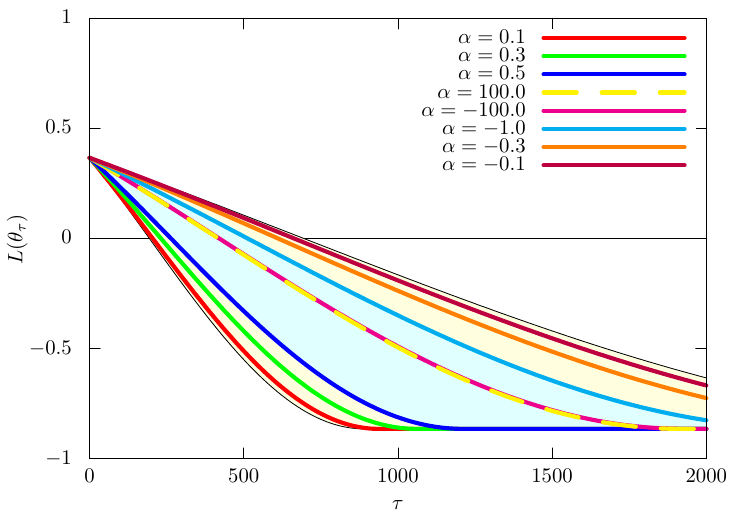}
	\includegraphics[scale=0.60]{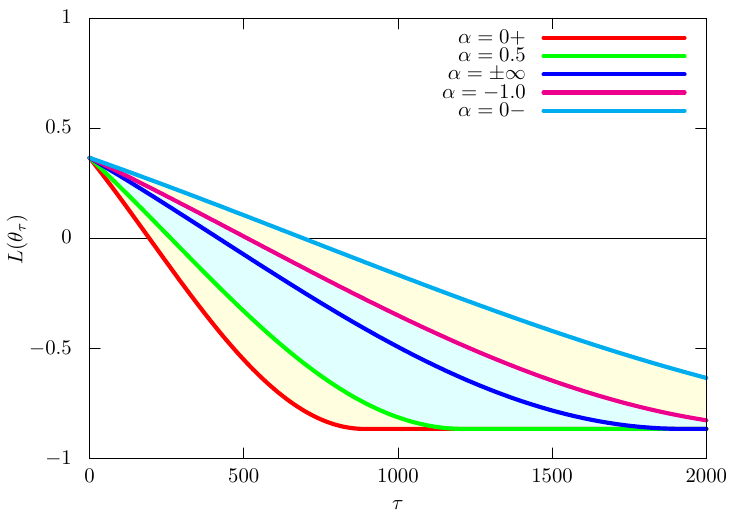}
	\caption{Cost functions in the case of $f_\alpha^\mathrm{sw} (t)$, Eq.~\eqref{main_eq_def_f_alpha_001_001}, for several $\alpha$ for the case of Eq.~\eqref{main_eq_update_theta_001_001}. We set $\epsilon = 1.0 \times 10^{-6}$, $\xi = 1.0 \times 10^{-3}$, and $\delta = 1.0 \times 10^{-3}$. The regimes of the monotone metrics and the rescaled sandwiched quantum R\'enyi divergence are highlighted by light cyan and light yellow, respectively.}
	\label{main_fig_performance_QNG_sandwiched_rot-states_002_001}
\end{figure}

\begin{figure}[t]
	\centering
	\includegraphics[scale=0.60]{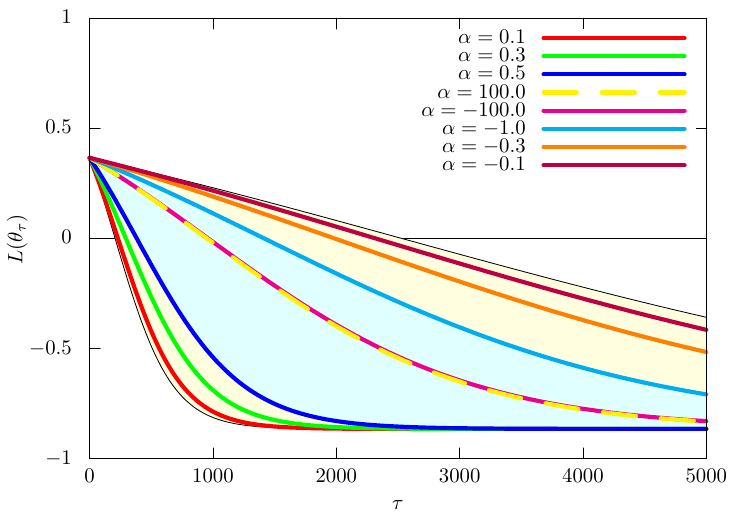}
	\includegraphics[scale=0.60]{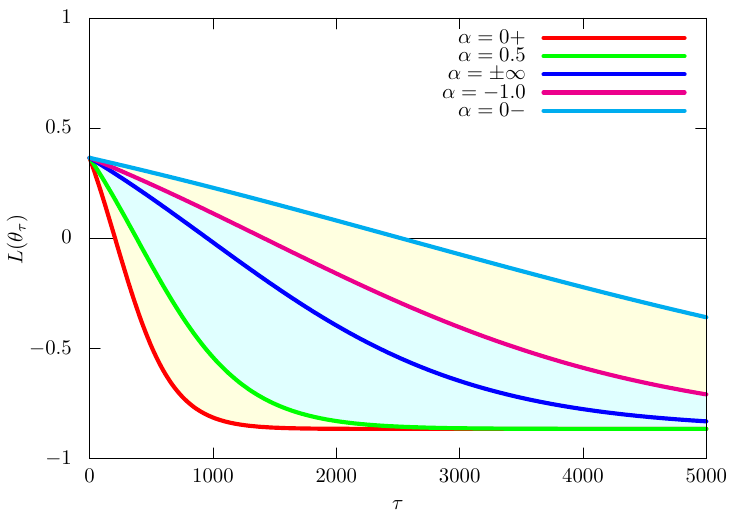}
	\caption{Cost functions in the case of $f_\alpha^\mathrm{sw} (t)$, Eq.~\eqref{main_eq_def_f_alpha_001_001}, for several $\alpha$ for the case of Eq.~\eqref{main_eq_update_theta_001_002}. We set $\eta = 1.0 \times 10^{-3}$, $\xi = 1.0 \times 10^{-3}$, and $\delta = 1.0 \times 10^{-3}$. The regimes of the monotone metrics and the rescaled sandwiched quantum R\'enyi divergence are highlighted by light cyan and light yellow, respectively.}
	\label{main_fig_performance_QNG_sandwiched_rot-states_002_002}
\end{figure}

First, we consider the Petz function associated with the rescaled sandwiched quantum R\'enyi divergence, Eq.~\eqref{main_eq_def_f_alpha_001_001}.
In Fig.~\ref{main_fig_performance_QNG_sandwiched_rot-states_002_001}, we plot the time evolution of the cost function for the case of Eq.~\eqref{main_eq_update_theta_001_001}.
In Fig.~\ref{main_fig_performance_QNG_sandwiched_rot-states_002_002}, we show the time evolution of the cost function for the case of Eq.~\eqref{main_eq_update_theta_001_002}.

\begin{figure}[t]
	\centering
	\includegraphics[scale=0.60]{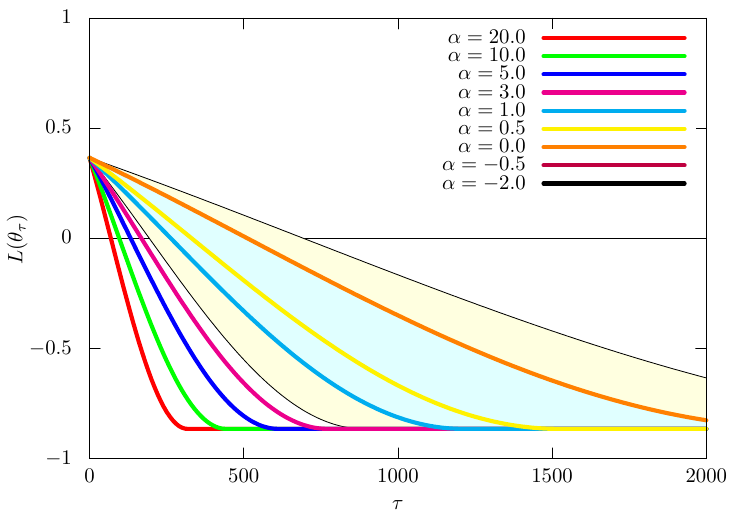}
	\caption{Cost functions in the case of $f_\alpha^\mathrm{lin} (t)$, Eq.~\eqref{main_eq_def_combined_Petz_function_001_001}, for several $\alpha$ for the case of Eq.~\eqref{main_eq_update_theta_001_001}. We set $\epsilon = 1.0 \times 10^{-6}$, $\xi = 1.0 \times 10^{-3}$, and $\delta = 1.0 \times 10^{-3}$. The regimes of the monotone metrics and the rescaled sandwiched quantum R\'enyi divergence are highlighted by light cyan and light yellow, respectively.}
	\label{main_fig_performance_QNG_linear_rot-states_002_001}
\end{figure}

\begin{figure}[t]
	\centering
	\includegraphics[scale=0.60]{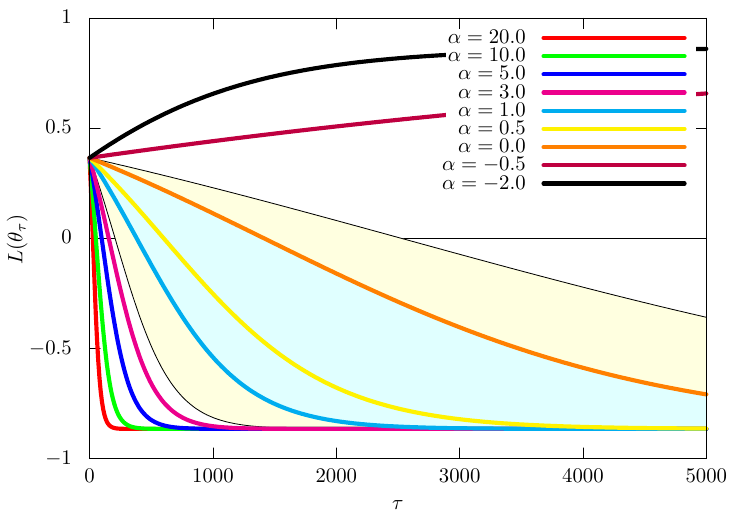}
	\caption{Cost functions in the case of $f_\alpha^\mathrm{lin} (t)$, Eq.~\eqref{main_eq_def_combined_Petz_function_001_001}, for several $\alpha$ for the case of Eq.~\eqref{main_eq_update_theta_001_002}. We set $\eta = 1.0 \times 10^{-3}$, $\xi = 1.0 \times 10^{-3}$, and $\delta = 1.0 \times 10^{-3}$. The regimes of the monotone metrics and the rescaled sandwiched quantum R\'enyi divergence are highlighted by light cyan and light yellow, respectively.}
	\label{main_fig_performance_QNG_linear_rot-states_002_002}
\end{figure}

Next, we investigate Eq.~\eqref{main_eq_def_combined_Petz_function_001_001} for the Petz function that defines the quantum Fisher metric.
In Fig.~\ref{main_fig_performance_QNG_linear_rot-states_002_001}, we plot the time evolution of the cost function for the case of Eq.~\eqref{main_eq_update_theta_001_001}.
In Fig.~\ref{main_fig_performance_QNG_linear_rot-states_002_002}, we show the time evolution of the cost function for the case of Eq.~\eqref{main_eq_update_theta_001_002}.

\subsection{Setup II}

Here a parameterized quantum state is assumed:
\begin{align}
	\hat{\rho}_\theta & \coloneqq \hat{U} (\theta) \hat{\rho}_\mathrm{ini} \hat{U}^\dagger (\theta).
\end{align}
where
\begin{align}
	\hat{\rho}_\mathrm{ini} & \coloneqq \hat{\rho}_1 \otimes \hat{\rho}_2 \otimes \hat{\rho}_3,
\end{align}
and
\begin{align}
	\hat{U} (\theta) & \coloneqq
	\begin{tikzcd}[row sep = 0.2cm, column sep = 0.2cm]
		& \gate{\hat{R}_3 (\theta^1)} & \ctrl{1}  & \qw          & \targ \qw     & \qw \\
		& \gate{\hat{R}_3 (\theta^2)} & \targ \qw & \ctrl{1} \qw & \qw           & \qw \\
		& \gate{\hat{R}_3 (\theta^3)} & \qw       & \targ \qw    & \ctrl{-2} \qw & \qw
	\end{tikzcd},
\end{align}
where $\theta \coloneqq [\theta^1, \theta^2, \theta^3]^\intercal$, and $\theta^i = [\theta^{i, 1}, \theta^{i, 2}, \theta^{i, 3}]^\intercal$ for $i = 1, 2, 3$.

We consider the following cost function:
\begin{align}
	\min_{\theta \in \mathbb{R}^{N_\theta}} L (\theta),
\end{align}
where
\begin{align}
	L (\theta)          & \coloneqq \mathrm{Tr} [\hat{\rho}_\theta \hat{H}_{J, \omega}],                                          \\
	\hat{H}_{J, \omega} & \coloneqq \sum_{i=1}^3 (\omega \hat{\sigma}_i^z + J \hat{\bm{\sigma}}_i \cdot \hat{\bm{\sigma}}_{i+1}).
\end{align}
Furthermore, we define, for $i = 1, 2, 3$,
\begin{align}
	\hat{\bm{\sigma}}_i & \coloneqq [\hat{\sigma}_i^x, \hat{\sigma}_i^y, \hat{\sigma}_i^z]^\intercal,
\end{align}
and
\begin{align}
	\hat{\bm{\sigma}}_4 & \coloneqq \hat{\bm{\sigma}}_1.
\end{align}

\subsection{Results II}

\begin{figure}[t]
	\centering
	\includegraphics[scale=0.60]{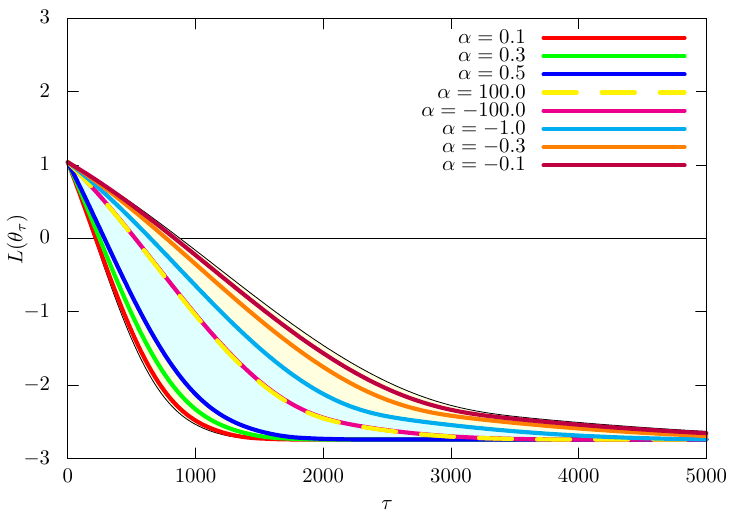}
	\includegraphics[scale=0.60]{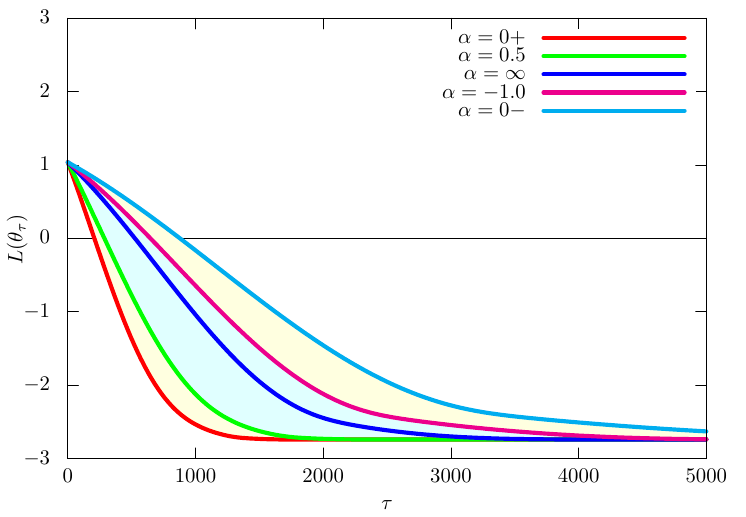}
	\caption{Cost functions in the case of $f_\alpha^\mathrm{sw} (t)$, Eq.~\eqref{main_eq_def_f_alpha_001_001}, for several $\alpha$ for the case of Eq.~\eqref{main_eq_update_theta_001_001}. We set $\epsilon = 1.0 \times 10^{-6}$, $\xi = 1.0 \times 10^{-3}$, and $\delta = 1.0 \times 10^{-3}$. The regimes of the monotone metrics and the rescaled sandwiched quantum R\'enyi divergence are highlighted by light cyan and light yellow, respectively.}
	\label{main_fig_performance_QNG_sandwiched_rot-states_003_001}
\end{figure}

\begin{figure}[t]
	\centering
	\includegraphics[scale=0.60]{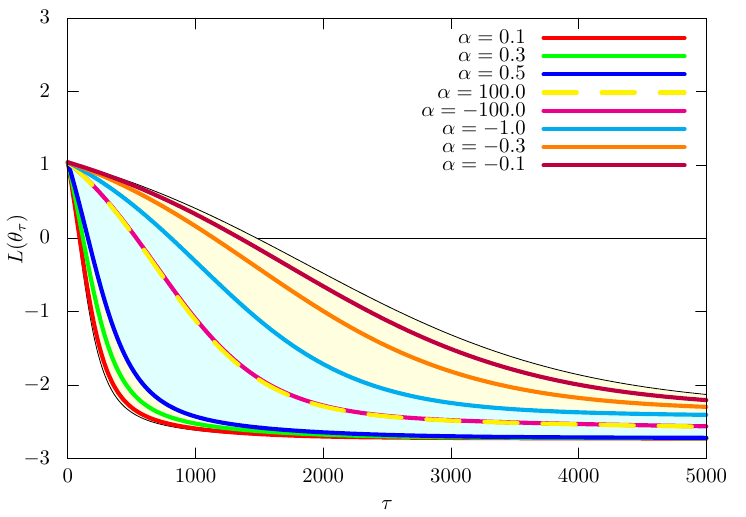}
	\includegraphics[scale=0.60]{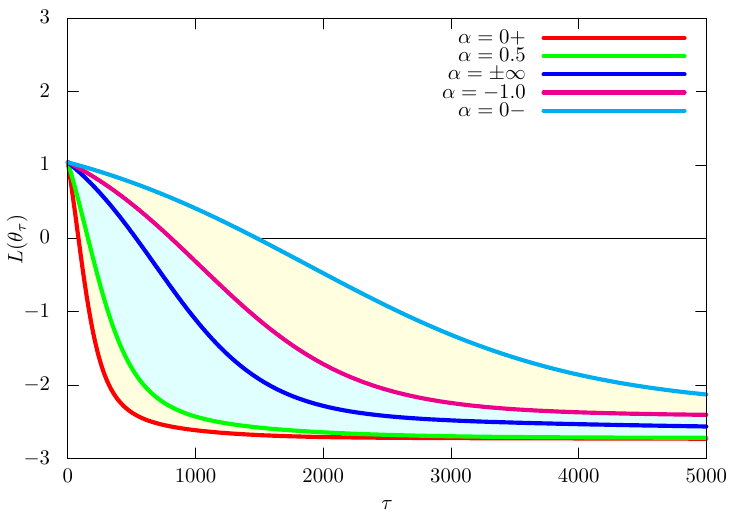}
	\caption{Cost functions in the case of $f_\alpha^\mathrm{sw} (t)$, Eq.~\eqref{main_eq_def_f_alpha_001_001}, for several $\alpha$ for the case of Eq.~\eqref{main_eq_update_theta_001_002}. We set $\eta = 1.0 \times 10^{-3}$, $\xi = 1.0 \times 10^{-3}$, and $\delta = 1.0 \times 10^{-3}$. The regimes of the monotone metrics and the rescaled sandwiched quantum R\'enyi divergence are highlighted by light cyan and light yellow, respectively.}
	\label{main_fig_performance_QNG_sandwiched_rot-states_003_002}
\end{figure}

First, we consider the Petz function associated with the rescaled sandwiched quantum R\'enyi divergence, Eq.~\eqref{main_eq_def_f_alpha_001_001}.
In Fig.~\ref{main_fig_performance_QNG_sandwiched_rot-states_003_001}, we plot the time evolution of the cost function for the case of Eq.~\eqref{main_eq_update_theta_001_001}.
In Fig.~\ref{main_fig_performance_QNG_sandwiched_rot-states_003_002}, we show the time evolution of the cost function for the case of Eq.~\eqref{main_eq_update_theta_001_002}.

\begin{figure}[t]
	\centering
	\includegraphics[scale=0.60]{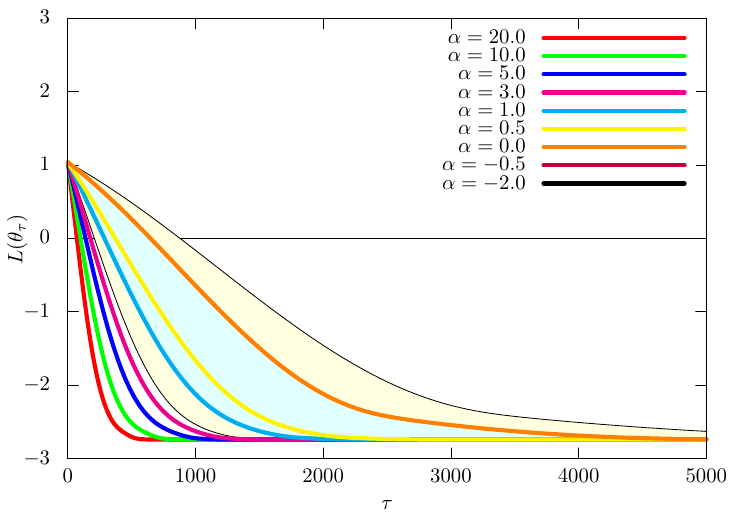}
	\caption{Cost functions in the case of $f_\alpha^\mathrm{lin} (t)$, Eq.~\eqref{main_eq_def_combined_Petz_function_001_001}, for several $\alpha$ for the case of Eq.~\eqref{main_eq_update_theta_001_001}. We set $\epsilon = 1.0 \times 10^{-6}$, $\xi = 1.0 \times 10^{-3}$, and $\delta = 1.0 \times 10^{-3}$. The regimes of the monotone metrics and the rescaled sandwiched quantum R\'enyi divergence are highlighted by light cyan and light yellow, respectively.}
	\label{main_fig_performance_QNG_linear_rot-states_003_001}
\end{figure}

\begin{figure}[t]
	\centering
	\includegraphics[scale=0.60]{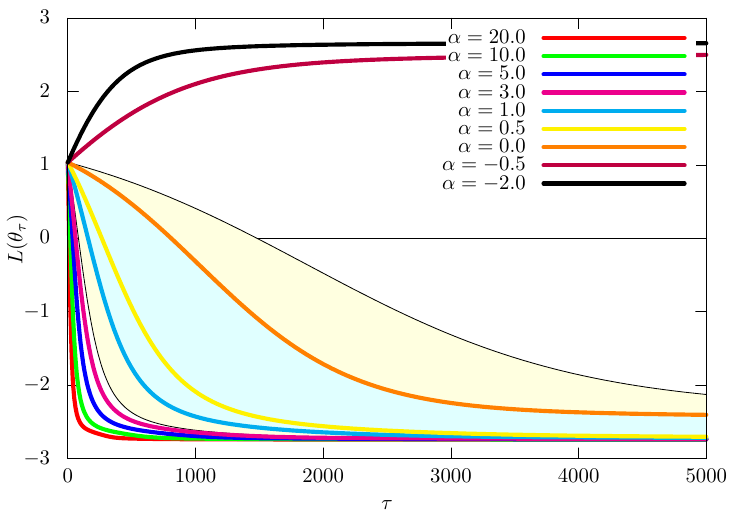}
	\caption{Cost functions in the case of $f_\alpha^\mathrm{lin} (t)$, Eq.~\eqref{main_eq_def_combined_Petz_function_001_001}, for several $\alpha$ for the case of Eq.~\eqref{main_eq_update_theta_001_002}. We set $\eta = 1.0 \times 10^{-3}$, $\xi = 1.0 \times 10^{-3}$, and $\delta = 1.0 \times 10^{-3}$. The regimes of the monotone metrics and the rescaled sandwiched quantum R\'enyi divergence are highlighted by light cyan and light yellow, respectively.}
	\label{main_fig_performance_QNG_linear_rot-states_003_002}
\end{figure}

Next, we investigate Eq.~\eqref{main_eq_def_combined_Petz_function_001_001} for the Petz function that defines the quantum Fisher metric.
In Fig.~\ref{main_fig_performance_QNG_linear_rot-states_003_001}, we plot the time evolution of the cost function for the case of Eq.~\eqref{main_eq_update_theta_001_001}.
In Fig.~\ref{main_fig_performance_QNG_linear_rot-states_003_002}, we show the time evolution of the cost function for the case of Eq.~\eqref{main_eq_update_theta_001_002}.

\bibliography{paper_IG_QNG_999_001}

\end{document}